\documentclass{jfm}
\usepackage{graphicx}
\usepackage{amsmath}
\usepackage[version=4]{mhchem}
\usepackage{siunitx}
\usepackage{longtable,tabularx}
\usepackage{color,soul}
\usepackage{nomencl}
\usepackage{subfigmat} 
\usepackage{bm}
\usepackage{relsize}
\usepackage{tablefootnote}
\usepackage{float}
\usepackage{epstopdf, epsfig}
\usepackage[dvipsnames]{xcolor}

\AppendGraphicsExtensions{.tiff}

\newcommand\figref[1]{Figure \ref{fig:#1}} 
\newcommand\eref[1]{Eq. (\ref{eq:#1})} 

\title{Turbulent boundary layer with strong favorable pressure gradient and curvature effects: Streamline coordinate and scaling analysis}

\author{Aviral Prakash\aff{1}
  \corresp{\email{aviral.prakash@colorado.edu}},
  Riccardo Balin\aff{2},
  John A. Evans\aff{1}
 \and Kenneth E. Jansen\aff{1}}

\affiliation{\aff{1} University of Colorado Boulder, Boulder, CO 80309, USA, \aff{2} Argonne National Laboratory, Lemont, IL 60439, USA}

\begin{document}

\maketitle

\begin{abstract}

Direct numerical simulation (DNS) of a turbulent boundary layer over the Gaussian (Boeing) bump is performed. This boundary layer exhibits a series of adverse and favorable pressure gradients and convex and concave curvature effects before separating. These effects on turbulent boundary layers are characterized and compared to a lower Reynolds number flow over the same geometry. The momentum budgets are analyzed in the streamline-aligned coordinate system upstream of the separation region. These momentum budgets allow the simplification of equations to facilitate an integral analysis. Integral analysis-based scalings for Reynolds stresses in the inner and outer regions of the boundary layer are also formulated. These proposed scalings exhibit a better collapse of Reynolds stress profiles compared to friction velocity scaling and Zagarola-Smits scaling in the strong favorable pressure gradient region and in the mild adverse pressure region that precedes it in this flow.

\end{abstract}


\section{Introduction}

Turbulent boundary layers are ubiquitous in engineering applications and geophysical flows of interest. A better understanding of turbulent 
boundary layers will allow for the design of more efficient airplanes, improved wind turbine performance characterization, and a better 
understanding of the atmospheric weather patterns that could subsequently inform the prediction and mitigation of wildfire propagation to list only a few examples. These 
flows have been studied intensely over the past century by both the engineering and geophysical communities. These studies have focused on 
characterizing and quantifying the behavior of turbulence in the inner and outer regions of the turbulent boundary layers, which has also led to 
advancements in turbulence models over the years. 

The presence of pressure gradients significantly affects the behavior of a turbulent boundary layer. Adverse pressure gradients (APGs) often precede the onset of turbulent flow separation, which is a major cause of higher aerodynamic drag for wings at high angles of attack. In the presence of a sufficiently strong APG, a flow exhibits an increase in Reynolds shear stresses, often observed as a secondary peak of these quantities in the outer region of the boundary layer. Using the regional similarity hypothesis, \cite{Perry1966} showed that the logarithmic law of the wall for the velocity distribution, hereafter referred to as the log-law, exists in the near-wall region, and a half-power law is observed in the transition to the wake region for a boundary layer that is far away from flow separation. These results are also aligned with the asymptotic analysis in \cite{Durbin1992}, where the role of APGs on a larger wake region was discussed. \cite{Marusic1995} showed that the turbulence intensity in the outer region increases in the presence of an APG when scaled by friction velocity, $u_{\tau}$. APGs also result in higher turbulence production due to high Reynolds shear stress and dissipation in the outer region of the turbulent boundary layer \citep{Skare1994}. Conversely, favorable 
pressure gradients (FPGs) stabilize a boundary layer and reduce turbulence intensity. In the presence of a FPG, a departure from the log-law is commonly observed \citep{Tsuji1976, Baskaran1987} which can be attributed to an increase in the thickness of the viscous wall region \citep{Narayanan1969, Blackwelder1972}. A reduction in Reynolds stresses, turbulent kinetic energy production, and dissipation is commonly observed in the presence of a FPG. A much more significant decrease in production and dissipation of turbulence is observed in the inner 
region of the boundary layer than in the outer region \citep{Bourassa2009}. Strong FPGs can even lead to a 
reverse transition of a turbulent boundary layer to a laminar state, a process often referred to as relaminarization. \cite{Patel1968} showed that in the presence of a substantial FPG, even a high Reynolds number turbulent boundary layer could relaminarize. This flow behavior is attributed to a slow response of the boundary layer to the 
strong FPG. During flow relaminarization, turbulence still exists in the outer region but has a passive influence on the downstream development of the boundary layer \citep{Launder1964, Jones1972} as the pressure forces dominate over Reynolds shear stress \citep{Narasimha1973}. A sequenceof events leading to relaminarization is well documented in \cite{Piomelli2013}. A turbulent boundary layer in the presence of pressure gradients also exhibits dependence on the flow history \citep{Bobke2017}. Furthermore, a shift from an APG to a FPG or from a FPG to an APG could trigger the formation of an internal boundary layer. The growth of these new boundary layers is dictated by the pressure gradient \citep{Tsuji1976, Baskaran1987}. The formation of an internal layer leads to the decoupling of the external boundary layer that behaves as a free-shear 
influenced by the local pressure gradients \citep{Baskaran1987, Balin_Jansen_2021}. More details on the behavior of turbulent boundary layers can be found in the recent review article \citep{Devenport2022}.

The curvature of streamlines also has a considerable influence on the physics of a turbulent boundary layer. Curvature effects are often tied to the curvature of the geometry over which the turbulent boundary develops, leading to the pressure gradient effects. Therefore, pressure gradient and curvature effects often accompany each other. There have been efforts where pressure gradients \citep{Spalart1997, Coleman2018} and curvature effects \citep{So1973, So1975} were isolated. Convex curvature has a stabilizing influence on the turbulent boundary layer, whereas concave curvature has a destabilizing influence on the turbulent boundary layer. \citep{Irwin1975} showed that the effect of curvature on the turbulence structure can be greater than the effect on the mean flow. This is reflected by the observation that the deviation of the velocity profile for a low curvature value ($\delta/R \approx 0.01$ where $\delta$ is the boundary layer thickness and $R$ is the radius of curvature) occurs after the log-law region \citep{Hunt1979}. Concave curvature results in increased turbulence intensity, Reynolds shear stresses, and turbulent kinetic energy in the outer region of the boundary layer \citep{So1975}. On the other hand, convex curvature reduces these quantities in the outer region of the boundary. A discontinuity in the surface curvature from concave to convex or vice-versa has been shown to trigger the formation of an internal layer \citep{Baskaran1987, Webster1996} which exhibits behavior similar to the internal layer formed due to a change in sign of pressure gradient. 

In this article, we assess the scaling of Reynolds stresses in both inner and outer regions for the turbulent boundary layer over the Gaussian (Boeing) bump \citep{Slotnick_SpeedBump_2019} at a Reynolds number ($Re_L$) of $2$ million in the pre-separation region. Several experimental \citep{Williams_ExpSpeedBump, Gray_ExpSpeedBump} and DNS campaigns \citep{Balin_Jansen_2021,Shur_SpeedBumpDNS,Uzun2021b} are underway to characterize and quantify the behavior of this turbulent boundary layer. The turbulent boundary layer over the bump exhibits a series of APGs and FPGs in conjunction with concave and convex curvature effects. The switch from APG to FPG triggers the formation of an internal layer \citep{Balin_Jansen_2021, Uzun2021b}. At $Re_L = 1$ million, the flow also experiences relaminarization \citep{Balin_Jansen_2021}. With this wide range of flow physics experienced by this turbulent boundary layer, it is a challenging test case for evaluating existing analysis and scaling methods. The goals of this article are: 1) provide new reference DNS data that the community could use to evaluate the performance of turbulence models, 2) assess the validity of commonly used velocity and Reynolds stress scaling and 3) propose new scalings for Reynolds shear stress and wall-normal normal stress based on integration of the mean momentum equation. Note that as in \cite{Uzun2021b}, a DNS at $Re_L = 2$ million was performed; we comment on the differences in the setup of the two DNS and extrapolate this to the observed differences in the results. An outline of the article is as follows. Section \ref{sec:Setup} describes the flow over the bump and details the simulation setup. In Section \ref{sec:Results_Quant}, we quantify the flow behavior and compare the results to those in \cite{Uzun2021b} and the $Re_L = 1$ million case in \cite{Balin_Jansen_2021}. In Section \ref{sec:SCSAnalysis}, we describe the streamline-aligned coordinate system and analyze the momentum budget statistics in this coordinate system. In Section \ref{sec:IntAnalysis}, we perform an integral analysis of the momentum equation to propose new scalings for Reynolds wall-normal normal and shear stresses in the inner and outer regions of the boundary layer. Finally, in Section \ref{sec:Conclusion}, we provide concluding remarks and avenues for future research.

\section{Simulation Setup}

\label{sec:Setup}
\subsection{Flow Description}
\label{sec:Setup_parta}

This work considers a prismatic extrusion of the Boeing bump, a Gaussian-shaped bump defined by \eqref{eq:GaussBump}.
\begin{equation}
y(x)=h\exp{\Big(-\big(x/x_0 \big)^2 \Big)} .
\label{eq:GaussBump}
\end{equation}
In this equation, the $x$ coordinate is horizontal and aligned with the freestream flow far upstream of the bump, the $y$ coordinate is normal to the freestream, $h$ is a parameter controlling the bump height, and $x_0$ controls the bump length.
The curve in \eref{GaussBump} is exactly the centerline of a three-dimensional (3D) bump developed at The Boeing Company \citep{Slotnick_SpeedBump_2019} and studied experimentally at the University of Washington \citep{Williams_ExpSpeedBump} and the University of Notre Dame \citep{Gray_ExpSpeedBump}.
To maintain similarities between the 3D and two-dimensional (2D) extruded bumps, the height and length parameters are matched at $h/L=0.085$ and $x_0/L=0.195$, where $L=3$ ft is the length of the square cross-section of the wind tunnel used for the 3D bump experiments.
With these values set, the bump 2D profile is precisely the one employed in DNS in \citet{Balin_Jansen_2021}, \citet{Shur_SpeedBumpDNS}, and \citet{Uzun2021b}, although either the Reynolds number or the full domain geometry herein is different from these previous studies.

\begin{figure}
\centering
  \includegraphics[width = 0.97\textwidth]{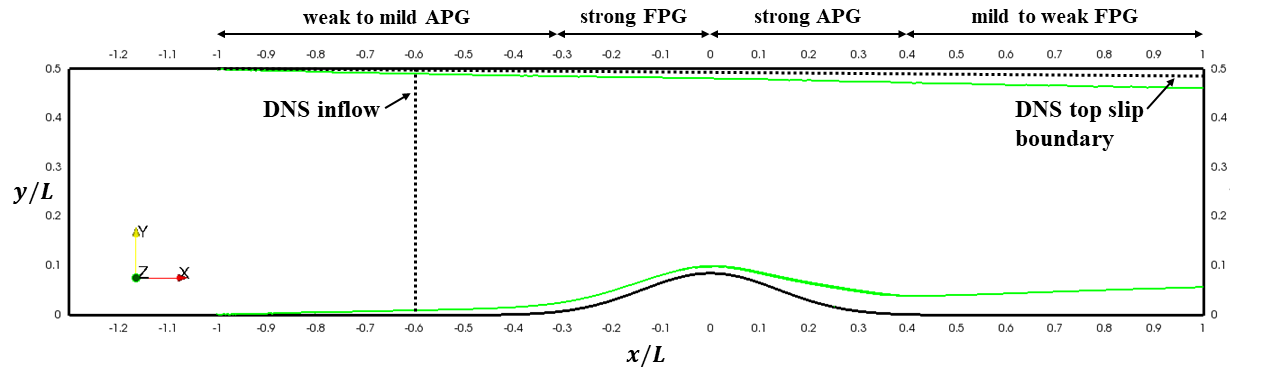}
  \caption{The solid black curves outline the full domain of the bump flow, while the green curves show the boundary layer thickness on both no-slip walls predicted by preliminary RANS. The black dotted lines mark the modified inflow and top boundaries used in the DNS.}
\label{fig:GaussBump}
\end{figure}

The setup of the flow problem and the remainder of the computational domain follow the DNS of \citet{Balin_Jansen_2021}, although updated for the twice larger Reynolds number, and are described by the schematic in \figref{GaussBump}.
Based on the freestream velocity of $U_\infty=32.80$ m/s, the flow has a Reynolds number of $Re_L=2.0 \times 10^6$, corresponding to $Re_h=170,000$ when measured against the bump height. Additionally, the flow was treated as incompressible due to the small Mach number of $M_\infty=0.09$ (computed using standard sea level conditions).
At the location of the inlet to the DNS, shown by the dotted vertical line in \figref{GaussBump}, the momentum thickness Reynolds number is approximately $Re_\theta=1,800$, and the boundary layer thickness is roughly $1/9$ of the bump height. 

While the solid curves in \figref{GaussBump} describe the entire flow domain computed with preliminary RANS, which includes the boundary layer origins on the bottom and top walls at $x/L=-1.0$, only a fraction of that domain can be feasibly computed by DNS. As outlined in detail in \citet{Balin_Jansen_2021}, the inflow to the DNS domain is moved downstream to $x/L=-0.6$, and the top boundary is slanted downwards according to a profile fitted to the displacement thickness of the RANS boundary layer on the top wall. The last modification was done to reproduce the constriction effects of the boundary layer growing on the top wall without resolving it.
Note that the RANS computations of the full domain were carried out using the Spalart-Allmaras (SA) one-equation model \citep{spalart1994one} augmented with the rotation and streamline curvature (SARC) correction \citep{Spalart_SARC,Shur_SARC} and the low-Reynolds number correction \citep{Spalart_lowRe}.

The following boundary conditions were implemented based on the DNS domain shown in \figref{GaussBump}. The bump surface was treated as a no-slip wall. It is worth mentioning that \eqref{eq:GaussBump} defines the entire lower surface of the domain, meaning that there is no flat-plate region on either side of the bump, and the curvature is continuous everywhere. The top surface was modeled as an inviscid wall with zero velocity component normal to the surface and zero tangential traction. At the outflow, total traction is weakly enforced to zero.  In addition, periodic boundary conditions were applied on a spanwise width of  $0.156L$. Finally, at the inflow, the synthetic turbulence generator (STG) of \citet{Shur_STG,pattersonAssessingImprovingAccuracy2021} was utilized to introduce unsteady flow into the domain and rapidly produce mature and realistic turbulence. This STG method was successfully used in the $Re_L=1$M DNS of \citet{Balin_Jansen_2021} and the mean velocity and stress profiles required for the method were obtained following the description therein.

\subsection{Flow Solver Description}

The simulations of the $Re_L=2$M Gaussian bump discussed herein were computed using a stabilized finite element method \citep{Whiting1999} and second-order accurate, fully implicit generalized-$\alpha$ time integration \citep{Jansen_GenAlpha_2000} to perform time-resolved simulations of the incompressible Navier-Stokes equations. The DNS was carried out using linear mesh elements to reduce the computational cost while still maintaining a high level of accuracy, as demonstrated for a channel flow in \citet{Trofimova_DNS_2009}, for a flat plate boundary layer in \citet{WrightSTGDNS_arxiv}, and for the Gaussian Bump flow at $Re_L=1$M in \citet{Balin_Jansen_2021}.
Furthermore, stabilization and time integration parameters chosen for this DNS follow the work outlined in \citet{Trofimova_DNS_2009} and were similar to those used in \citet{Balin_Jansen_2021}. The simulations were started from an initial condition with mean velocity and pressure from RANS and superposed fluctuations obtained from DNS at $Re_L = 1$ million. The initial transient part of the simulation was performed for a half-span domain first, and then the solution was transferred to the full domain using one instantaneous time step for half of the domain and another time step for the other half of the domain.  The second time step was chosen such that its re-attachment structures were the most  out-of-phase from the first to break any domain-width lock in present in the flow.  This full domain width initial condition necessarily breaks continuity on the center line and on the periodic planes but otherwise satisfies the governing equations and the flow solver quickly repaired these breaks.  More significant than the time to repair these continuity breaks, the  separation point and the re-attachment point did show an adjustment to the new domain width and this must be considered part of the transient. Then, after this transient on the full domain, statistics were collected. While DNS and LES articles often discuss statistics collected in terms of flow through times, this is of less relevance to the bump flow considered for two reasons.  First, the flow is not periodic and also has no global circulation to come to equilibrium and is thus fully determined by the inflow which, via the STG boundary condition, is statistically stationary on a time scale minimized through the choice of random numbers \citep{pattersonAssessingImprovingAccuracy2021}. Second, the more relevant time scale to turbulent boundary layers is the eddy turnover time $\delta(s) / u_e(s)$ which, in this flow varies by a factor of 10 between the most rapid turbulence at the bump peak and the slowest time scale in the thick boundary layer within the recovery region.   As this paper is concerned only with the region before separation, the range of eddy turnover times  is less dramatic (a factor of 2 between the peak of the weak APG and the bump apex).  At the peak of the weak APG, the flow requires about 600 time steps per eddy turnover time.  For the plots in this article, we have used statistics windows of at least 32,000 (most used 159,000) time steps which means that this slowest region averages at least 53.33 eddy turnover times. Note further that, to properly model the separated flow region, the domain width is set based on the boundary layer thickness there.  In the region of interest to this paper, the boundary layer thickness reaches a local maximum at the peak of the weak APG at a value of 0.015. Thus our domain width is more than 10 local boundary layer thicknesses wide within the domain discussed in this paper.  Because we are employing spanwise averaging in our statistics, this means each time step carries roughly 3 times the statistical power of a typical DNS that has a 3 boundary layer thickness domain width.  Taken together, this gives at least 160 eddy turnover times in the statistics shown which is well above that of a typical DNS \citep{Spalart1988DirectSO}. Most of the results shown used 159,000 time steps and thus at least 795 eddy turnover times.  This assessment of adequate statistics and sufficient passing of all transients was confirmed by comparing all plots shown on two half-time windows to visually confirm the convergence of statistics.

\subsection{Mesh Description}
\label{sec:Setup_partb}
Considering the differences in mesh requirements relative to our simulation at $Re_L = 1$M \citep{Balin_Jansen_2021}, the mesh for the simulation at $Re_L = 2$M requires an even higher resolution in the near wall region and a much larger volume of the domain with turbulent structures due to a much thicker boundary layer downstream of separation. These factors raise the number of grid points required, especially if structured grids are used.  To address this challenge, a new unstructured grid mesh generation technique was developed to
locally match the grid size to a fixed multiple of the Kolmogorov length scale (e.g., $\Delta=2\eta$).  This approach is regularly carried out for simulations of isotropic turbulence where the isotropic grid size is selected to match the {\it a priori} known Kolmogorov length scale that a given forcing and Reynolds number will produce.

However, the situation differs for a complex flow like the bump. Our approach was to make use of prior flat plate boundary layer DNS \citep{Spalart1988DirectSO}
which established a relationship $\eta^+=f(n^+)$ where $n$ is the wall-normal direction making use of predictions of dissipation ($\epsilon$) profiles in the wall-normal coordinate. This idea is also not completely new as it was applied to pipe and channel flows by \cite{PIROZZOLI2021110408} independently and concurrently with this effort. The application to boundary layers has a greater opportunity for savings by exploiting the larger growth of $\eta$ in the outer part of the boundary layer. Even larger gains come from the large variation in wall shear stress in this flow, unlike the channels and pipes studied in \cite{PIROZZOLI2021110408}. Here, with 
this wall-normal variation in dissipation (and thus $\eta$) known, we are able to first set the $\Delta_n$ spacing for any stream-wise location $s$ so long as an estimate for $u_\tau$ is available for each $s$ location. We make use of RANS simulations to obtain this estimate. While the RANS prediction of $u_\tau$ is known to have some error, it is usually on the high side, which is conservative.  

While the $n$ point distribution optimization described above yields significant savings, significantly more savings can be obtained in the streamwise ($s$) and spanwise ($z$) spacing.  These savings are only available to unstructured grids since a structured grid must satisfy the most strict grid resolution in each direction and propagate that spacing 
through an $ijk$ grid. Some relief from this constraint is possible for overset grids but this choice faces challenges presented by large jumps in grid size across
overset grid interfaces. For this simulation, we have developed and applied mesh generation techniques that 
smoothly coarsen not only the $n$ direction but also $s$ and $z$. In practice, we employ a wall grid that uses a 
local stream-wise spacing of 15 plus units ($\Delta_s^+=u_\tau(s) \Delta_s(s) /\nu=15$ at the wall) and a local spanwise spacing of 6 plus units ($\Delta_z^+=u_\tau(s) \Delta_z(s) /\nu=6$ at the wall) which was shown to be adequate in \citet{Balin_Jansen_2021}.
This requires a triangulated surface mesh to satisfy and realize a growth in spanwise spacing where $u_\tau$ is lower (fewer points in the span).  As the first point in the normal
direction off of the wall is very small ($\Delta^+_n=u_\tau(s) \Delta_n(s) /\nu=0.3$ at the wall) to
resolve the high gradients there, wedge elements are used to extrude this wall resolution (starting with aspect ratios of 50:1:20 for $(\Delta_s:\Delta_n:\Delta_z)/\Delta_n$) ) 
normal to the surface.  We employ a growth factor of 1.025 until the wall-normal spacing catches up to the desired multiple of $\eta(s)$. At that point, the normal spacing grows with the local Kolmogorov spacing (in this case $2\eta(s,n)$). This is built into the normal spacing described above. Eventually, the growth of the wall-normal spacing catches up to the spanwise spacing. At this point, the extrusion of wedge elements from the surface triangle can be stopped. Subsequent layers then coarsen in the spanwise direction, matching the spanwise spacing to that of normal spacing. This spanwise coarsening can be accomplished with an unstructured grid (and most smoothly accomplished with tetrahedral elements). Therefore, these layers have normal and spanwise spacing matching the desired multiple of Kolmogorov units. The stream-wise spacing in these layers is still fixed to what was set by the wall spacing, which continues until the wall-normal (and spanwise) spacing grows larger than the stream-wise wall spacing, after which the stream-wise spacing also grows.  
When following this approach, the resulting grid has one-third of the nodes that an equivalent structured grid would 
have. Reiterating, it accomplishes this by matching a multiple of the local Kolmogorov spacing in all three directions wherever that spacing is larger than 
the wall plus unit spacing dictated by the local wall shear. As the dissipation and thus Kolmogorov spacing is smooth, so is the gradation of element size in all three directions
which is critical to the success of a DNS. Note that while the dissipation function of wall normal variation was taken from prior DNSs of flat plates, this choice is conservative for FPGs where the dissipation
is reduced and thus $\eta(n)$ grows faster than the flat plate profile assumed.  Conversely, in an APG, dissipation is enhanced, making $\eta(n)$ grows slower than the assumed flat plate profile. However, within the weak APG region preceding the strong FPG, $\eta(s,n)$ was confirmed to 
remain under three which is within DNS requirements for the pre-separation regions considered in this paper.


\section{Results and Analysis}

\subsection{Boundary Layer Characterization}
\label{sec:Results_Quant}

\begin{figure}
    \centering
    \subfigure[\label{fig:Cp_DNSComp}]{\includegraphics[width=0.49\textwidth]{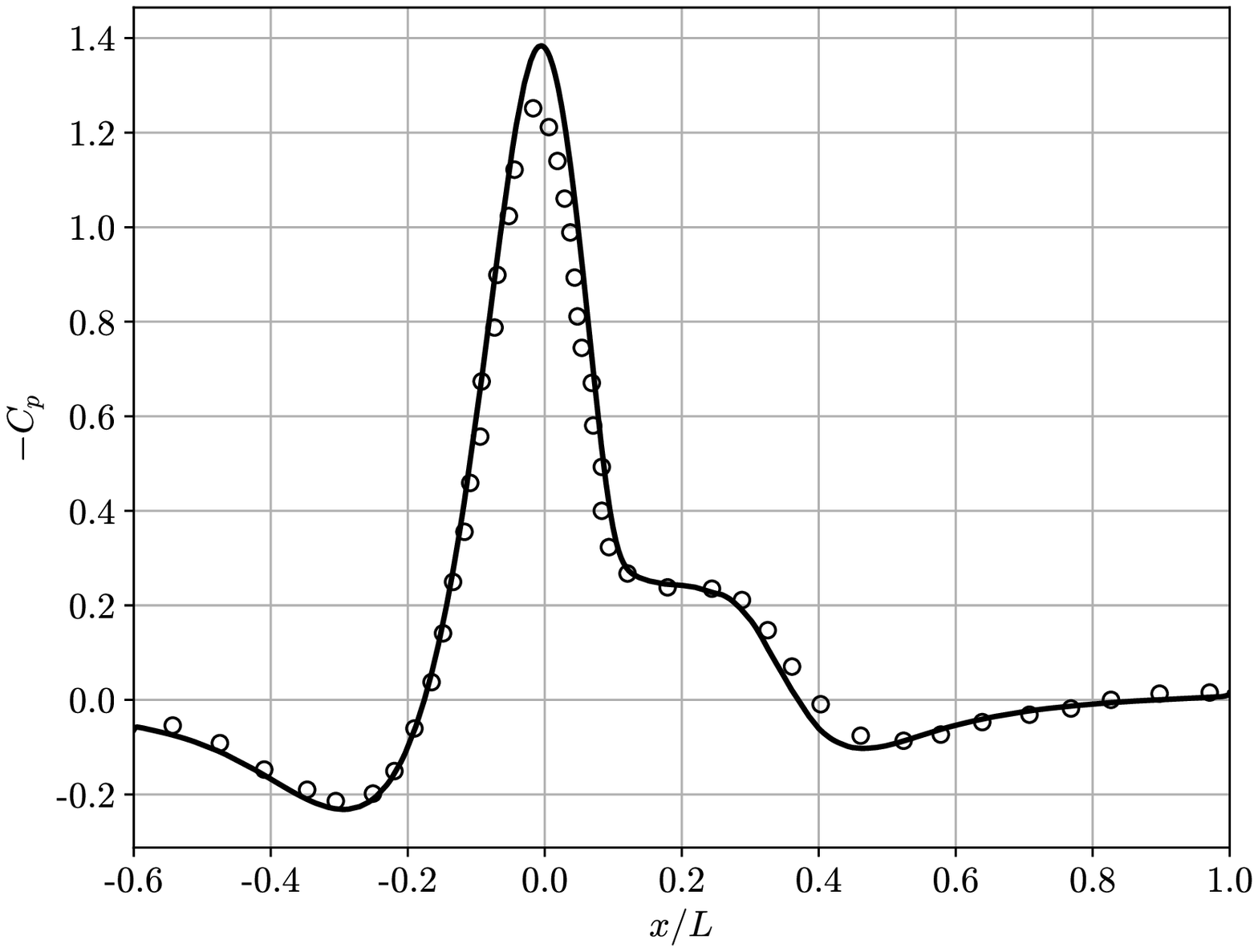}} 
    \subfigure[\label{fig:Cf_DNSComp}]{\includegraphics[width=0.49\textwidth]{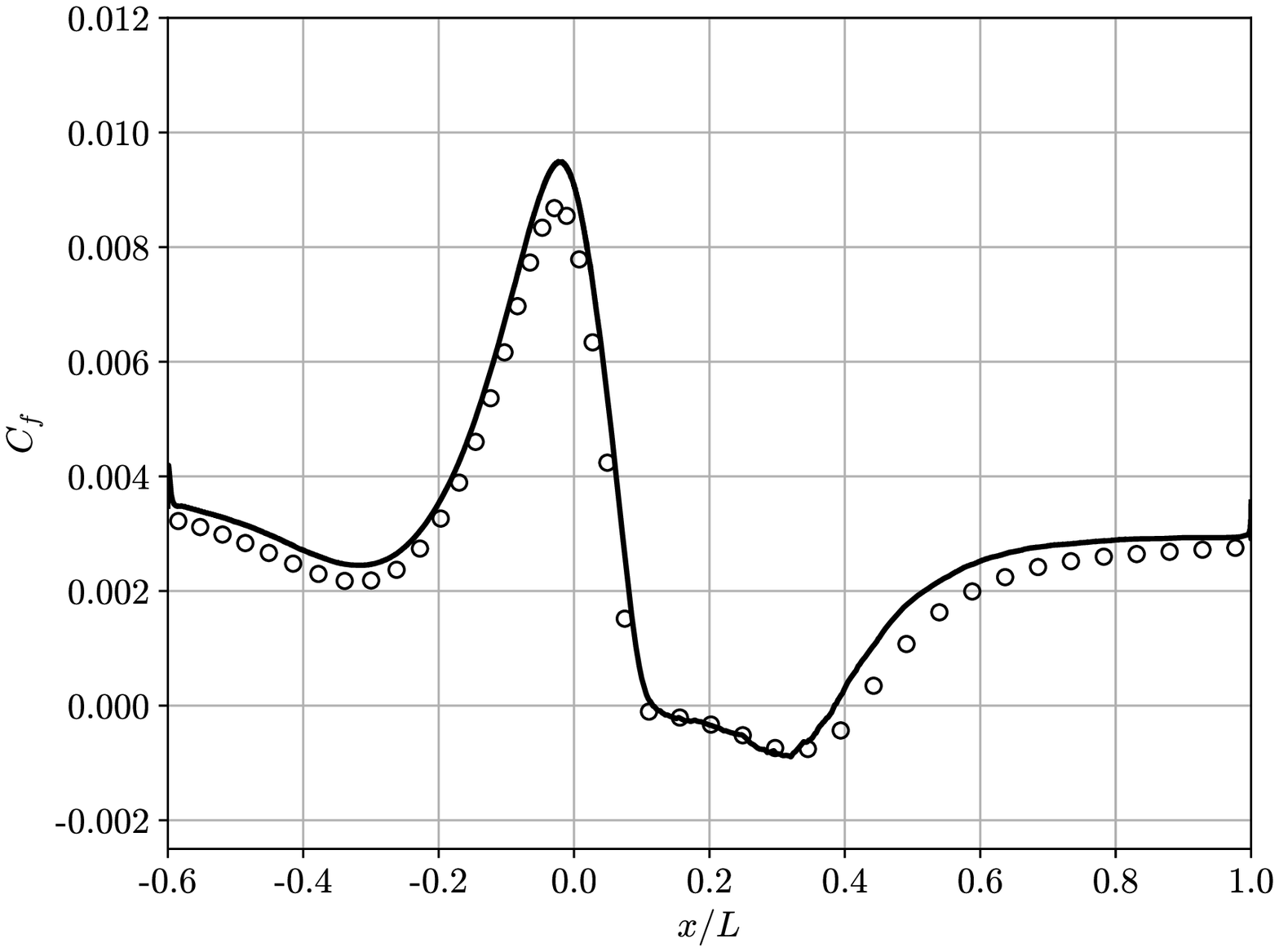}}     
    \caption{\label{fig:U&MComp} (a) $C_p$ and (b) $C_f$ for the bump flow at $Re_L = 2$ million. Line types: \bm{$\circ$} , DNS in \cite{Uzun2021b}; $\bm{-}$, Present DNS.}  
\end{figure}

As the experiment's three-dimensionality, even on the center line, is well documented \citep{Williams_ExpSpeedBump,Gray_ExpSpeedBump}, a limited comparison to the DNS of \cite{Uzun2021a} is appropriate. However, their DNS had the following differences: 1) location and type of top boundary condition (Riemann BC at $y=2L$), 2) more narrow domain width ($L_z=0.08L$), 3) compressible solver, and 4) recycling inflow boundary condition. 
To both compare and understand the differences in the two simulations, we first define the coefficient of pressure ($C_p$) and skin-friction coefficient ($C_f$) as follows,
\begin{equation}
    C_p = \frac{p - p_{\infty}}{\frac{1}{2} \rho U_{\infty}^2}, \quad     C_f = \frac{\tau_w}{\frac{1}{2} \rho U_{\infty}^2}
\end{equation}
\noindent where $p_{\infty}$ is the freestream pressure, $U_{\infty}$ is the freestream speed and $\tau_w$ is the wall shear stress. We compare $C_p$ and $C_f$ for the two DNSs in \figref{U&MComp}. The results indicate that our DNS predicts a slightly lower $-C_p$ than the other DNS in the mild APG region, $-0.6 \leq x/L \leq -0.29$. In the FPG region, $-0.29 \leq x/L \leq 0$, $-C_p$ is higher for our DNS than the other DNS, especially at the bump peak. These results indicate that the resulting pressure gradient experienced by the boundary layer is different between the two DNS. In \citet{Prakash2022}, we showed that these differences are primarily due to the top boundary condition difference, which can be expected (ours is more confined leading to strong pressure gradients) and that these differences in pressure gradients are also the key reason for the differences in $C_f$. 
We obtain a higher $C_f$ than the other DNS until the point of separation. The boundary layer in our DNS separates slightly later and reattaches earlier than the other DNS. The different spanwise lengths also likely influence the location of the boundary layer reattachment point. Similar to the other DNS, a bimodal shape of $C_f$ and $C_p$ in the separated region is observed in our DNS. These results indicate that even though the flow is quantitatively different, 
the resulting separation flow physics appears to be similar. 
A more detailed analysis of the differences in simulation setup and the resulting differences in velocity and stress profiles between the two DNSs are discussed in Appendix \ref{sec:AppA}. 

\begin{figure}
    \centering
    \subfigure[\label{fig:Beta}]{\includegraphics[width=0.32\textwidth]{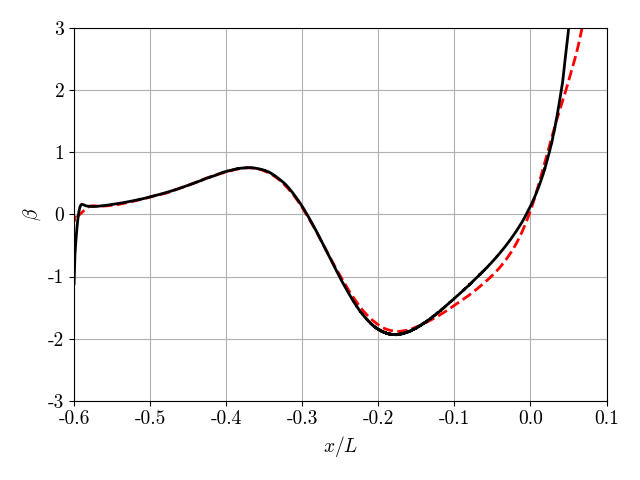}} 
    \subfigure[\label{fig:BLThick}]{\includegraphics[width=0.32\textwidth]{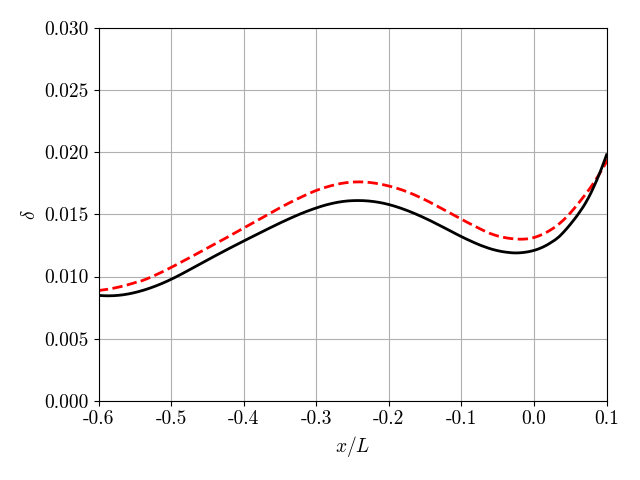}}
    \subfigure[\label{fig:deltaR}]{\includegraphics[width=0.32\textwidth]{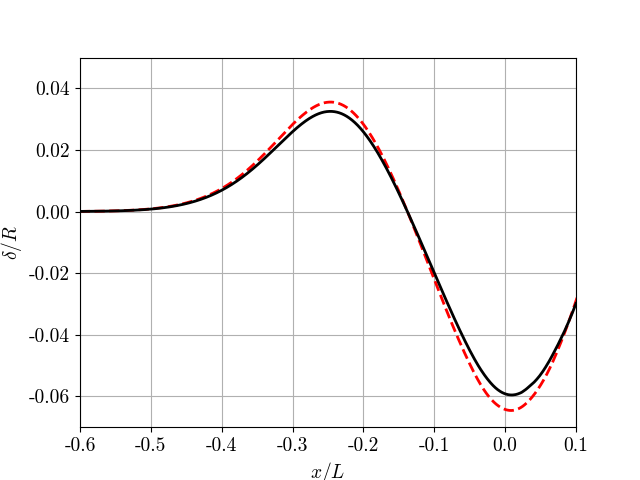}}     
    \caption{Comparison of (a) Clauser's pressure gradient parameter, (b) Boundary Layer thickness, and (b) Curvature parameter for the bump flow at $Re_L = 1$ million and $2$ million. Line types: $\color{red}{\bm{- -}}$, $Re_{L} = 1M$; $\bm{-}$, $Re_L = 2M$.}  
\end{figure}

The turbulent boundary layer over the bump experiences varied flow physics. The flow is influenced by changing pressure gradients, from adverse to favorable and back to adverse before flow separation, and varying curvature effects, from concave to convex curvature. In this article, we will keep the discussion on pressure gradient and curvature effects limited to the pre-separation region of the flow. The effect of pressure gradient on a turbulent boundary layer is commonly quantified using Clauser's pressure gradient parameter \citep{Clauser1954},
\begin{equation}
    \beta = \frac{\delta^*}{\tau_w} \frac{\p p}{\p s},
\end{equation}
\noindent where $\delta^*$ is the displacement thickness computed based on the integrated vorticity method \citep{Lighthill1963, Spalart1993}, $\tau_w$ is the wall shear stress, and $\partial p/\p s$ is the streamwise pressure gradient at the bottom surface. To understand the influence of Reynolds number on these and other parameters, we switch from comparing our results to \citep{Uzun2021a} and instead compare to the flow at $Re_L = 1$M \citep{Balin_Jansen_2021} which as noted earlier matches all boundary conditions except half the inflow speed. The variation of $\beta$ for DNSs at $Re_L = 1$M and $2$M are shown in \figref{Beta} where it is observed that both flows exhibit a relatively mild APG ($\beta > 0$; maximum $\beta$ is $0.8$ at $x/L \approx -0.35$) and a relatively stronger FPG ($\beta < 0$; minimum $\beta$ is $-2$ at $x/L \approx -0.17$). The difference in $\beta$ for the $Re_L = 1$ million and $2$ million are small, indicating similar pressure gradient effects for both Reynolds numbers. We compare the boundary layer thickness, $\delta$, computed using the integrated vorticity method specified in \cite{Lighthill1963, Spalart1993} for the two $Re_L$ in \figref{BLThick}. For the two Reynolds number flows, the variation in the BL thickness is similar: an increase in $\delta$ is observed in the mild APG region, and a decrease in $\delta$ is observed in the strong FPG region.

\begin{figure}
    \centering
    \includegraphics[width=0.49\textwidth]{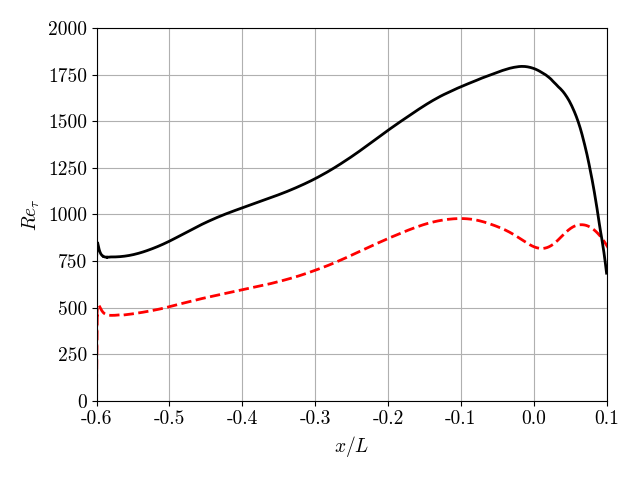}
    \caption{$Re_{\tau}$ comparison for the bump flow at $Re_L = 1$ and $2$ million. Line types: $\color{red}{\bm{- -}}$, $Re_{L} = 1M$; $\bm{-}$, $Re_L = 2M$.}
    \label{fig:ReTau}
\end{figure}

The effect of curvature on a turbulent boundary layer is quantified using the curvature parameter ($\delta/R$), where $R$ is the radius of curvature of the surface. A positive value of $\delta/R$ corresponds to concave curvature, whereas a negative value of $\delta/R$ implies convex curvature. As shown in \figref{deltaR}, the flow exhibits an initial concave curvature until $x/L \approx -0.14$ and then switches to convex curvature. The negative peak of $\delta/R$ is twice higher than the positive peak indicating that the boundary layer experiences more substantial convex curvature effects than the concave curvature effects. This behavior is because $R$ is smaller at $x/L = 0$ than its value at $x/L = -0.25$, which offsets the lower value of $\delta$ at $x/L = 0$ compared to its value at $x/L = -0.25$.  The flow at $Re_L = 1$ million experiences slightly stronger curvature effects than the flow at $Re_L = 2$ million due to its slightly larger boundary layer thickness for the same geometric curvature.

\begin{figure}
    \centering
    \subfigure[\label{fig:Lambda}]{\includegraphics[width=0.49\textwidth]{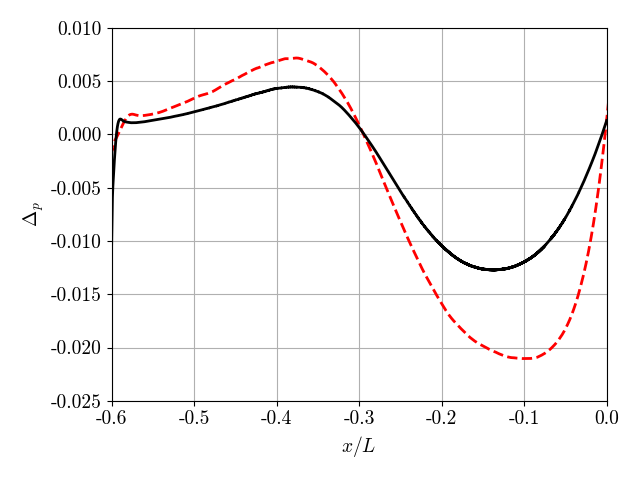}}    
    \subfigure[\label{fig:Kaccel}]{\includegraphics[width=0.49\textwidth]{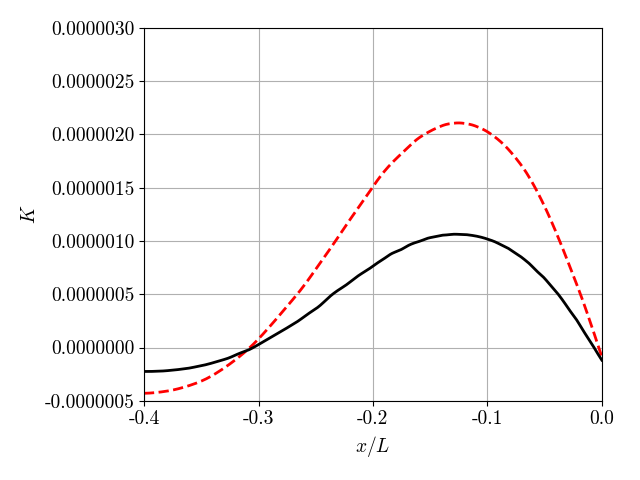}}    
    \caption{Relaminarization parameters (a) $\Delta_P$ and (b) K for turbulent boundary layer over bump flow at $Re_L = 1$ and $2$ million. Line types: $\color{red}{\bm{- -}}$, $Re_{L} = 1M$; $\bm{-}$, $Re_L = 2M$.}
    \label{fig:Accel_param}
\end{figure}

Although $\beta$ and $\delta/R$ variations are qualitatively similar, the turbulent boundary layer exhibits different flow physics for the two Reynolds numbers. These differences are highlighted by comparing friction Reynolds number for the two Reynolds numbers: $Re_{\tau} = u_{\tau} \delta/\nu$, where $u_{\tau} = \sqrt{\tau_w/\rho}$ is the friction velocity. As shown in \figref{ReTau}, we observe a similar trend for both Reynolds numbers with an increase in $Re_{\tau}$ until $x/L \approx -0.11$. Downstream of that location, $Re_{\tau}$ for $Re_L = 1$ million decreases due to a relaminarizing boundary layer.  The monotonic increase in $Re_{\tau}$ in the FPG region for $Re_L = 2$ million qualitatively indicates that the pressure gradient effects are not strong enough to trigger relaminarization and decrease $Re_{\tau}$ within the FPG region. The onset of relaminarization of a turbulent boundary layer is often quantified based on parameters \citep{Launder1964, Kline1967, Patel1968},
\begin{equation}
    \Delta_p = \frac{\nu}{\rho u_{\tau}^3} \frac{\p p}{\p s}, \quad \text{and} \quad K = \frac{\nu}{\bar{u}_{s,e}^2} \frac{\p \bar{u}_{s,e}}{\p s},
\end{equation}
\noindent where $\bar{u}_{s,e}$ is the wall-aligned velocity at the edge of the boundary layer. Several threshold values for these parameters signifying the onset of the relaminarization process have been proposed over the years \citep{Launder1964, Kline1967, Patel1968, Sreenivasan1982}. A commonly used value for indicating relaminarization is $\Delta_p = -0.025$ \citep{Patel1968} or $K = 3 \text{x} 10^{-6}$ \citep{Kline1967}. The variation of $\Delta_p$ and $K$ for the two Reynolds numbers is shown in \figref{Accel_param}. We observe that the flow for both Reynolds numbers does not cross the relaminarization threshold. The results for the flow at $Re_L = 1$ million, presented in \cite{Balin_Jansen_2021} and \cite{Uzun2021a}, gave evidence of an incomplete relaminarization process. For $Re_L = 2$ million, the values of $\Delta_p$ and $K$ are much lower than the common threshold value, and the results for a similar flow \citep{Uzun2021b} indicated that the flow does not undergo relaminarization. As discussed in \cite{Sreenivasan1982}, an accelerated boundary layer exhibits a region of departure from equilibrium scaling laws despite the fully turbulent nature of the flow. This flow region was termed laminarescent \citep{Schraub1965, Sreenivasan1982}.

The onset of laminarescent behavior of turbulent boundary layers was shown to be at $\Delta_p = -0.005$ \citep{Patel1965}. This value aligns with the results discussed in \cite{Narasimha1973}, where a corrected threshold value of $\Delta_p = -0.004$ was mentioned. We observe that for $Re_L = 2$ million, this threshold value is crossed $x/L = -0.25$. These results indicate that the higher Reynolds number flow avoids relaminarization as also observed for a similar flow in \cite{Uzun2021b}. However, the flow shows signs of exhibiting a laminarescent state. In the laminarescent state, the flow is still fully turbulent. In \cite{Sreenivasan1982}, the laminarescent region of the boundary layer is further classified into two categories: 1) a local flow equilibrium state and 2) a non-equilibrium state. In the local flow equilibrium state, a local similarity analysis could be performed to determine local scaling laws, whereas, in the non-local flow equilibrium state, such a local similarity analysis may not be employed. 

\subsection{Streamline-Aligned Coordinate System (SCS) Analysis}

\label{sec:SCSAnalysis}

The two-dimensional statistical nature of the flow allows us to consider a local streamline-aligned coordinate system (SCS) for analyzing the flow behavior. \cite{Morse2021} considered this coordinate system for analysis of the axisymmetric Reynolds-averaged Navier-Stokes equations to study pressure gradient effects on a submarine hull. Analysis of the momentum budget in the SCS will enable us to isolate the effects of momentum budget terms along and perpendicular to streamlines leading to acceleration along these directions. 

\subsubsection{Streamline-Aligned Coordinate System (SCS)}

We work with a $\psi - \phi - z$ coordinate system, where the $\psi$ direction is locally aligned to be in the direction of the mean velocity vector, the $z$ direction is same as the $z$ direction in the Cartesian coordinate system, and the $\phi$ direction is orthogonal to the $\psi$ and $z$ directions based on the right-hand rule. Representing relevant statistics in this coordinate system allows us to understand the flow behavior in directions parallel and perpendicular to the streamlines. The complex flow physics results in deviations of the streamline towards and away from the wall. 
This behavior leads to the evolution of the streamline-aligned coordinate system (SCS) locally in space in contrast to the $s-n-z$ wall-aligned coordinate system mentioned in \cite{Bradshaw1973}, which does not vary in the wall-normal direction. From here on, the subscript indicates the direction of the vector component; for example, $\bar{u}_{\psi}$ is the component of averaged velocity in the $\psi$ direction. If we were to represent the continuity and momentum equations in the SCS, a simple rotation of terms would not suffice as the local coordinate system also changes in space. Therefore, the derivation of continuity and momentum equations in the SCS involves principles from differential geometry. This problem was chased in \cite{Finnigan1983}, and the physical interpretation of different terms were discussed. As the streamline normal velocity ($\bar{u}_{\phi}$) is zero in the SCS, the mean continuity equation the continuity equation is trivially satisfied in the SCS for incompressible flows  \citep{Finnigan1983}. In the absence of gravity forces, the steady mean momentum equations in the streamline coordinate system \citep{Finnigan1983} are given as follows,
\begin{multline}
    \bar{u}_{\psi} \frac{\p \bar{u}_{\psi}}{\p \psi} = -\frac{1}{\rho} \frac{\p \bar{p}}{\p \psi} - \frac{\p \overline{u^2_{\psi}}}{\p \psi} - \frac{\p \overline{u_{\psi} u_{\phi}}}{\p \phi} + \frac{\overline{u^2_{\psi}} - \overline{u^2_{\phi}}}{L_a} + 2\frac{\overline{u_{\psi} u_{\phi}}}{R} \\
    + \nu \Big[ \frac{\p^2 \bar{u}_{\psi}}{\p \psi^2} + \frac{\p^2 \bar{u}_{\psi}}{\p \phi^2} - \frac{2}{L_a} \frac{\p \bar{u}_{\psi}}{\psi} - \frac{1}{R} \frac{\p \bar{u}_{\psi}}{\p \phi} - \frac{\bar{u}_{\psi}}{R^2} \Big],
    \label{eq:Budget_Psi}
\end{multline}    

\begin{multline}
    \frac{\bar{u}_{\psi}^2}{R} = -\frac{1}{\rho} \frac{\p \bar{p}}{\p \phi} - \frac{\p \overline{u_{\phi} u_{\psi}}}{\p \psi} - \frac{\p \overline{u^2_{\phi}}}{\p \phi} + 2\frac{\overline{u_{\psi} u_{\phi}}}{L_a} + \frac{ \overline{u_{\psi}^2} - \overline{u_{\phi}^2}}{R} \\
    + \nu \Big[ - \frac{\p^2 \bar{u}_{\psi}}{\p \psi \p \phi} + \frac{1}{L_a} \frac{\p \bar{u}}{\p \phi} + \frac{\p }{\p \psi} \Big( \frac{\bar{u}_{\psi}}{R} \Big) + \frac{\bar{u}_{\psi}}{R L_a} \Big],
    \label{eq:Budget_Phi}
\end{multline}
\noindent where $\zeta$ is an integrating factor, $L_a$ is known as the 'e-folding' distance that characterizes the length scale for streamwise acceleration and $R$ is the local radius of curvature of the streamline. These two quantities can be computed as follows,
\begin{equation}
    \frac{\bar{u}_{\psi}}{L_a} = \frac{\p \bar{u}_{\psi}}{\p \psi}, \qquad \frac{1}{R} = \frac{1}{\bar{u}_{\psi}} \Big( \Omega + \frac{\p \bar{u}_{\psi}}{\p \phi} \Big),
\end{equation}
\noindent where $\Omega$ is the magnitude of mean vorticity based on the right-hand coordinate system. In \eref{Budget_Psi}, the first term on the right can be combined with the term on the left resulting in a total pressure gradient term. With this change, the equation indicates that the total pressure is conserved along the streamlines when the viscous and turbulent momentum fluxes are negligible, which aligns with Bernoulli's principle.

\subsubsection{$\Psi$-Momentum Budget in the SCS}
\label{sec:SCS_budget_Psi}

\begin{figure}
    \centering
    \subfigure[\label{fig:PsiMom_APG_inner}]{\includegraphics[width=0.49\textwidth]{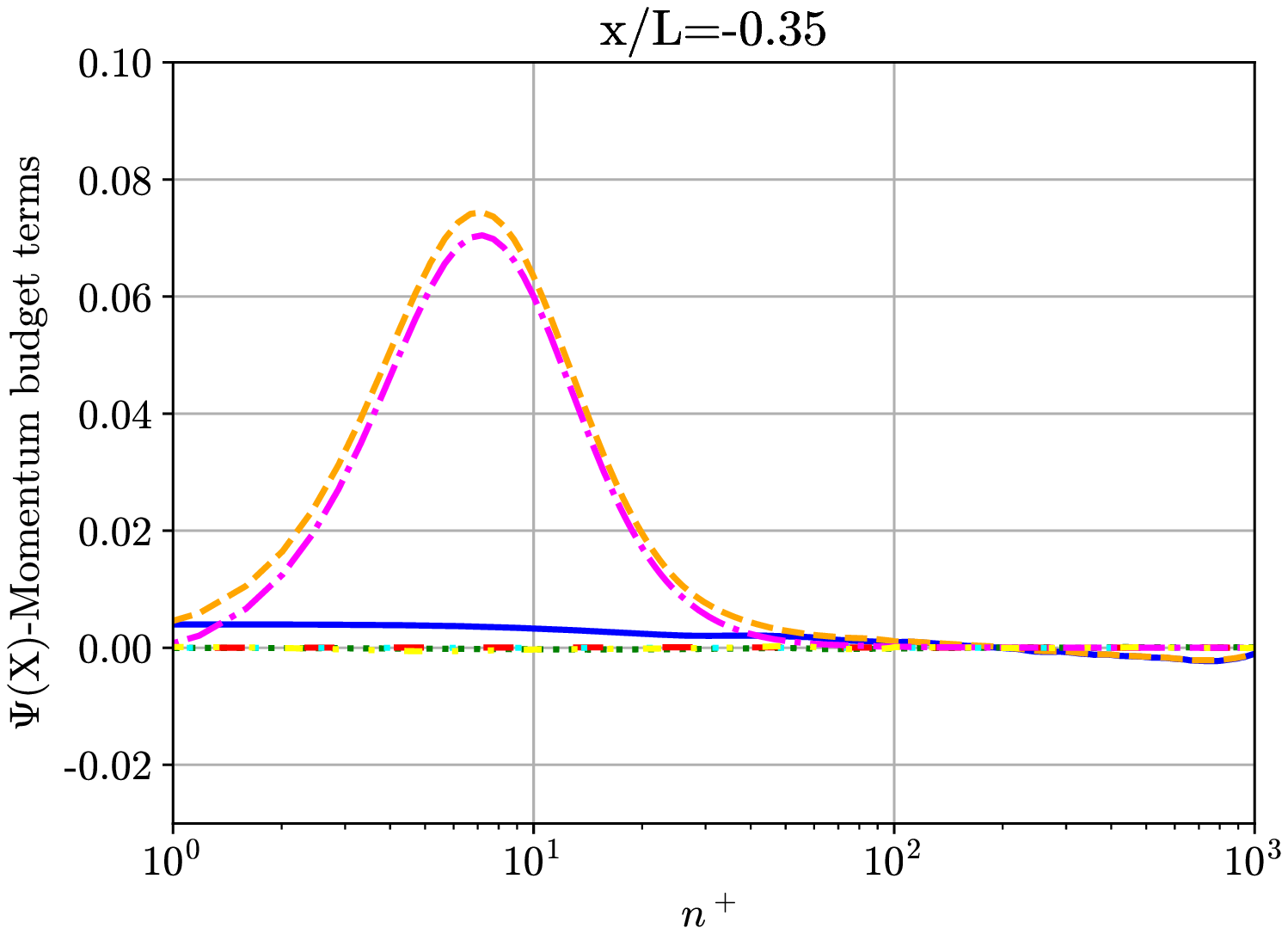}}
    \subfigure[\label{fig:PsiMom_APG_outer}]{\includegraphics[width=0.49\textwidth]{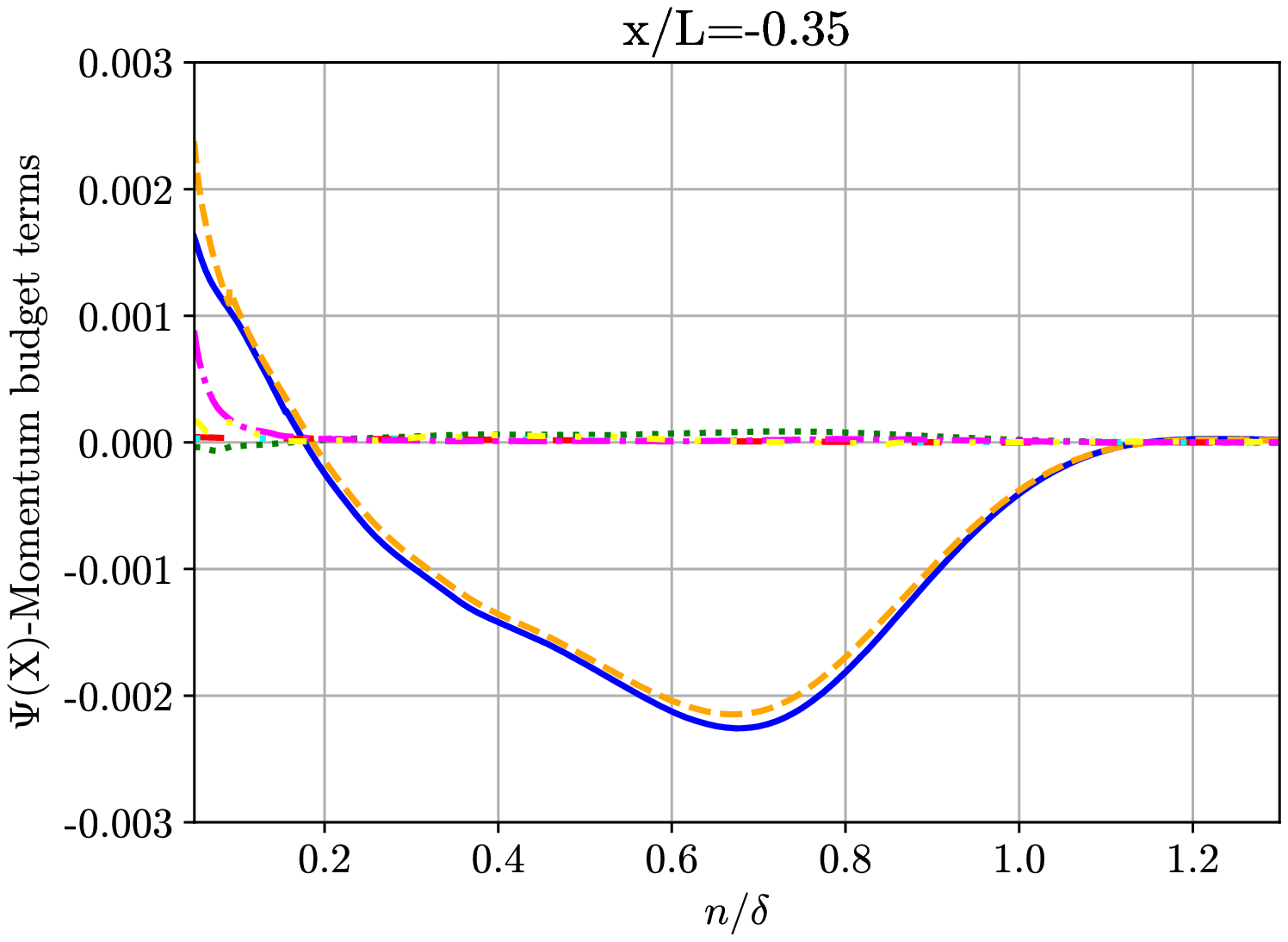}}

    \caption{$\Psi$-momentum budget stations at $x/L = -0.35$ in APG region. The x-axis is scaled in (a) inner units and (b) outer units, whereas the y-axis is scaled in inner units. Line types: $\textcolor{blue}{\bm{-}}$ , $U_{\psi} d U_{\psi}/d\psi + dp/d\psi$; ${\color{ForestGreen}{\bm{\cdot}}}$ , $d\overline{u^2_{\psi}}/d\psi$; $\textcolor{orange}{\bm{--}}$ , - $d\overline{u_{\psi} u_{\phi}}/d\phi$; $\textcolor{red}{\bm{- \; \; -}}$ , $- (\overline{u_{\psi}^2 - u_{\phi}^2})/L_a$; $\textcolor{cyan}{\bm{\cdot \; \; \cdot}}$ , $-2\overline{u_{\psi} u_{\phi}}/R$; $\textcolor{magenta}{\bm{- \cdot -}}$ , negative of viscous terms; $\textcolor{yellow}{\bm{-\cdot\cdot- }}$ , budget balance. Note that the difference in $y$-axis limits for (a) and (b) highlight the difference in magnitude of these terms in the inner and outer regions of the flow.}
    \label{fig:PsiMom_APG}
\end{figure}

Assessing the momentum equation budget in the SCS allows us to characterize the dominant forces acting on a fluid packet following the streamlines. In \figref{PsiMom_APG_inner}, we show the $\Psi$-momentum budget in the near-wall region at $x/L = -0.35$, the location of the maximum APG. We observe that the momentum transfer due to the turbulent and viscous fluxes (defined in the second line of \eref{Budget_Psi}) is the most dominant. The small difference between the two momentum fluxes is balanced by the total pressure gradient close to the wall. The advection term is much smaller than the pressure gradient near the wall within the total pressure gradient term. The $\Psi$-momentum budget far from the wall is shown in \figref{PsiMom_APG_outer}. We observe that momentum fluxes due to viscous effects are negligible in this region. The momentum flux due to Reynolds shear stress is almost balanced by the total pressure gradient term, and the slight differences are attributed to the momentum flux due to streamline-normal normal stresses. Within the total pressure gradient term, a net negative value of advection momentum transfer is observed, indicating deceleration of the streamlines in the mild APG region. The budget behavior in both the inner and outer regions of the boundary layer is similar at other locations in the APG region. 

\begin{figure}
    \centering
    \subfigure[\label{fig:PsiMom_m0p29_FPG_inner}]{\includegraphics[width=0.32\textwidth]{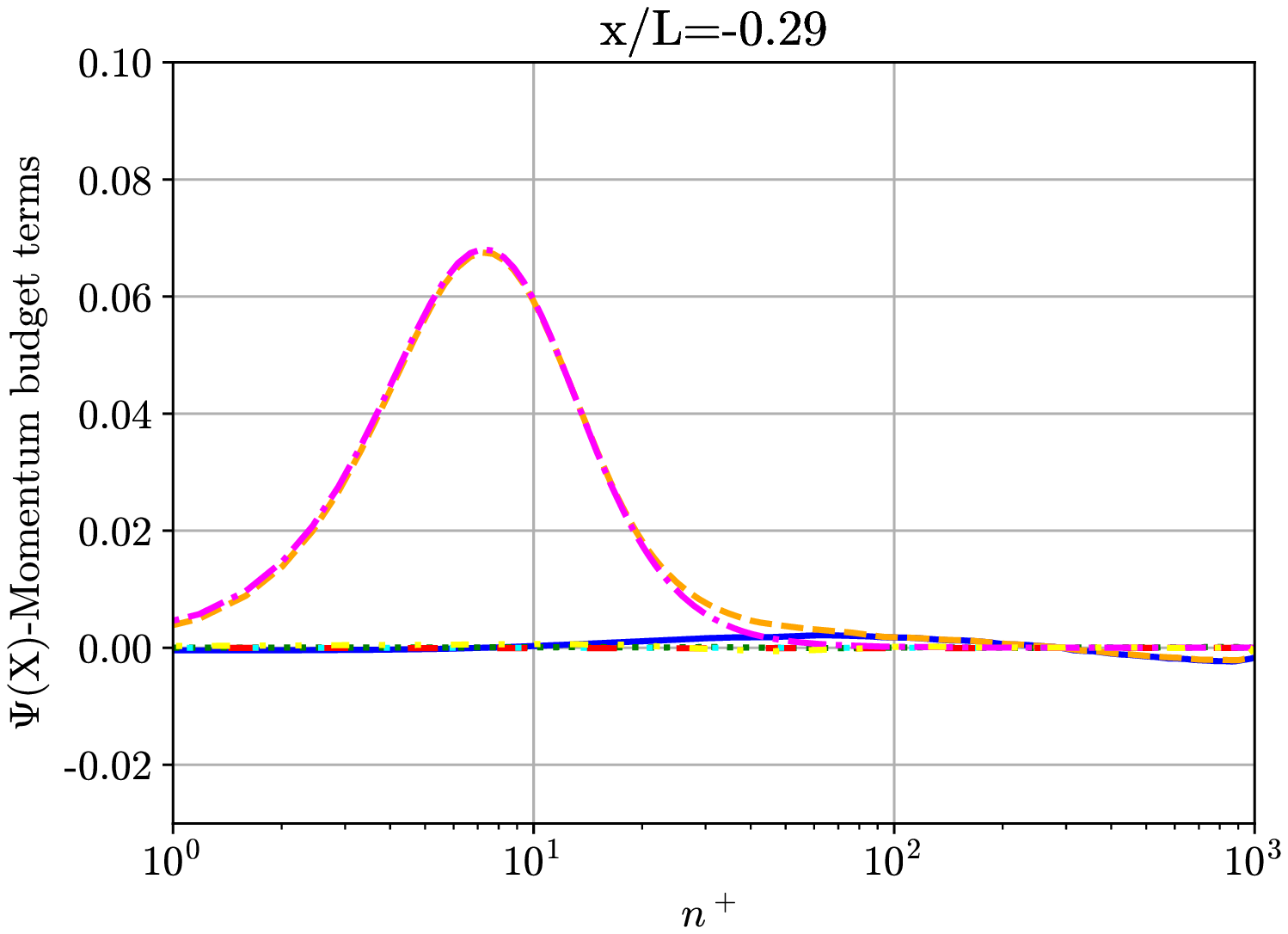}}
    \subfigure[\label{fig:PsiMom_m0p15_FPG_inner}]{\includegraphics[width=0.32\textwidth]{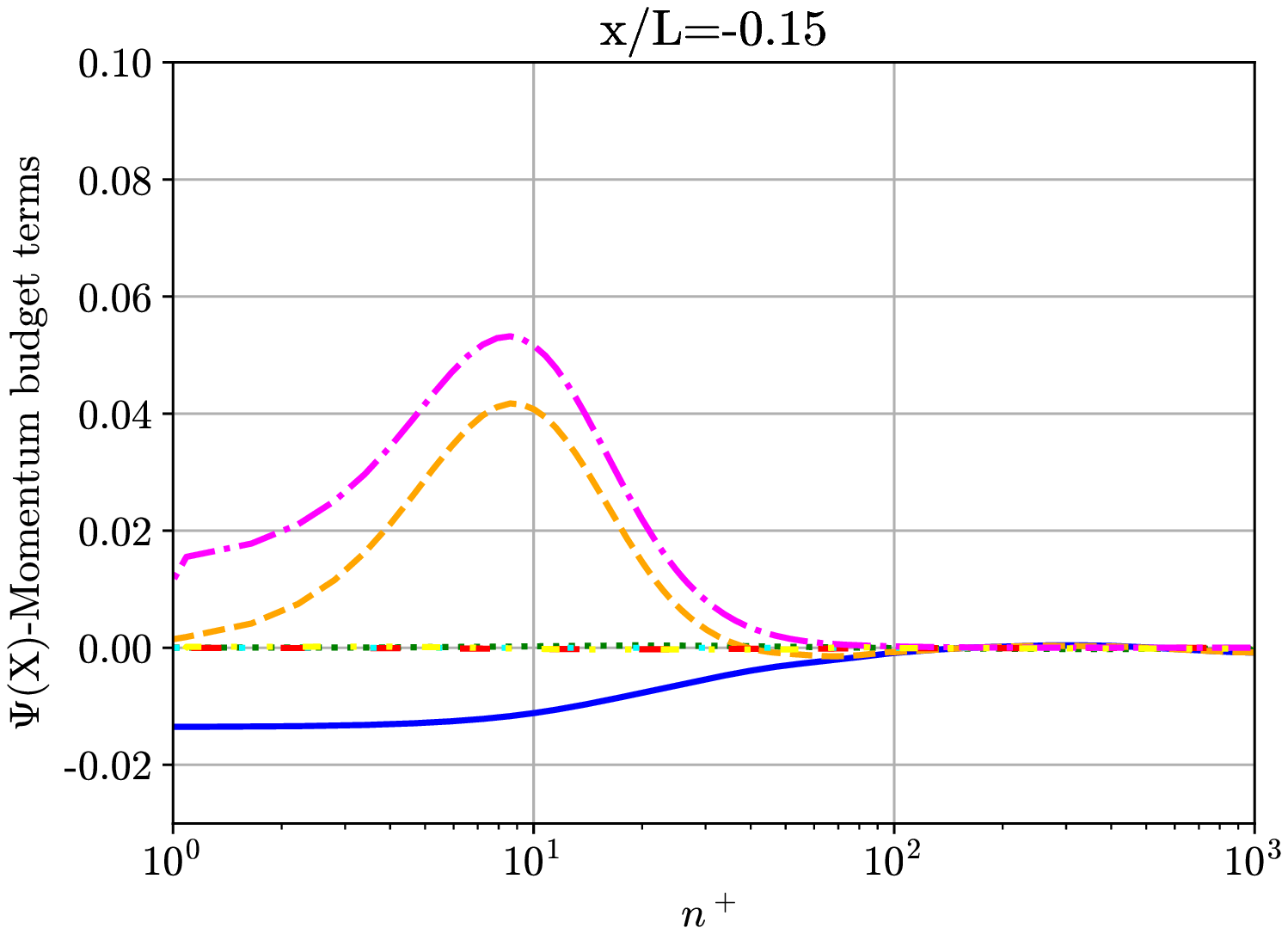}}
    \subfigure[\label{fig:PsiMom_m0p05_FPG_inner}]{\includegraphics[width=0.32\textwidth]{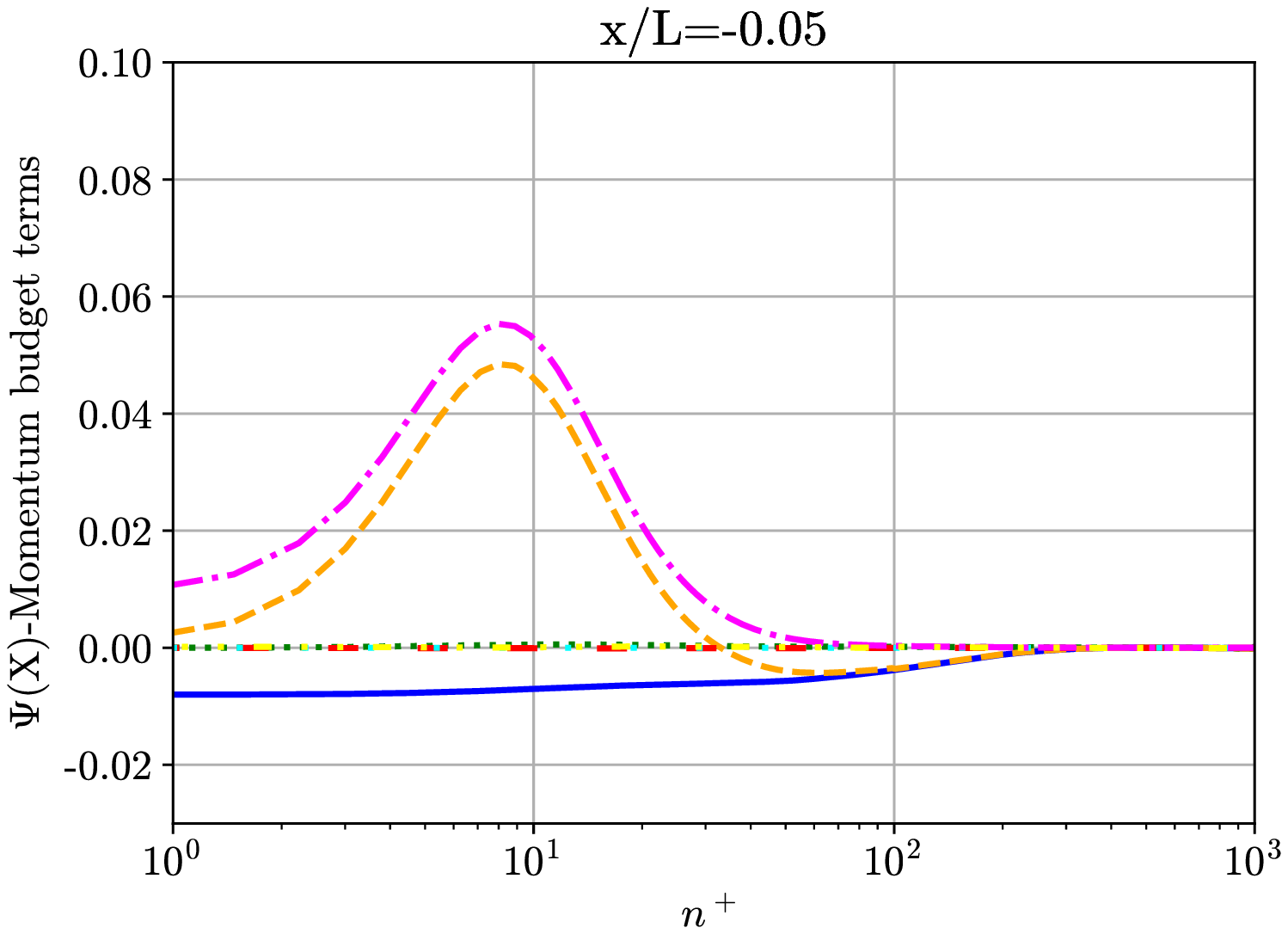}}

    \caption{$\Psi$-momentum budget at several $x/L$ stations in the FPG region. Both axes are scaled by local inner units. Line types: $\textcolor{blue}{\bm{-}}$ , $U_{\psi} d U_{\psi}/d\psi + dp/d\psi$; ${\color{ForestGreen}{\bm{\cdot}}}$ , $d\overline{u^2_{\psi}}/d\psi$; $\textcolor{orange}{\bm{--}}$ , - $d\overline{u_{\psi} u_{\phi}}/d\phi$; $\textcolor{red}{\bm{- \; \; -}}$ , $- (\overline{u_{\psi}^2 - u_{\phi}^2})/L_a$; $\textcolor{cyan}{\bm{\cdot \; \; \cdot}}$ , $-2\overline{u_{\psi} u_{\phi}}/R$; $\textcolor{magenta}{\bm{- \cdot -}}$ , negative of viscous terms; $\textcolor{yellow}{\bm{-\cdot\cdot- }}$ , budget balance.}
    \label{fig:PsiMom_FPG_Inner}
\end{figure}

\begin{figure}
    \centering
    \subfigure[\label{fig:PsiMom_m0p29_FPG_outer}]{\includegraphics[width=0.32\textwidth]{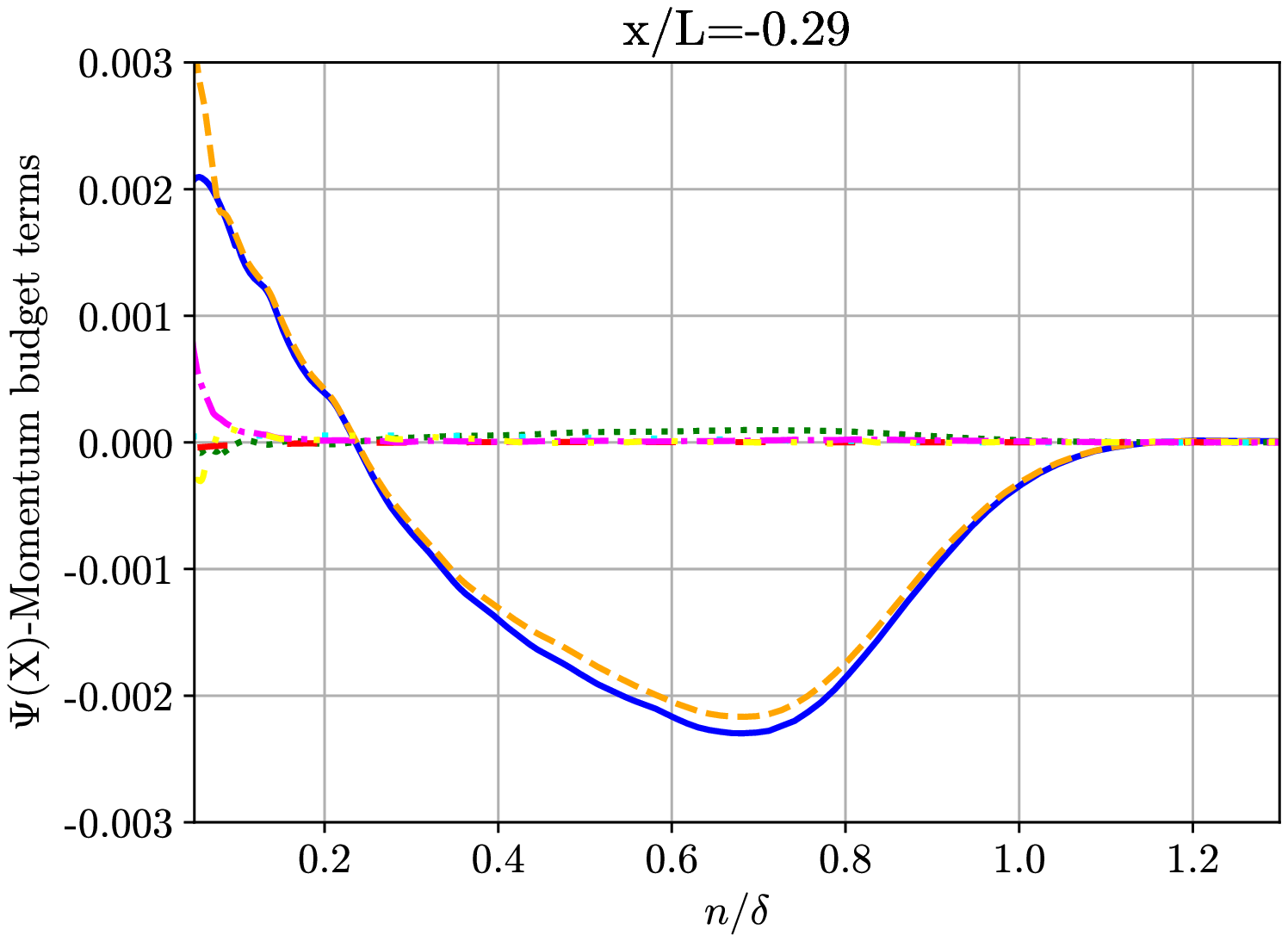}}
    \subfigure[\label{fig:PsiMom_m0p15_FPG_outer}]{\includegraphics[width=0.32\textwidth]{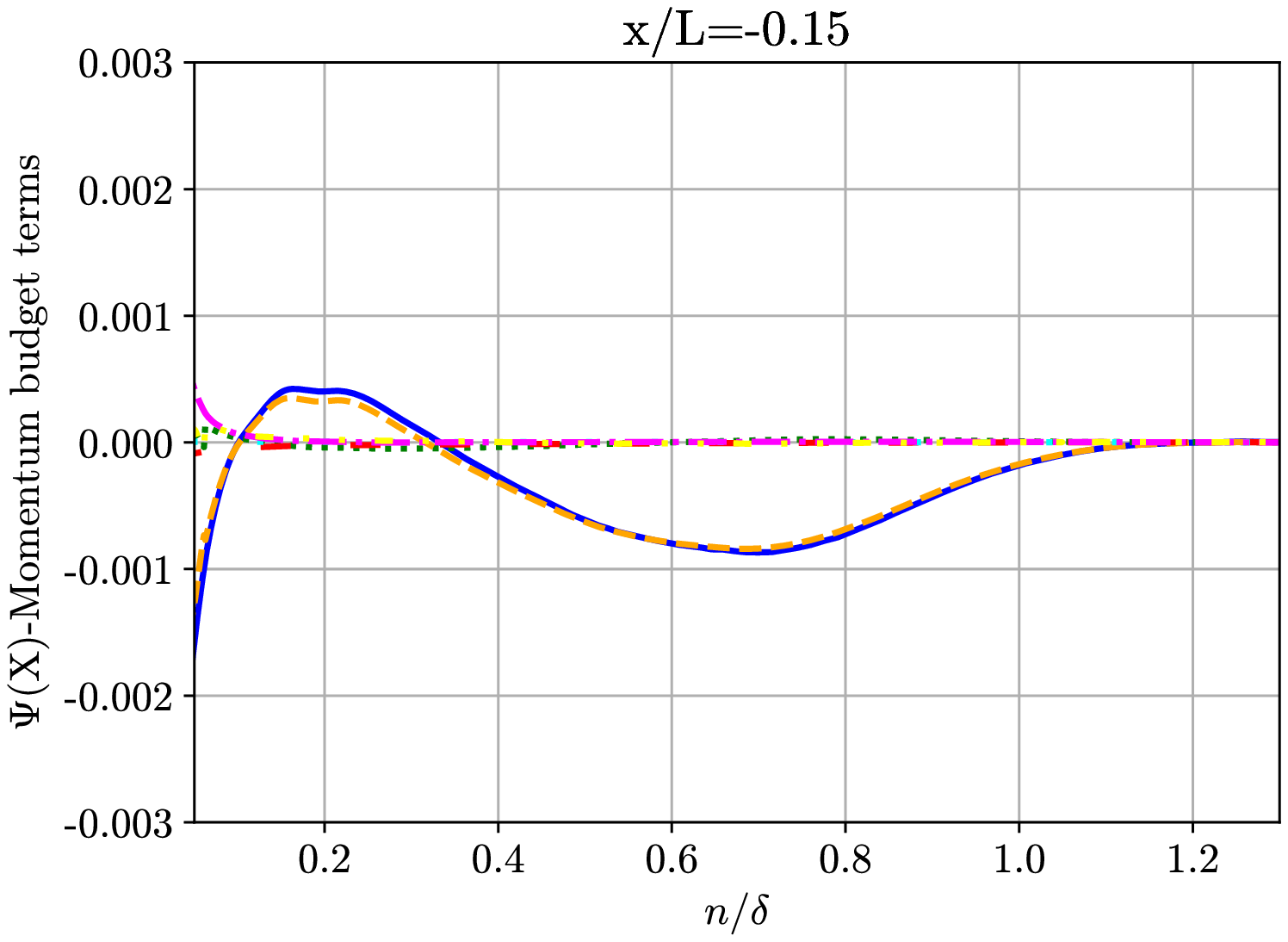}}
    \subfigure[\label{fig:PsiMom_m0p05_FPG_outer}]{\includegraphics[width=0.32\textwidth]{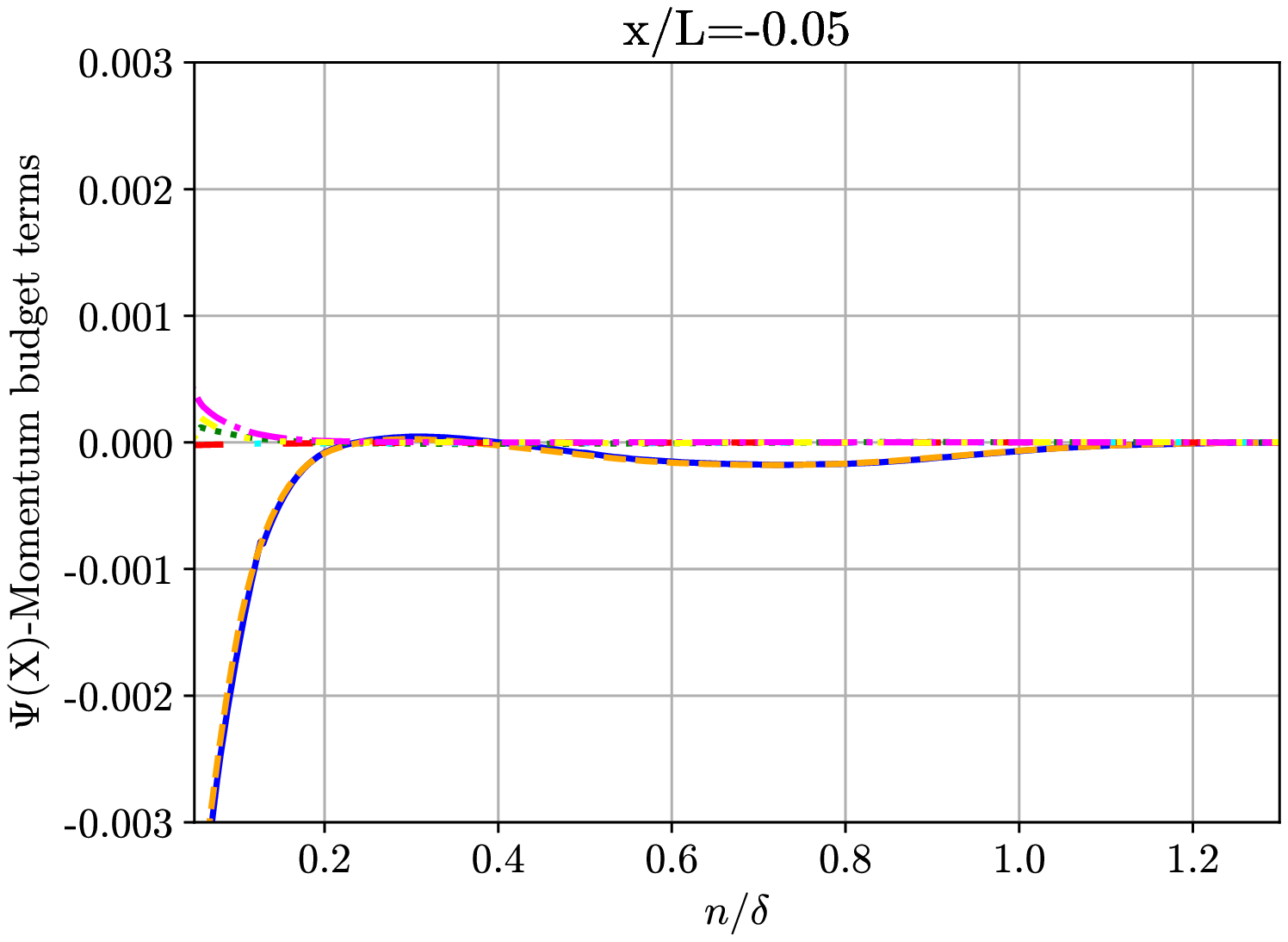}}

    \caption{$\Psi$-momentum budget at several stations in the FPG region. Both axes are scaled by local outer units. Line types: $\textcolor{blue}{\bm{-}}$ , $U_{\psi} d U_{\psi}/d\psi + dp/d\psi$; ${\color{ForestGreen}{\bm{\cdot}}}$ , $d\overline{u^2_{\psi}}/d\psi$; $\textcolor{orange}{\bm{--}}$ , - $d\overline{u_{\psi} u_{\phi}}/d\phi$; $\textcolor{red}{\bm{- \; \; -}}$ , $- (\overline{u_{\psi}^2 - u_{\phi}^2})/L_a$; $\textcolor{cyan}{\bm{\cdot \; \; \cdot}}$ , $-2\overline{u_{\psi} u_{\phi}}/R$; $\textcolor{magenta}{\bm{- \cdot -}}$ , negative of viscous terms; $\textcolor{yellow}{\bm{-\cdot\cdot- }}$ , budget balance.}
    \label{fig:PsiMom_FPG_Outer}
\end{figure}

In \figref{PsiMom_FPG_Inner}, we show $\Psi$-momentum budget in the near-wall region for different $x/L$ locations corresponding to the start of the FPG region, close to the location of the maximum FPG and near the end of the FPG region. We observe that the momentum fluxes due to viscous forces, turbulence, and total pressure gradient balance very close to the wall. The decrease in the total pressure gradient term after $n^+ > 10$ is due to the growing importance of the advection term. The behavior in the outer region of the boundary layer is shown in \figref{PsiMom_FPG_Outer}. The outer region of the boundary layer is dominated by momentum flux due to the total pressure gradient and turbulence terms. In this region, there is a positive value of the advection momentum flux, which indicates a net acceleration of the streamlines in the FPG region. Like the APG region, the total pressure gradient and Reynolds shear stress term are mostly balanced by the streamline-normal normal stress term in the outer region during the early parts of the FPG region. From these results, we observe that the dominant terms of the momentum equation budget in both APG and FPG regions are: viscous flux, turbulence normal and shear fluxes, and total pressure gradient flux. 

\subsubsection{$\Phi$-Momentum Budget in the SCS}

\label{sec:SCS_budget_Phi}

\begin{figure}
    \centering
    \subfigure[\label{fig:PhiMom_APG_Inner}]{\includegraphics[width=0.49\textwidth]{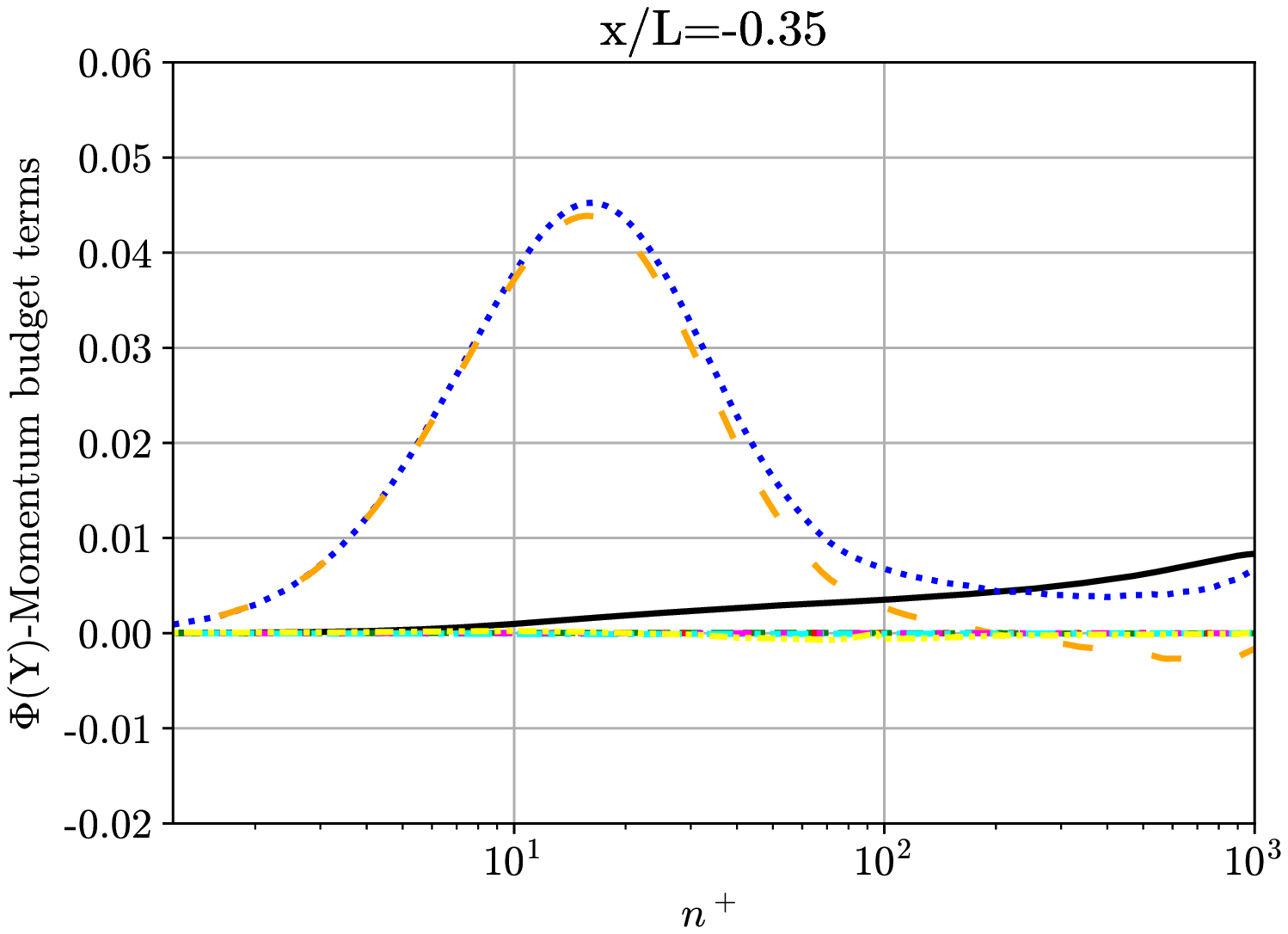}}
    \subfigure[\label{fig:PhiMom_APG_Outer}]{\includegraphics[width=0.49\textwidth]{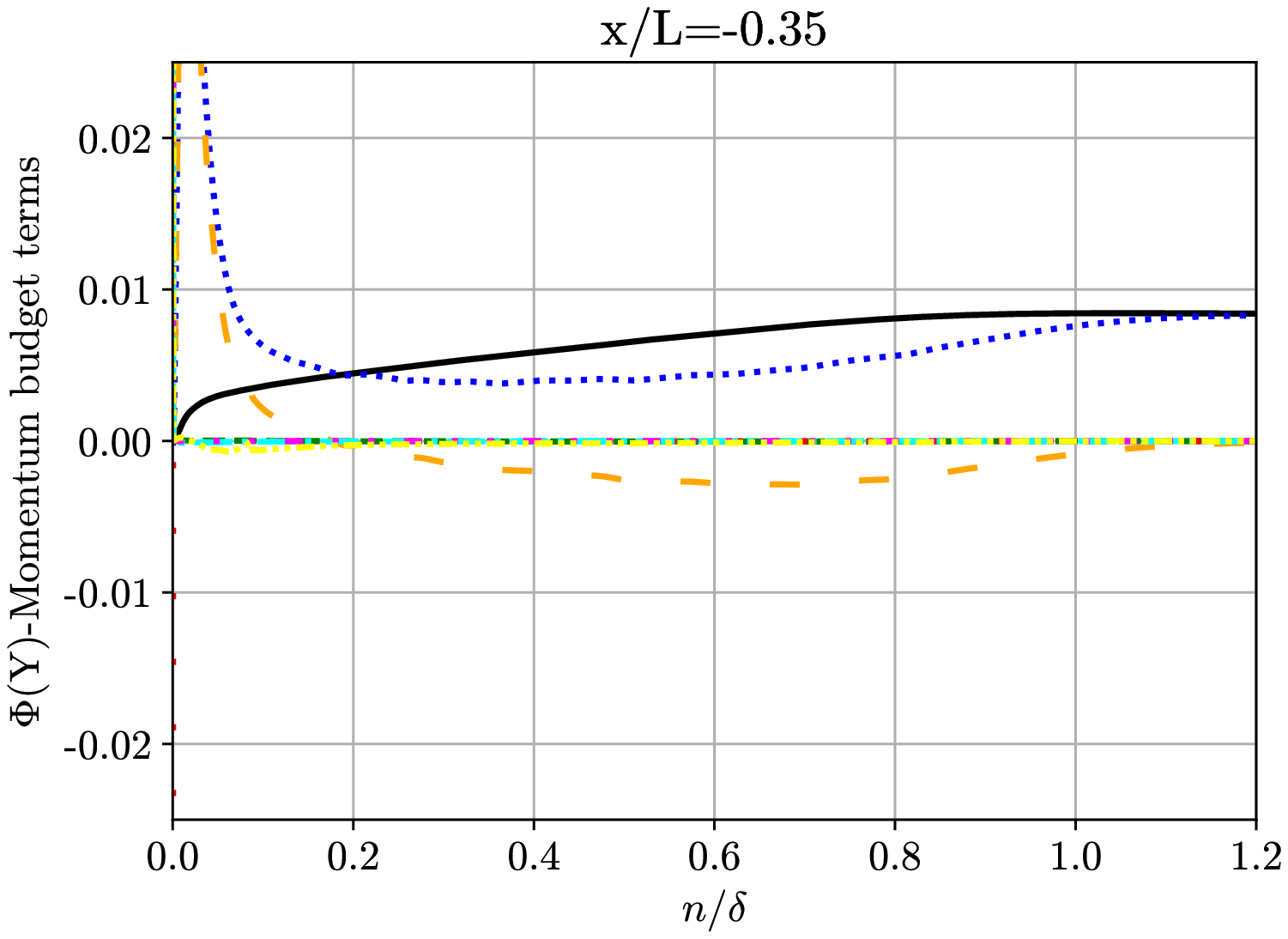}}

    \caption{$\Phi$-momentum budget at $x/L = -0.35$ in the APG region. The x-axis is scaled in (a) inner units and (b) outer units, whereas the y-axis is scaled in inner units. Line types: $\textcolor{black}{\bm{-}}$ , $U_{\psi}^2/R$; ${\color{blue}{\bm{\cdot}}}$ , - $dp/dx_{\phi}$; ${\textcolor{lime}{\bm{- -}}}$ , - $d \overline{u_{\psi} u_{\phi}}/d x_{\psi}$; $\textcolor{orange}{\bm{- \; \; -}}$ , $d\overline{u^2_{\phi}}/d x_{\phi}$; $\textcolor{red}{\bm{\cdot \; \; \cdot}}$ , $-2\overline{u_{\psi}u_{\phi}}/L_a$; $\textcolor{cyan}{\bm{- \cdot-}}$, $-\overline{u^2_{\psi} - u^2_{\phi}}/R$; $\textcolor{magenta}{\bm{-\;\cdot\;\cdot\;\;-}}$ , negative of viscous terms; $\textcolor{yellow}{\bm{-\cdot\cdot -}}$ , budget balance.}
    \label{fig:PhiMom_APG}
\end{figure}

The $\Phi$-momentum budget enables the assessment of forces normal to the streamlines acting on a fluid packet moving along the streamlines. In \figref{PhiMom_APG_Inner}, we show the $\Phi$-momentum budget in the near-wall region for the APG region of the flow. At $x/L = -0.35$, we observe fluxes due to streamline-normal normal stresses and pressure gradient balance each other until $n^+ \approx 10$. The effects of curvature are observed from $n^+ \geq 10$ onwards in \figref{PhiMom_APG_Inner} and \figref{PhiMom_APG_Outer}.
We observe that the difference between the momentum flux due to the pressure gradient and streamline-normal normal stresses is non-zero. The net difference is equal to the advection normal to the streamlines.

\begin{figure}
    \centering
    \subfigure[\label{xL_0p29_phiInner}]{\includegraphics[width=0.32\textwidth]{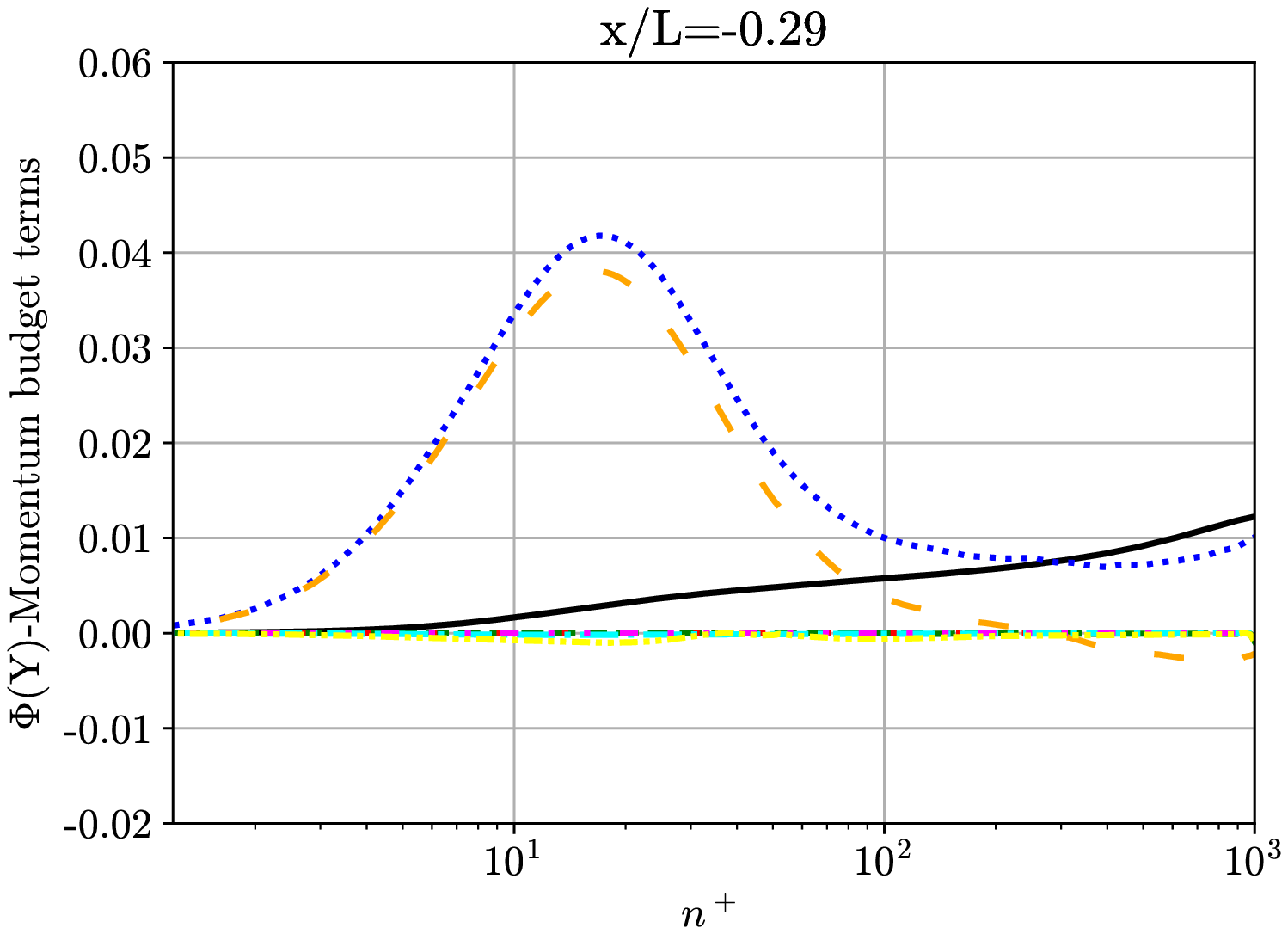}}
    \subfigure[\label{xL_0p15_phiInner}]{\includegraphics[width=0.32\textwidth]{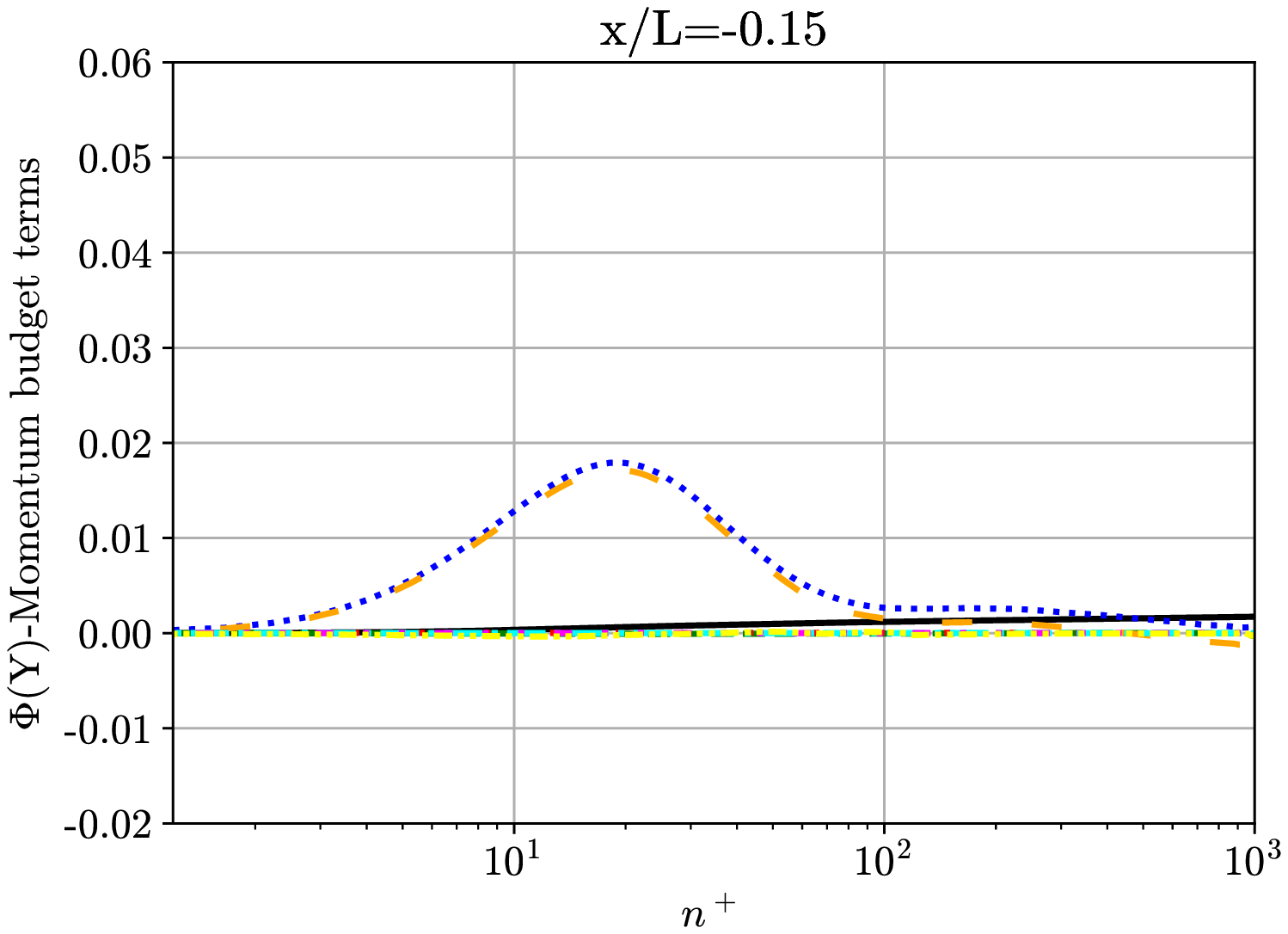}}
    \subfigure[\label{xL_0p05_phiInner}]{\includegraphics[width=0.32\textwidth]{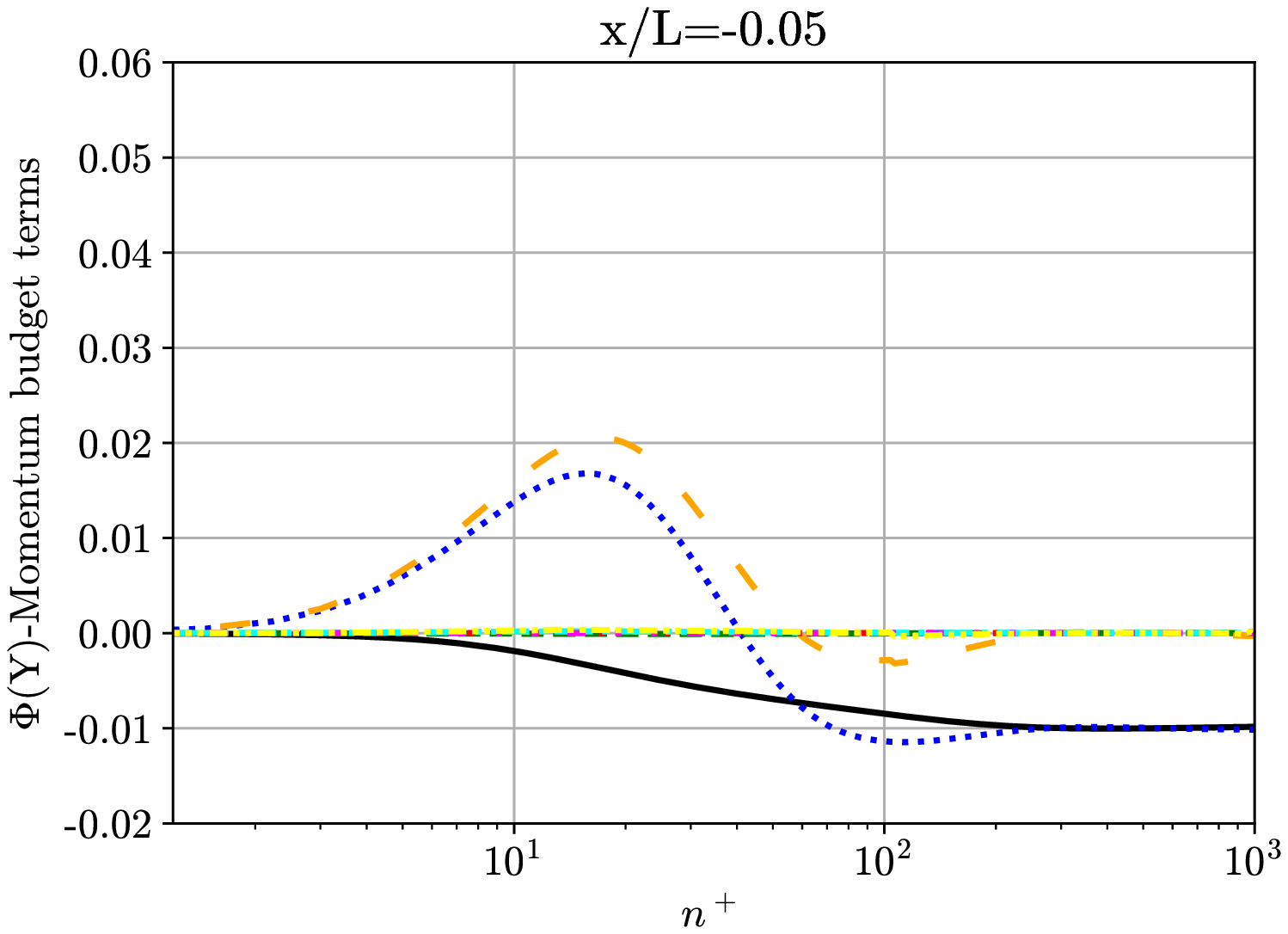}}    
    \caption{$\Phi$-momentum budget at several $x/L$ stations in FPG region. Both axes are scaled by local inner units. Line types: $\textcolor{black}{\bm{-}}$ , $U_{\psi}^2/R$; ${\color{blue}{\bm{\cdot}}}$ , - $dp/dx_{\phi}$; ${\textcolor{lime}{\bm{- -}}}$ , - $d \overline{u_{\psi} u_{\phi}}/d x_{\psi}$; $\textcolor{orange}{\bm{- \; \; -}}$ , $d\overline{u^2_{\phi}}/d x_{\phi}$; $\textcolor{red}{\bm{\cdot \; \; \cdot}}$ , $-2\overline{u_{\psi}u_{\phi}}/L_a$; $\textcolor{cyan}{\bm{- \cdot-}}$, $-\overline{u^2_{\psi} - u^2_{\phi}}/R$; $\textcolor{magenta}{\bm{-\;\cdot\;\cdot\;\;-}}$ , negative of viscous terms; $\textcolor{yellow}{\bm{-\cdot\cdot -}}$ , budget balance.}
    \label{fig:PhiMom_FPG_Inner}
\end{figure}

\begin{figure}
    \centering    \subfigure[\label{xL_0p29_phiOuter}]{\includegraphics[width=0.32\textwidth]{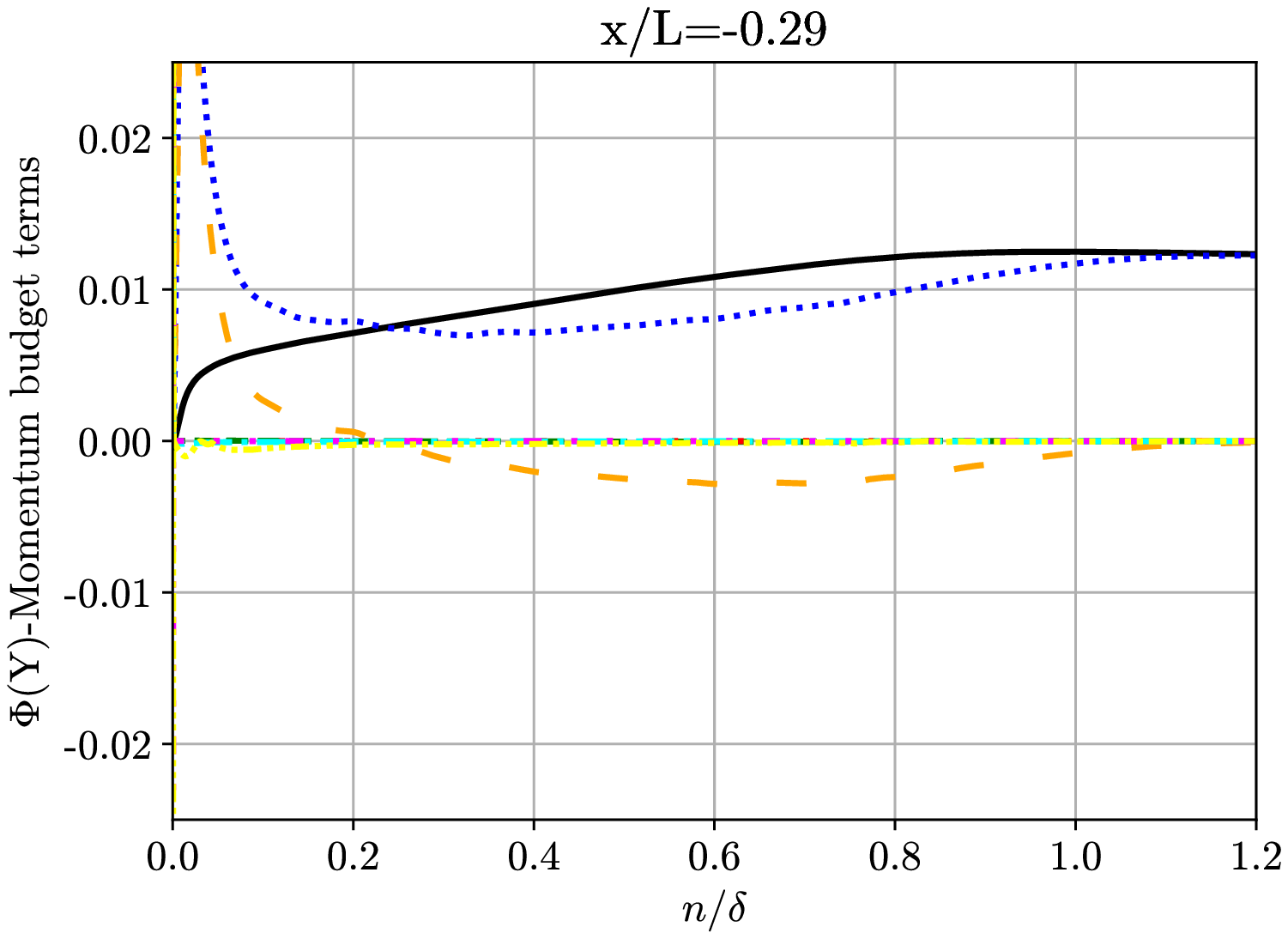}}
    \subfigure[\label{xL_0p15_phiOuter}]{\includegraphics[width=0.32\textwidth]{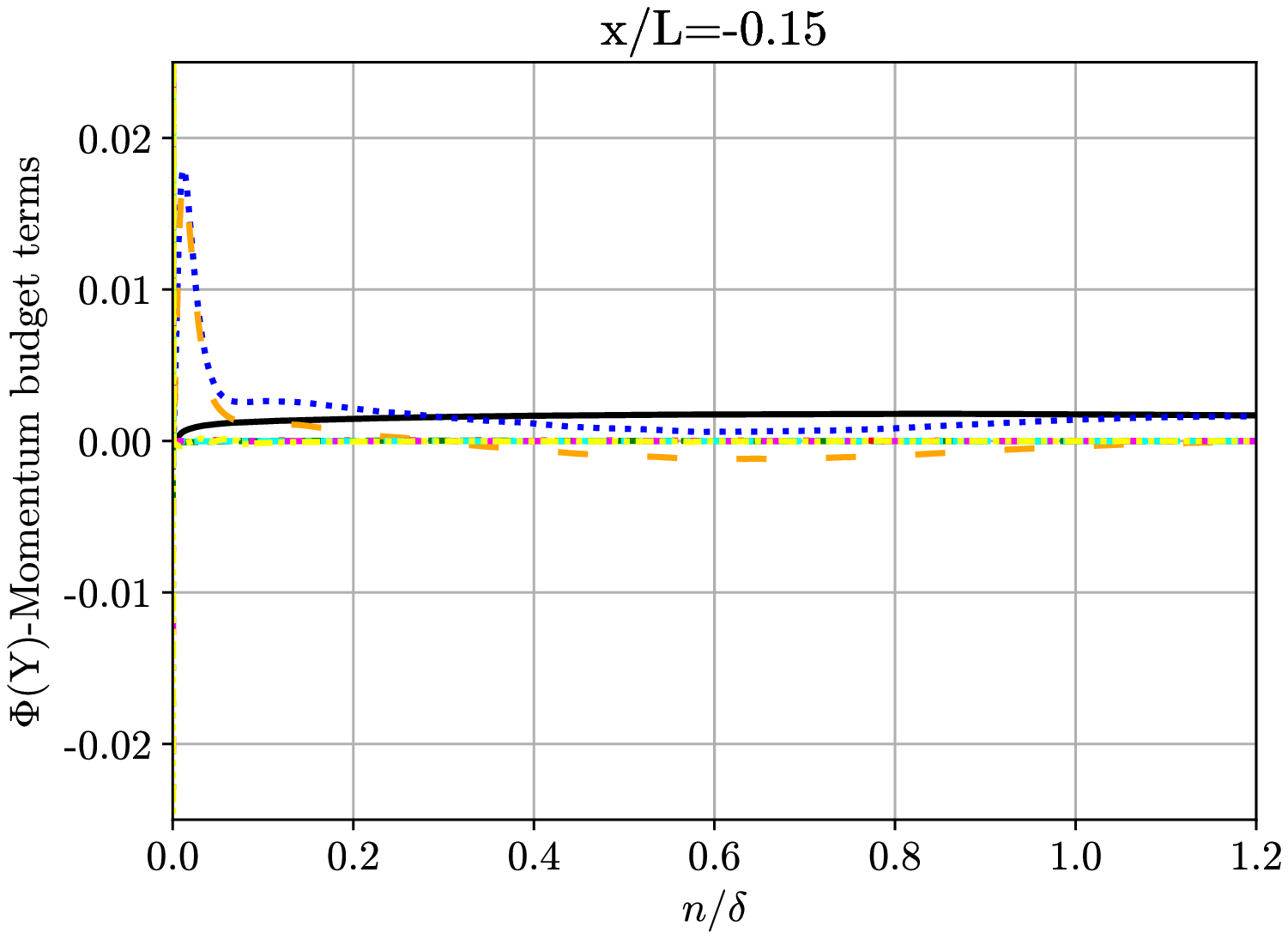}}
    \subfigure[\label{xL_0p05_phiOuter}]{\includegraphics[width=0.32\textwidth]{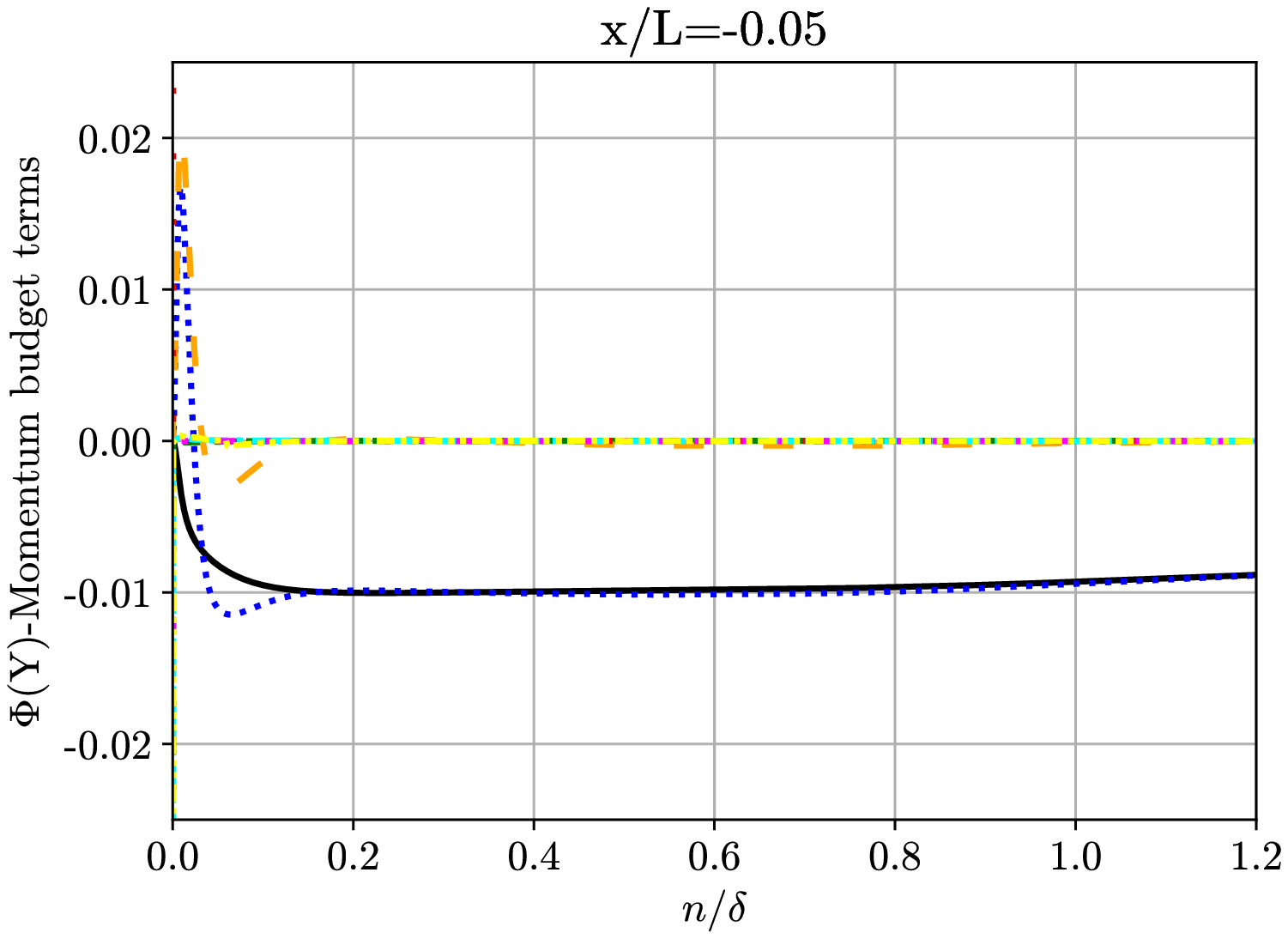}}    
    \caption{$\Phi$-momentum budget at several $x/L$ stations in FPG region. $y$-axis is scaled by local inner units whereas $x$-axis is scaled by $\delta$. Line types: $\textcolor{black}{\bm{-}}$ , $U_{\psi}^2/R$; ${\color{blue}{\bm{\cdot}}}$ , - $dp/dx_{\phi}$; ${\textcolor{lime}{\bm{- -}}}$ , - $d \overline{u_{\psi} u_{\phi}}/d x_{\psi}$; $\textcolor{orange}{\bm{- \; \; -}}$ , $d\overline{u^2_{\phi}}/d x_{\phi}$; $\textcolor{red}{\bm{\cdot \; \; \cdot}}$ , $-2\overline{u_{\psi}u_{\phi}}/L_a$; $\textcolor{cyan}{\bm{- \cdot-}}$, $-\overline{u^2_{\psi} - u^2_{\phi}}/R$; $\textcolor{magenta}{\bm{-\;\cdot\;\cdot\;\;-}}$ , negative of viscous terms; $\textcolor{yellow}{\bm{-\cdot\cdot -}}$ , budget balance.}
    \label{fig:PhiMom_FPG_Outer}
\end{figure}

In \figref{PhiMom_FPG_Inner} and \figref{PhiMom_FPG_Outer}, we show the $\Phi$-momentum budget at different $x/L$ locations in the FPG region of the flow. We observe that the FPG region shows similar behavior as the APG region until $x/L = -0.14$. For $-0.14 \leq x/L \leq 0$, the difference between momentum fluxes due to streamline-normal normal stress and pressure gradient results in a net acceleration of streamlines. This behavior reduces the momentum in the outer region of the boundary layer and brings it into the inner region of the boundary layer leading to the decay of turbulence in the  outer region of the boundary layer. From the $\Phi$-momentum budget, we observe that the pressure gradient normal to the streamlines and streamline-normal normal stress term are dominant components for the momentum budget that governs the acceleration of streamlines towards or away from the wall.

\subsection{Integral Analysis-Based Reynolds Stress Scaling}
\label{sec:IntAnalysis}

Simplified momentum equations were integrated normal to the wall in \cite{Knopp2015} and \cite{Romero2022} to determine a scaling for turbulent statistics in the inner region units of turbulent boundary layers. In this section, we extend this analysis to consider the physics exhibited by the bump flow where the advection and centrifugal acceleration terms become prominent. Further, this analysis is also extended to the outer region of the turbulent boundary layer to derive an appropriate scaling for the Reynolds stresses.

\subsubsection{Inner Region}
\label{sec:IntAnalysis_Inner}

We draw insights from the $\Psi$-momentum budget to obtain a non-local scaling for the Reynolds shear stress. From the $\Psi$-momentum budgets shown in Section \ref{sec:SCS_budget_Psi}, we observe that only a few terms contribute to the momentum fluxes: viscous stress, Reynolds shear stress, and pressure gradient, thereby leading to a net acceleration or inertia along the streamlines. Therefore, the $\Psi$-momentum budget, \eref{Budget_Psi}, reduces to:
\begin{equation}
    \bar{u}_{\psi} \frac{\p \bar{u}_{\psi}}{\p \psi} = -\frac{1}{\rho} \frac{\p \bar{p}}{\p \psi} - \frac{\p \overline{u_{\psi} u_{\phi}}}{\p \phi} + \nu \frac{\p^2 \bar{u}_{\psi}}{\p \phi^2}.
    \label{eq:Budget_Psi_reduced_int}
\end{equation}  
\noindent Rearranging the equation and integrating the equation along $\hat{\phi}$ results in:
\begin{equation}
    -\overline{u_{\psi} u_{\phi}} + \nu \frac{\p \bar{u}_{\phi}}{\p \phi} = \int_{0}^{\phi} \Big(  \bar{u}_{\psi} \frac{\p \bar{u}_{\psi}}{\p \psi} + \frac{1}{\rho} \frac{\p \bar{p}}{\p \psi} \Big) d \phi' + \frac{\tau_w}{\rho}.
    \label{eq:Budget_Psi_int}
\end{equation}
\noindent Sufficiently far away from the wall, the viscous term is negligible, and the equation simplifies to:
\begin{equation}
    -\overline{u_{\psi} u_{\phi}} = \int_{0}^{\phi} \Big(  \bar{u}_{\psi} \frac{\p \bar{u}_{\psi}}{\p \psi} + \frac{1}{\rho} \frac{\p \bar{p}}{\p \psi} \Big) d \phi' + \frac{\tau_w}{\rho}.
    \label{eq:Budget_Psi_int_2}
\end{equation}

\begin{figure}
    \centering
    \subfigure[]{\includegraphics[width=0.49\textwidth]{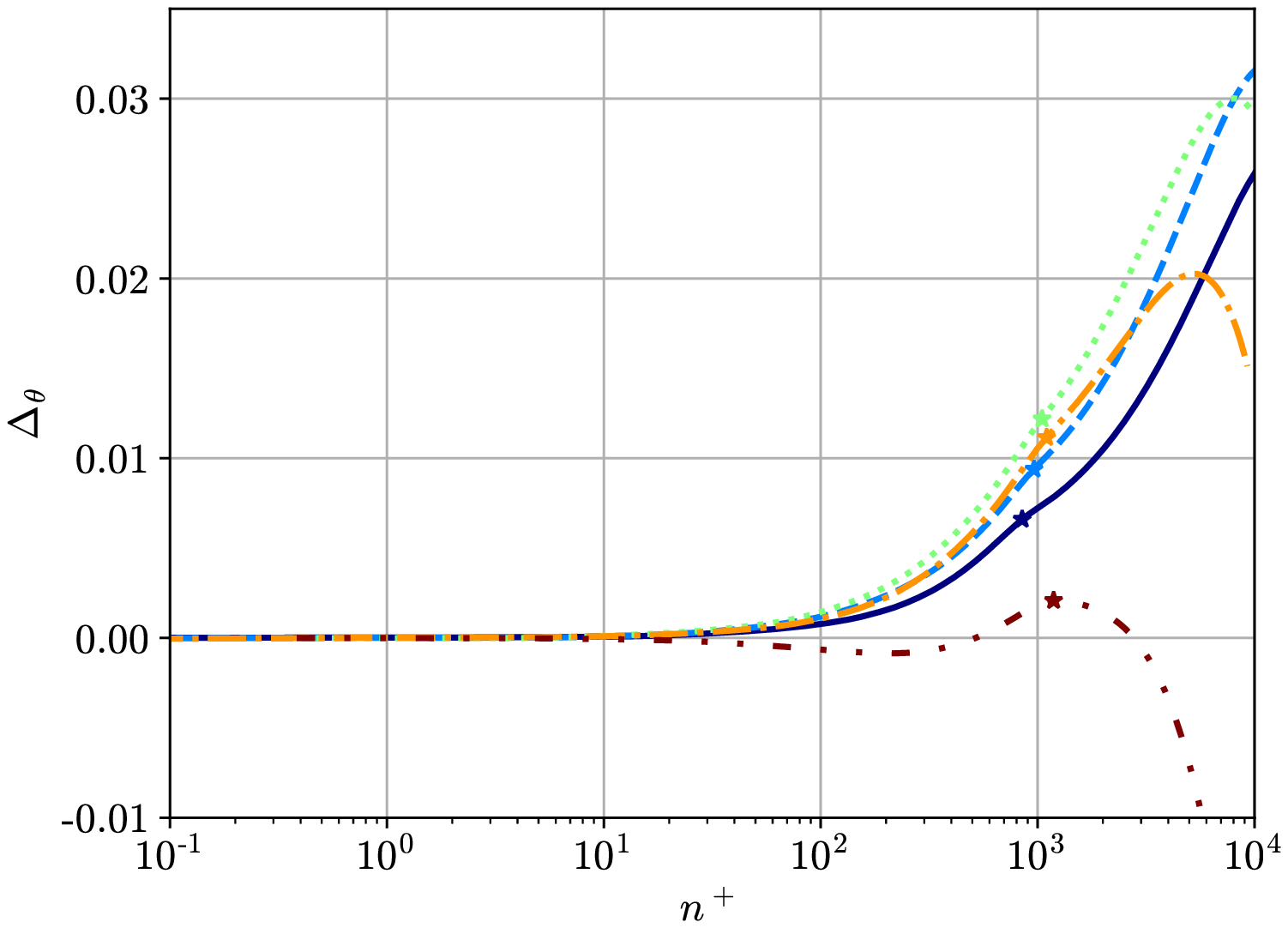}}
    \subfigure[]{\includegraphics[width=0.49\textwidth]{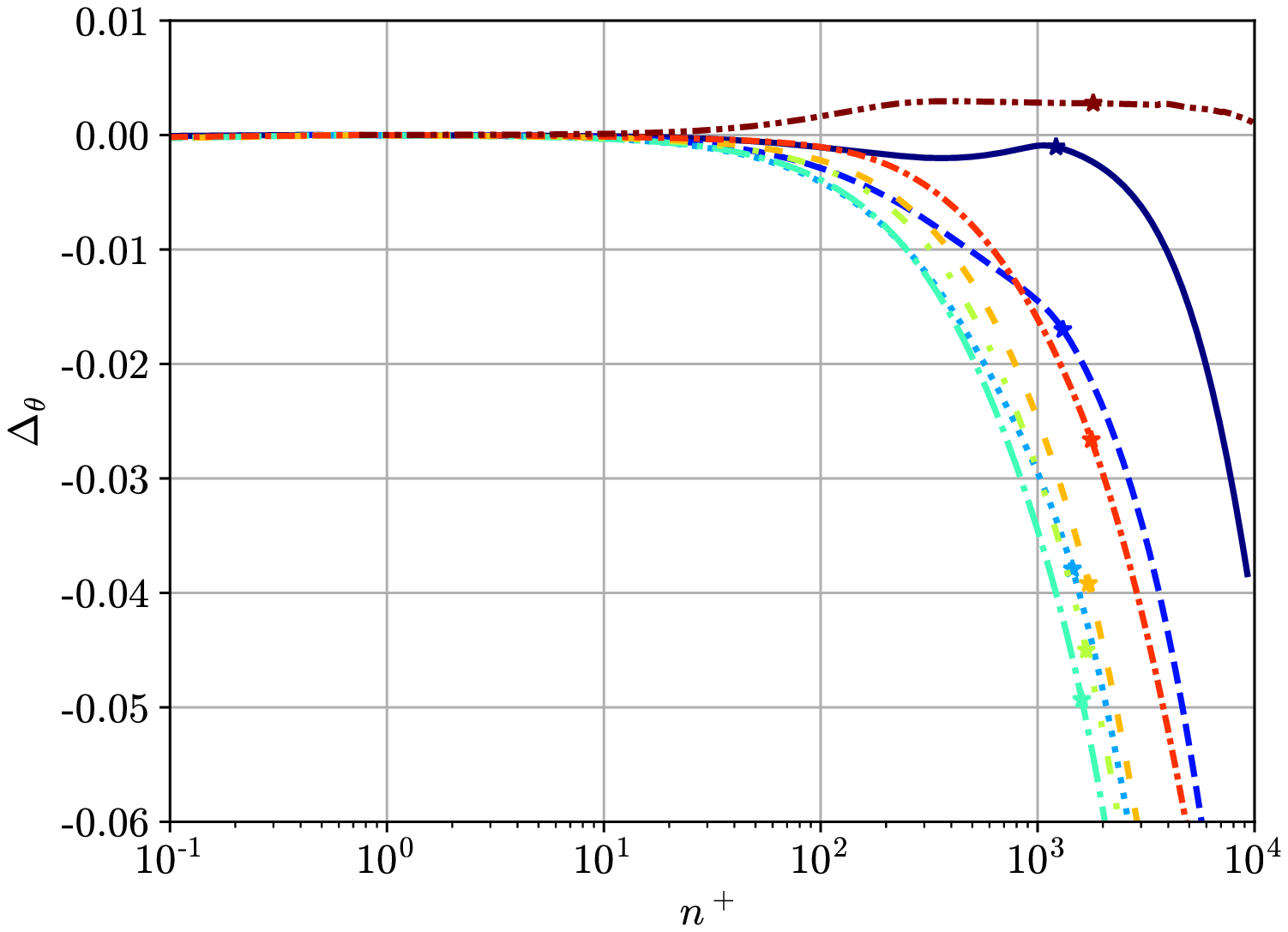}}    
    
    \caption{The angle of deflection of streamlines from the bump surface in (a) APG region and (b) FPG region. $\star$ indicates the edge of the boundary layer. Line types for APG region at $x/L$: ${\color{Violet}\bm{-}}$, $-0.5$; ${\color{Cerulean}{\bm{--}}}$, $-0.45$; ${\color{green}{\bm{\cdot}}}$, $-0.39$; ${\color{orange}{\bm{-\cdot-}}}$, $-0.35$; ${\color{Mahogany}{\bm{-\cdot\cdot-}}}$, $-0.3$. Line types for FPG region at $x/L$: ${\color{Violet}{{\bm{-}}}}$, $-0.29$; ${\color{RoyalBlue}{{\bm{--}}}}$, $-0.25$; ${\color{Cyan}{\bm{\cdot}}}$, $-0.2$; ${\color{SeaGreen}{\bm{-\; \cdot\; -}}}$, $-0.15$; ${\textcolor{green}{\bm{-\;\;\cdot\;\;\cdot\;\;-}}}$, $-0.1$ ;
    ${\color{YellowOrange}{\bm{-\;\;- }}}$, $-0.079$ ; ${\color{RedOrange}{\bm{-\cdot-}}}$, $-0.05$ ; ${\color{Mahogany}{\bm{-\cdot\cdot-}}}$, $-0.006$. }
    \label{fig:DeltaTheta_Inner}
\end{figure}

 \noindent Integrating along $\hat{\phi}$ is quite challenging. Therefore, we assume that streamlines are parallel to the bump surface to transform the integral to instead be along the wall-normal direction. The validity of the above assumption is assessed by computing the deviation of streamlines from the wall. In \figref{DeltaTheta_Inner}, we plot the angle between the streamlines and the wall, $\Delta_{\theta}$. In the mild APG region, the streamlines within the boundary layer deviate away from the wall. As the pressure gradient relaxes to zero near the APG-FPG switch, streamlines become nearly parallel to the surface in the boundary layer. In the FPG region, the streamlines deflect toward the wall. We observe that in both APG and FPG regions, the streamlines deviate from the surface although this deviation is not significant until the end of the inner region. In the inner region of the boundary layer, the assumption of alignment of streamwise-normal and wall-normal directions still seems to hold, and therefore \eref{Budget_Psi_int_2} simplifies to:
\begin{equation}
    -\overline{u_{\psi} u_{\phi}} = \int_{0}^{n} \Big(  \bar{u}_{\psi} \frac{\p \bar{u}_{\psi}}{\p \psi} + \frac{1}{\rho} \frac{\p \bar{p}}{\p \psi} \Big) d n' + \frac{\tau_w}{\rho}.
    \label{eq:Budget_Psi_int_3}
\end{equation}
\noindent The first term on the right-hand side of the above equation accounts for the effect of streamline acceleration, and the second term on the right-hand side accounts for pressure gradient effects. We can define a new velocity scale, 
\begin{equation}
    u_{*,i} = \Big( \int_{0}^{n} \Big(  \bar{u}_{\psi} \frac{\p \bar{u}_{\psi}}{\p \psi} + \frac{1}{\rho} \frac{\p \bar{p}}{\p \psi} \Big) d n' + \frac{\tau_w}{\rho} \Big)^{1/2}, 
\end{equation}
\noindent which could normalize shear stress in SCS as:
\begin{equation}
    \frac{-\overline{u_{\psi} u_{\phi}}}{u_{*,i}^2} \approx 1.
    \label{eq:Budget_Psi_norm_1}
\end{equation}
\noindent The above equation indicates that the scaled shear stress in SCS is independent of Reynolds number, pressure gradient, and acceleration effects. By assuming that the misalignment of SCS and $s-n-z$ coordinate systems is sufficiently small, we can extend this scaling to Reynolds stresses in the $s-n-z$ coordinate system.
\begin{equation}
    \frac{-\overline{u_{s} u_{n}}}{u_{*,i}^2} \approx 1.
    \label{eq:Budget_Psi_norm_sncoord}
\end{equation}
\noindent If the acceleration and pressure gradient terms are negligible: $u_{*,i} \approx u_{\tau}$, indicating the Reynolds shear stress in the wall-aligned coordinate system scales independently of the Reynolds number and follows the classical Reynolds shear stress scaling used for zero pressure gradient turbulent boundary layers. When the acceleration term is negligible but pressure gradient effects are still be significant in the inner region of the boundary layer, the velocity scale reduces to 
\begin{equation}
u_{*,i} \approx \Big( \int_{0}^{n} \frac{1}{\rho} \frac{\p \bar{p}}{\p \psi} d n' + \frac{\tau_w}{\rho} \Big)^{1/2}. 
\end{equation}
\noindent Assuming that the pressure gradient is constant in the wall-normal direction, we attain the following:
\begin{equation}
    u_{*,i} \approx \Big(\frac{n}{\rho} \frac{\p \bar{p}}{\p \psi} + u_{\tau}^2 \Big)^{1/2},
    \label{eq:u_hyb}
\end{equation}
\noindent which is similar to the hybrid velocity scale presented in \citep{Sekimoto2019, Romero2022}. These results indicate that $u_{*,i}$ scaling reduces to another commonly used scaling for Reynolds shear stress under a particular set of assumptions.

\begin{figure}
    \centering
    \subfigure[\label{UV_utau_APG}]{\includegraphics[width=0.49\textwidth]{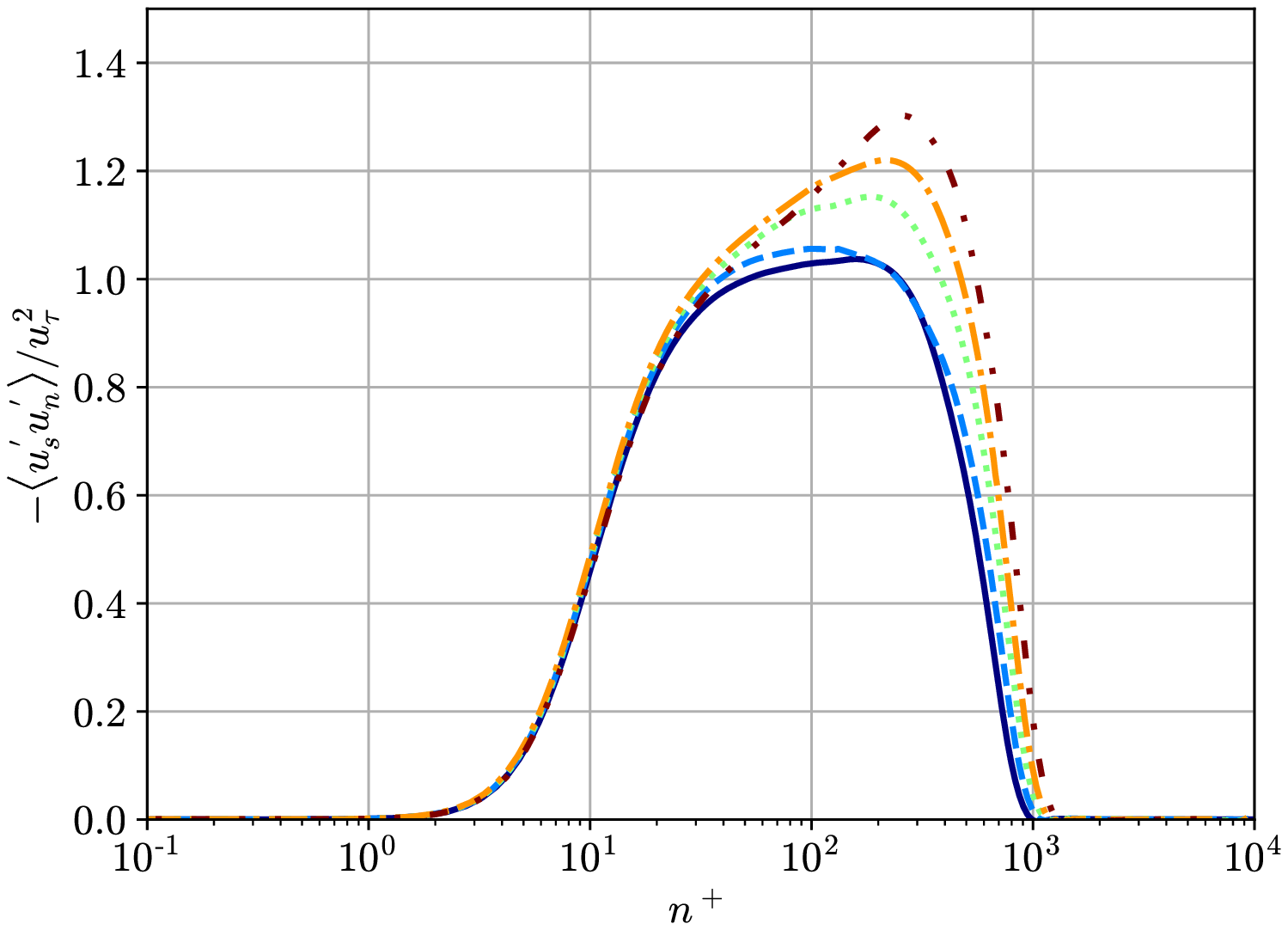}}
    \subfigure[\label{UV_uStar_APG}]{\includegraphics[width=0.49\textwidth]{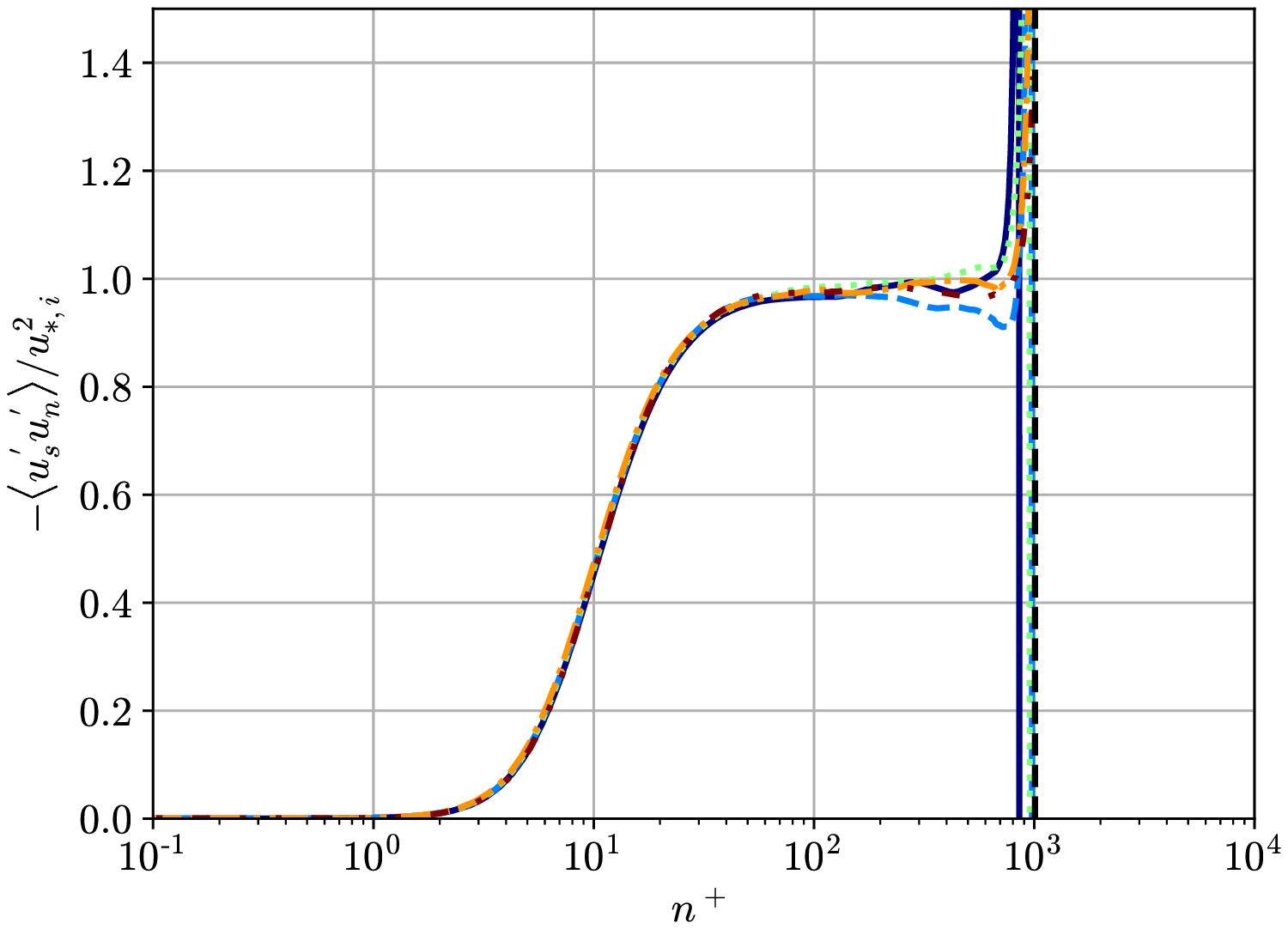}}    
    
    \caption{Reynolds shear stress in the APG region using (a) $u_{\tau}$ and (b) $u_{*,i}$. Vertical lines indicate that the quantity asymptotically goes to $\pm \infty$. Line types for APG region at $x/L$: ${\color{Violet}\bm{-}}$, $-0.5$; ${\color{Cerulean}{\bm{--}}}$, $-0.45$; ${\color{green}{\bm{\cdot}}}$, $-0.39$; ${\color{orange}{\bm{-\cdot-}}}$, $-0.35$; ${\color{Mahogany}{\bm{-\cdot\cdot-}}}$, $-0.3$.}
    \label{fig:UV_ScaleComp_APG}
\end{figure}

We compare the Reynolds shear stress profiles in the APG region with different scalings in \figref{UV_ScaleComp_APG}. We observe that friction velocity scaling does not collapse at all $x/L$ stations. At $x/L = -0.5$ and $-0.45$, the pressure gradient is slight, and the Reynolds shear stress profiles collapse and agree well with theoretical estimates of $-\overline{u_{s} u_{n}}/u_{\tau}^2 \approx 1$ sufficiently far away from the wall in the inner region of the boundary layer. As the APG increases for later downstream locations, and as $u_{\tau}$ does not account for it, the results deviate from theoretical estimates and the curves do not collapse. On the other hand, the use of $u_{*,i}$ for scaling accounts for the effect of pressure gradients and the resulting acceleration, thereby collapsing the velocity profiles to the theoretical values $-\overline{u_{s} u_{n}}/u_{*,i}^2 \approx 1$ in the inner region of the boundary layer. This analysis indicates that the Reynolds shear stress scaling in the inner region of the boundary layer depends on streamwise acceleration, streamwise pressure gradient, and wall shear stress. Upstream history or memory effects commonly attributed to non-equilibrium boundary layers subject to non-zero pressure gradients \citep{Bobke2017} are incorporated into the scaling through the streamwise acceleration, which depends on the upstream flow. Unfortunately, $u_{*,i}$ decreases faster than the rate for $-\overline{u_{s} u_{n}}$  in the outer region of the boundary layer, forming a singularity that makes the scaling ineffective far away from the wall. Furthermore, we also tested the scaling in \eref{u_hyb} and observed that it gives a good collapse of stress profiles in the inner region in the early parts of the mild APG region. However, it fails to collapse the stress profile at $x/L \approx -0.3$. This observation indicates that the advection term is essential to account for upstream history effects, which must be included in the scaling. 

\begin{figure}
    \centering
    \subfigure[\label{UV_utau_FPG}]{\includegraphics[width=0.49\textwidth]{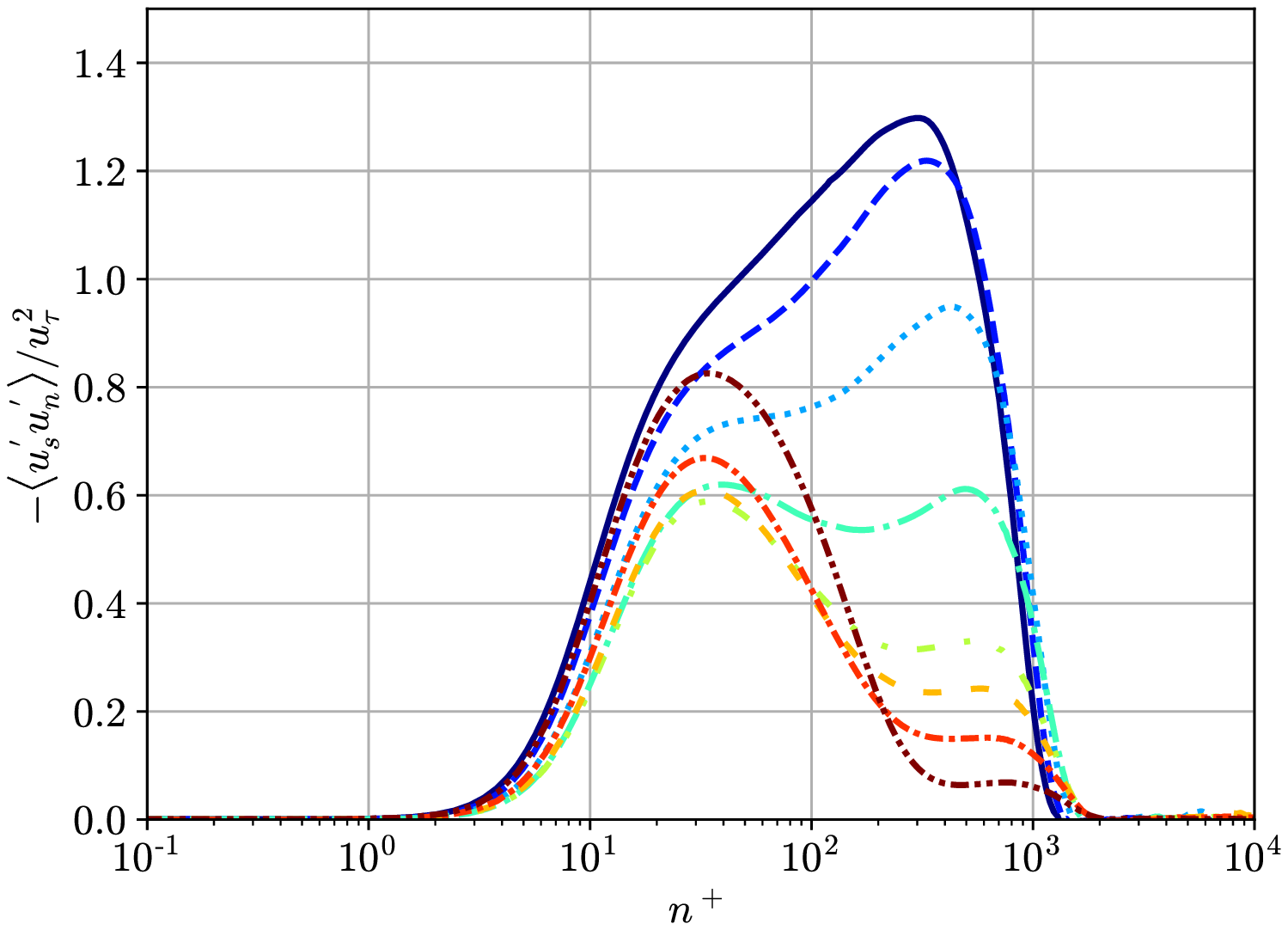}}
    \subfigure[\label{UV_uStar_FPG}]{\includegraphics[width=0.49\textwidth]{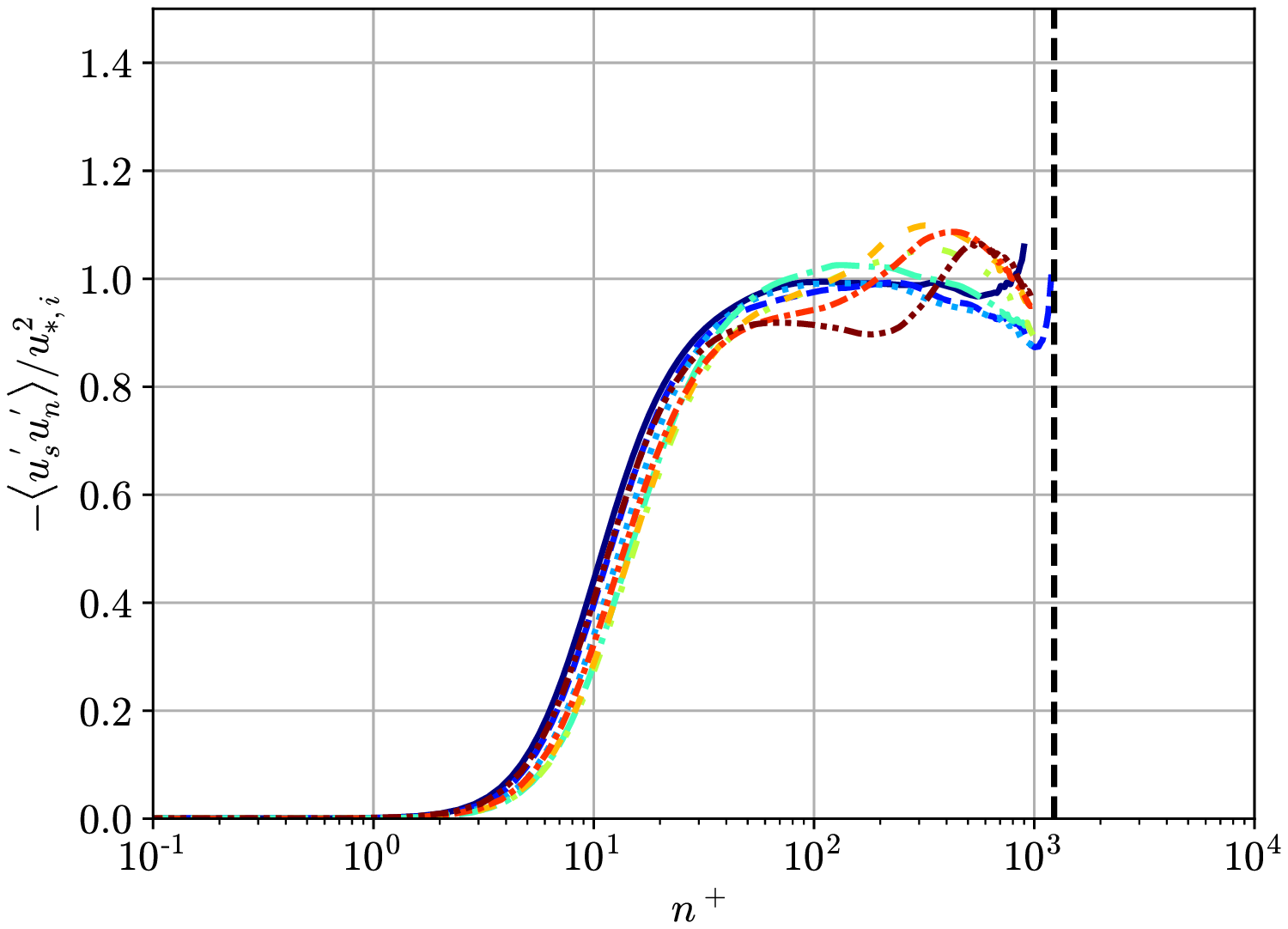}}       
    
    \caption{Reynolds shear stress in the FPG region using (a) $u_{\tau}$ and (b) $u_{*,i}$.  Vertical lines indicate that the quantity asymptotically goes to $\pm \infty$. Line types for FPG region at $x/L$: ${\color{Violet}{{\bm{-}}}}$, $-0.29$; ${\color{RoyalBlue}{{\bm{--}}}}$, $-0.25$; ${\color{Cyan}{\bm{\cdot}}}$, $-0.2$; ${\color{SeaGreen}{\bm{-\; \cdot\; -}}}$, $-0.15$; ${\textcolor{green}{\bm{-\;\;\cdot\;\;\cdot\;\;-}}}$, $-0.1$ ;
    ${\color{YellowOrange}{\bm{-\;\;- }}}$, $-0.079$ ; ${\color{RedOrange}{\bm{-\cdot-}}}$, $-0.05$ ; ${\color{Mahogany}{\bm{-\cdot\cdot-}}}$, $-0.006$. }
    \label{fig:UV_ScaleComp_FPG}
\end{figure}

We compare the Reynolds shear stress profiles in the FPG region with different scalings in \figref{UV_ScaleComp_FPG}. We observe that for all $x/L$ stations in the FPG, the friction velocity scaling does not collapse the shear stress profiles in the inner region of the boundary layer. On the other hand, the new velocity scaling leads to a much better collapse of the Reynolds shear stress profiles in the inner region. Note that the flow is experiencing pressure gradients leading to acceleration, the presence of an internal layer, and concave and convex curvature effects in this region. Even with such complex flow physics, the scaling collapses the Reynolds shear stress profiles well. Even though the collapse is good, we expected it to be better in the buffer region $n^+ < 30$. While deriving the velocity scale, we assumed that the viscous effects are negligible sufficiently far from the wall. This assumption may not hold in the FPG region where the strong pressure gradient effects lead to the laminarescent stage, where the boundary layer adjusts towards starting the relaminarization process. Therefore, we would expect viscous effects to be important even further away from the wall, and these effects should also be incorporated into the scaling. From \eref{Budget_Psi_int} and assuming the misalignment between the SCS and s-n-z coordinate systems is sufficiently small, we attain:
\begin{equation}
    -\overline{u_{\psi} u_{\phi}} = \int_{0}^{n} \Big(  \bar{u}_{\psi} \frac{\p \bar{u}_{\psi}}{\p \psi} + \frac{1}{\rho} \frac{\p \bar{p}}{\p \psi} \Big) d n' + \frac{\tau_w}{\rho} - \nu \frac{\p \bar{u}_{\phi}}{\p \phi},
    \label{eq:Budget_Psi_int_new}
\end{equation}
\noindent leading to a new viscous velocity scale, $u_{**}$:
\begin{equation}
    \frac{ -\overline{u_{s} u_{n}}}{u^2_{**,i}}  \approx 1, \quad \text{where}  \quad u_{**,i} = \Big(\int_{0}^{n} \Big(  \bar{u}_{\psi} \frac{\p \bar{u}_{\psi}}{\p \psi} + \frac{1}{\rho} \frac{\p \bar{p}}{\p \psi} \Big) d n' + \frac{\tau_w}{\rho} - \nu \frac{\p \bar{u}_{\phi}}{\p \phi} \Big)^{1/2}.
\end{equation}

\begin{figure}
    \subfigure[\label{UV_uStarStar_APG}]{\includegraphics[width=0.49\textwidth]{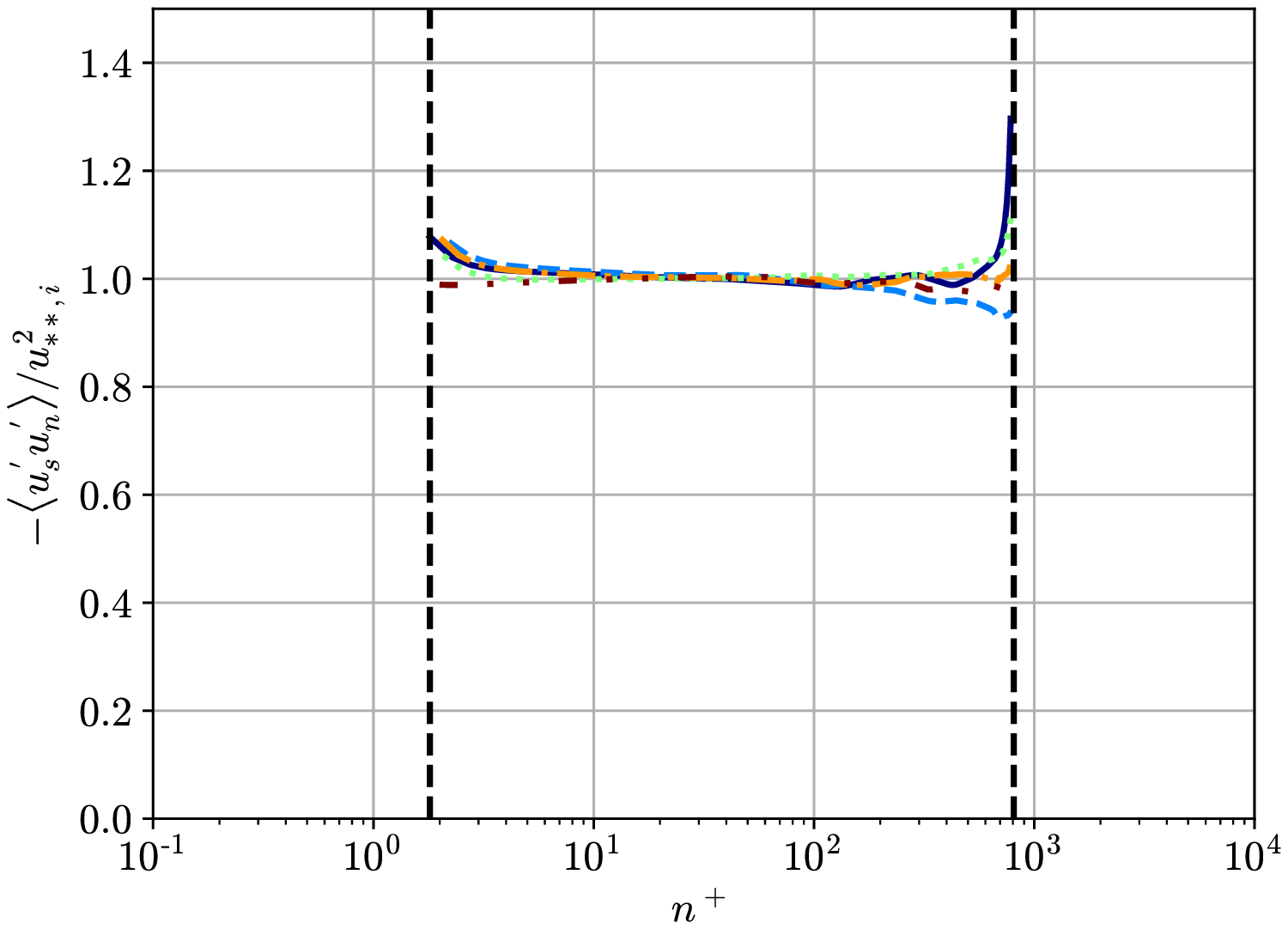}}
    \subfigure[\label{UV_uStarStar_FPG}]{\includegraphics[width=0.49\textwidth]{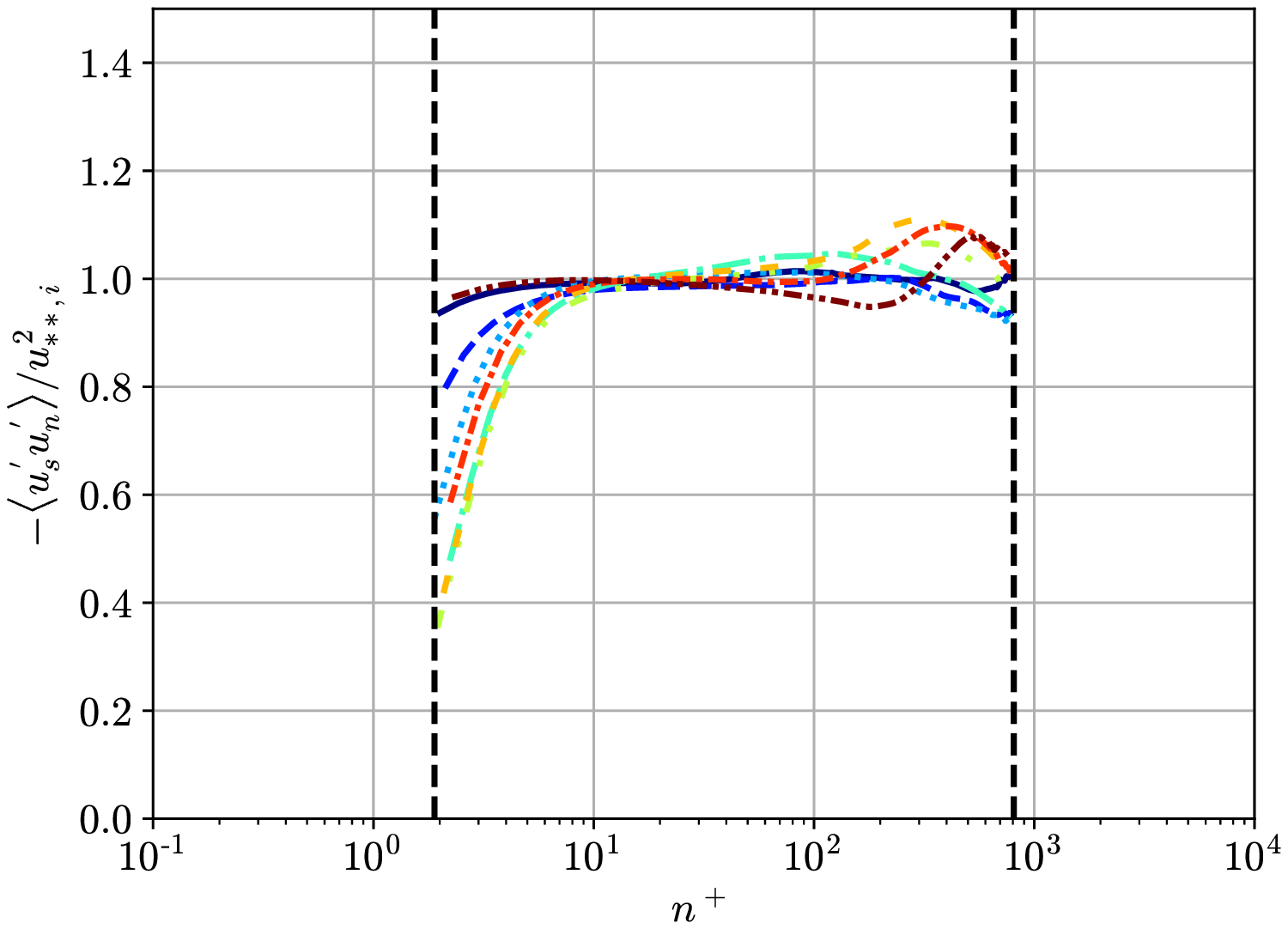}}
    \caption{Reynolds shear stress in the (a) APG region and (b) FPG region scaled by $u_{**,i}$. Vertical lines indicate that the quantity asymptotically goes to $\pm \infty$. Line types for APG region at $x/L$: ${\color{Violet}\bm{-}}$, $-0.5$; ${\color{Cerulean}{\bm{--}}}$, $-0.45$; ${\color{green}{\bm{\cdot}}}$, $-0.39$; ${\color{orange}{\bm{-\cdot-}}}$, $-0.35$; ${\color{Mahogany}{\bm{-\cdot\cdot-}}}$, $-0.3$. Line types for FPG region at $x/L$: ${\color{Violet}{{\bm{-}}}}$, $-0.29$; ${\color{RoyalBlue}{{\bm{--}}}}$, $-0.25$; ${\color{Cyan}{\bm{\cdot}}}$, $-0.2$; ${\color{SeaGreen}{\bm{-\; \cdot\; -}}}$, $-0.15$; ${\textcolor{green}{\bm{-\;\;\cdot\;\;\cdot\;\;-}}}$, $-0.1$ ;
    ${\color{YellowOrange}{\bm{-\;\;- }}}$, $-0.079$ ; ${\color{RedOrange}{\bm{-\cdot-}}}$, $-0.05$ ; ${\color{Mahogany}{\bm{-\cdot\cdot-}}}$, $-0.006$. }
    \label{fig:UV_ViscScale_FPG}
\end{figure}

We show the Reynolds shear stress using $u_{**,i}$ scaling in \figref{UV_ViscScale_FPG}. The $u_{**,i}$ scaling does not significantly affect the collapse of stress profiles in the APG region compared to $u_{*,i}$. However, it substantially improves the collapse of Reynolds shear stress profiles in the FPG region in the buffer and log regions ($10 \leq n^+ \leq 100$) of the boundary layer compared to $u_{*,i}$ scaling. These results indicate that for flows where viscous effects become substantial, the scaling of Reynolds shear stress must be adjusted to accommodate the thick near-wall viscous region. Furthermore, with a better collapse of Reynolds shear stress profiles with this proposed scaling, it appears that the upstream history effects and development of internal layers are also accounted for.

Even though these new velocity scales work well for Reynolds shear stress, they do not work well for Reynolds normal stresses. Therefore, we follow a similar procedure for deriving scaling Reynolds shear stresses to derive the scaling for wall-normal normal stress components. In this case, we consider the $\Phi$-momentum budget, \eref{Budget_Phi} and remove all the equation components that had negligible influence as observed in Section \ref{sec:SCS_budget_Psi}, 
\begin{equation}
    \frac{\bar{u}_{\psi}^2}{R} = -\frac{1}{\rho} \frac{\p \bar{p}}{\p \phi} - \frac{\p \overline{u^2_{\phi}}}{\p \phi}.
    \label{eq:Budget_Phi_reduced}
\end{equation}
\noindent Rearranging this equation and integrating it along $\hat{\phi}$, we obtain:
\begin{equation}
    \overline{u^2_{\phi}} = - \int_{0}^{\phi} \Big( \frac{\bar{u}_{\psi}^2}{R} + \frac{1}{\rho} \frac{\p \bar{p}}{\p \phi} \Big) d \phi' = - \int_{0}^{\phi} \frac{\bar{u}_{\psi}^2}{R} d \phi' + \frac{\bar{p}_w - p}{\rho}.
    \label{eq:Budget_Phi_reduced_2}
\end{equation}
\noindent To simplify the calculation, we assume alignment between $\hat{\phi}$ and $\hat{n}$, which seems to be valid in the inner region as shown in \figref{DeltaTheta_Inner}. 
\begin{equation}
    \overline{u^2_{\phi}} = - \int_{0}^{n} \frac{\bar{u}_{\psi}^2}{R} d n' + \frac{\bar{p}_w - p}{\rho}.
    \label{eq:Budget_Phi_reduced_3}
\end{equation}
\noindent Therefore, we can define a new velocity scale for the streamline-normal normal stress,
\begin{equation}
u_{\gamma,i} = \Big( - \int_{0}^{n} \frac{\bar{u}_{\psi}^2}{R} d n' + \frac{\bar{p}_w - p}{\rho} \Big)^{1/2},    
\end{equation}
\noindent which satisfies following equation in the near-wall region:
\begin{equation}
    \frac{ \overline{u^2_{\phi}}}{u^2_{\gamma,i}} \approx \frac{ \overline{u^2_{n}}}{u^2_{\gamma,i}} \approx 1.
\end{equation}
\noindent The expression for $u_{\gamma,i}$ indicates that the streamline-normal normal stress component dynamically depends on the centrifugal force and local pressure. In the absence of the streamline curvature effects, the centrifugal force term is expected to be negligible, resulting in a dependence on the difference between the local pressure and the pressure at the wall.

\begin{figure}
    \centering
    \subfigure[\label{VV_utau_APG}]{\includegraphics[width=0.49\textwidth]{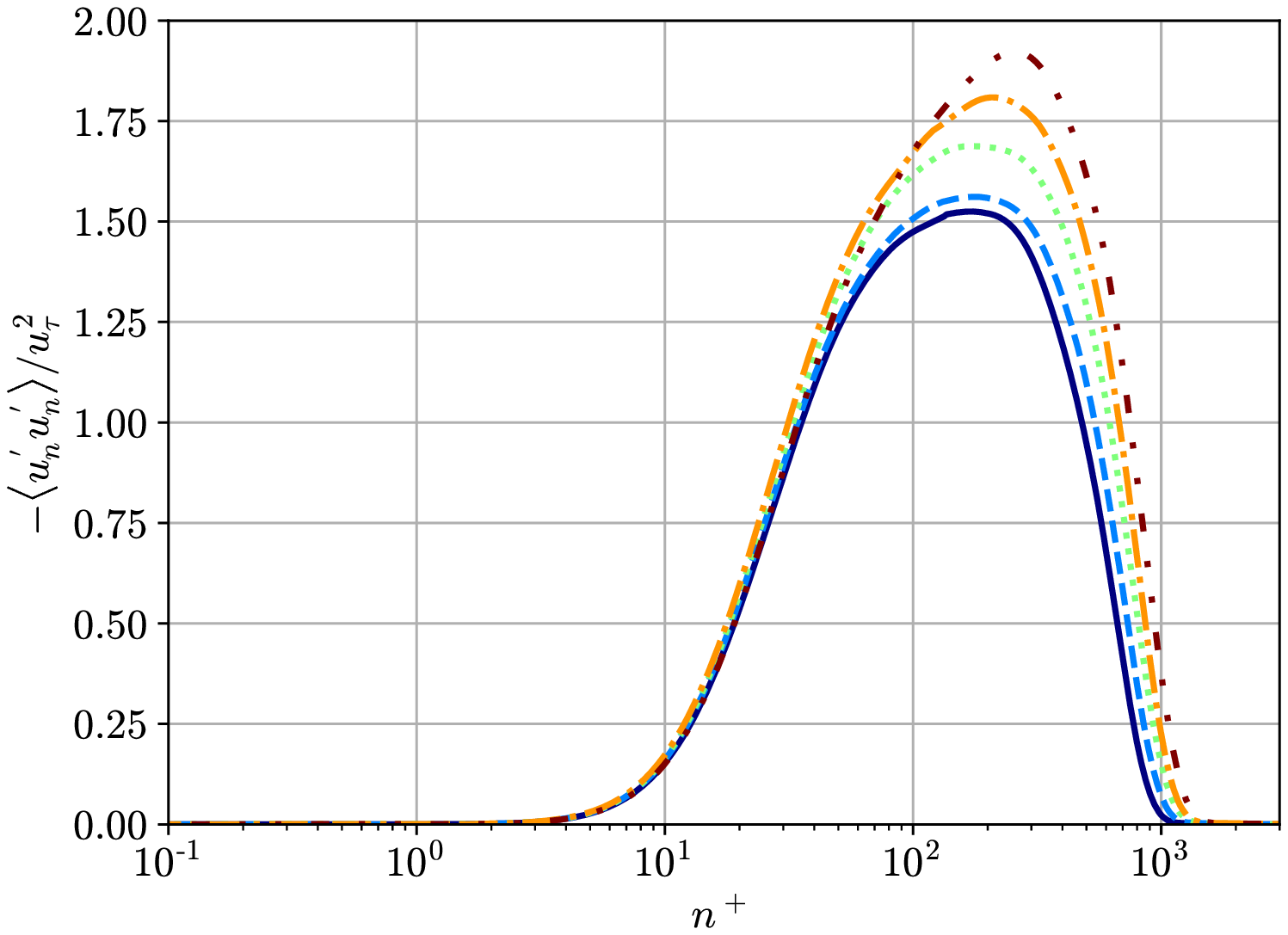}}
    \subfigure[\label{VV_ugamma_APG}]{\includegraphics[width=0.49\textwidth]{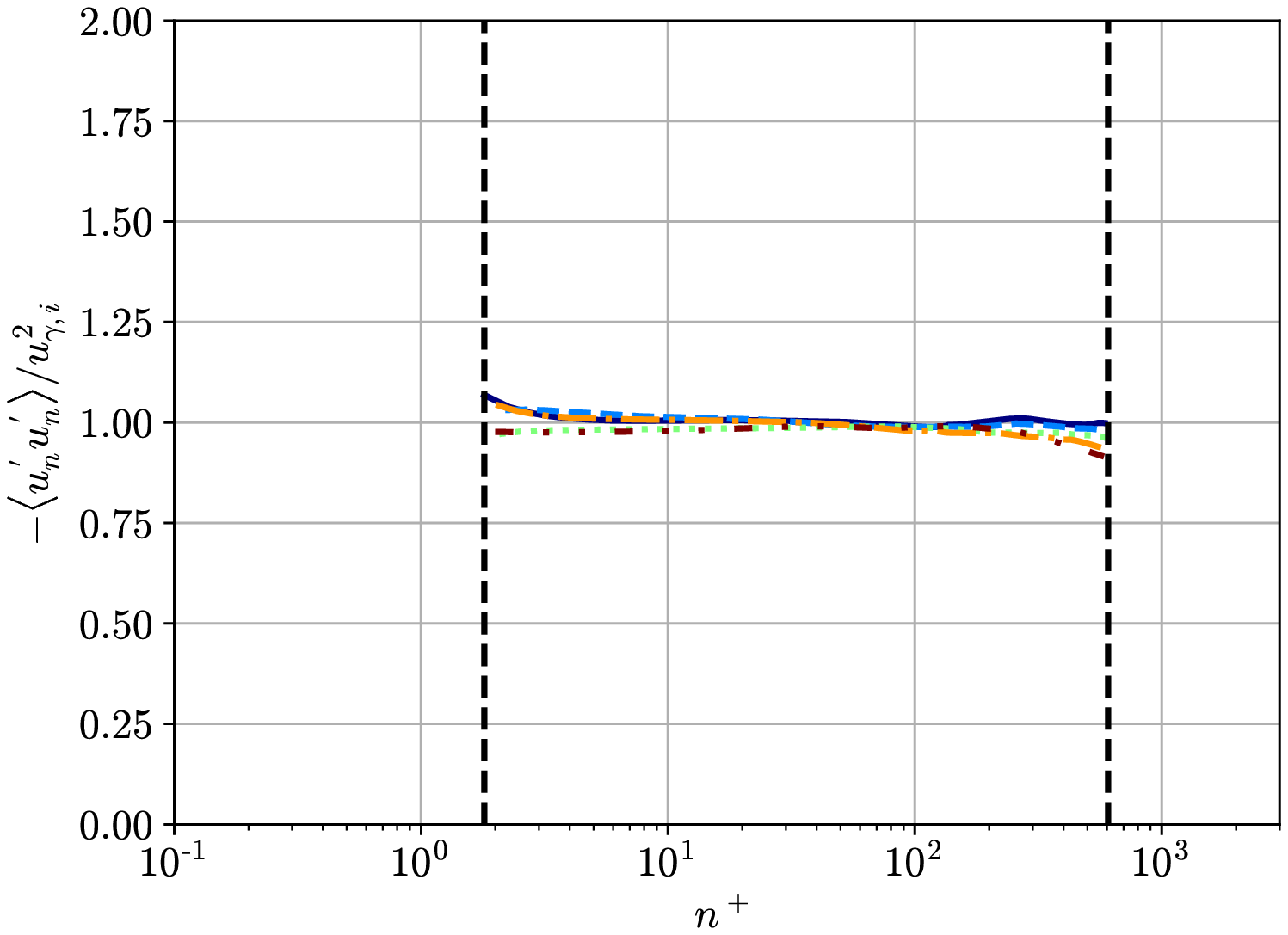}}    
    
    \caption{Wall-normal normal stress in APG region scaled by  (a) $u_{\tau}$ and (b) $u_{\gamma,i}$. Vertical lines indicate that the quantity asymptotically goes to $\pm \infty$. Line types for APG region at $x/L$: ${\color{Violet}\bm{-}}$, $-0.5$; ${\color{Cerulean}{\bm{--}}}$, $-0.45$; ${\color{green}{\bm{\cdot}}}$, $-0.39$; ${\color{orange}{\bm{-\cdot-}}}$, $-0.35$; ${\color{Mahogany}{\bm{-\cdot\cdot-}}}$, $-0.3$.}
    \label{fig:VV_ScaleComp_APG}
\end{figure}

\begin{figure}
    \centering
    \subfigure[\label{VV_utau_FPG}]{\includegraphics[width=0.49\textwidth]{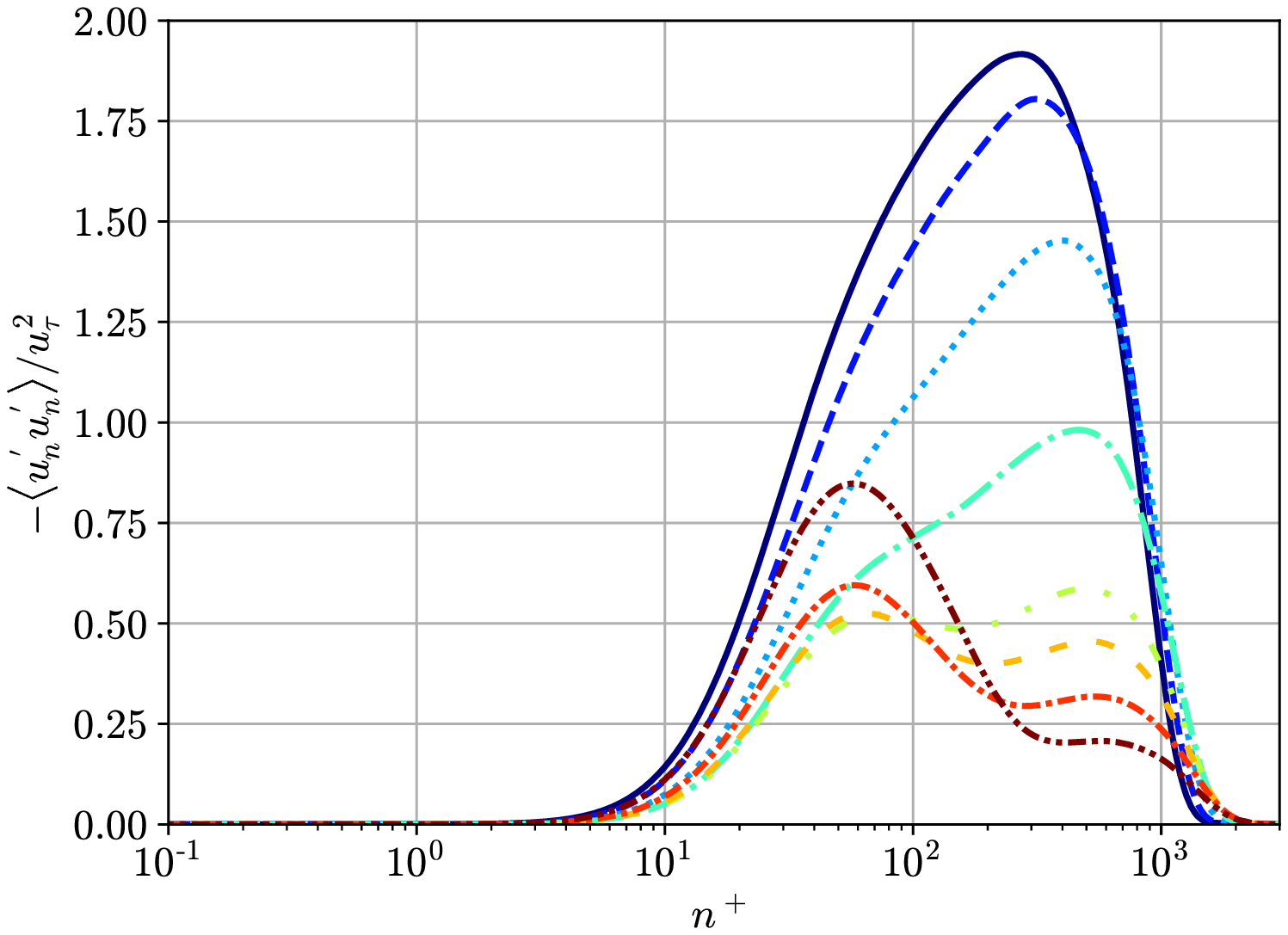}}
    \subfigure[\label{VV_ugamma_FPG}]{\includegraphics[width=0.49\textwidth]{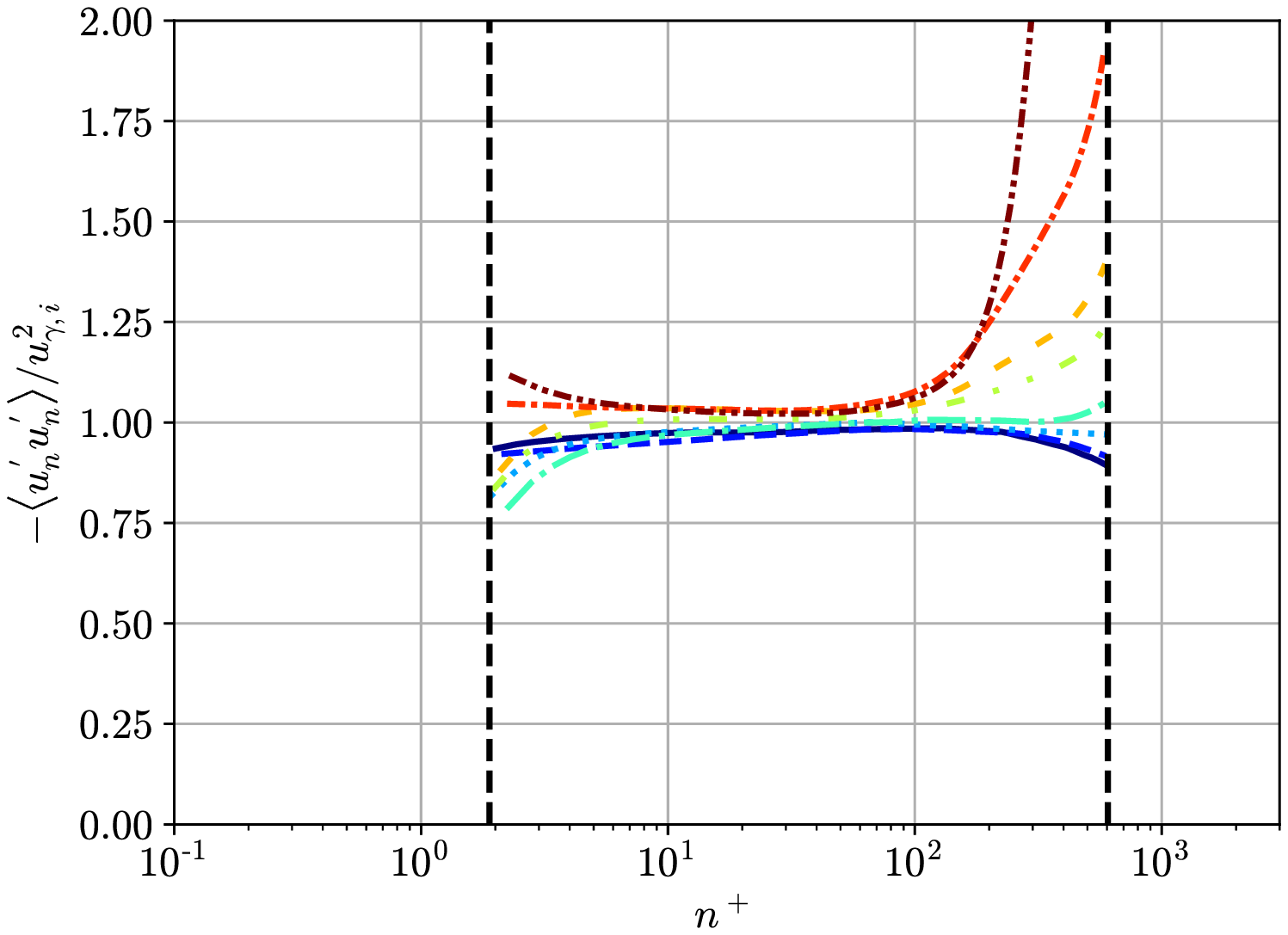}}    
        
    \caption{Wall-normal normal stress in FPG region scaled by  (a) $u_{\tau}$ and (b) $u_{\gamma,i}$. Vertical lines indicate that the quantity asymptotically goes to $\pm \infty$. Line types for FPG region at $x/L$: ${\color{Violet}{{\bm{-}}}}$, $-0.29$; ${\color{RoyalBlue}{{\bm{--}}}}$, $-0.25$; ${\color{Cyan}{\bm{\cdot}}}$, $-0.2$; ${\color{SeaGreen}{\bm{-\; \cdot\; -}}}$, $-0.15$; ${\textcolor{green}{\bm{-\;\;\cdot\;\;\cdot\;\;-}}}$, $-0.1$ ;
    ${\color{YellowOrange}{\bm{-\;\;- }}}$, $-0.079$ ; ${\color{RedOrange}{\bm{-\cdot-}}}$, $-0.05$ ; ${\color{Mahogany}{\bm{-\cdot\cdot-}}}$, $-0.006$. }
    \label{fig:VV_ScaleComp_FPG}
\end{figure}

The scaled wall-normal normal stress for the APG and FPG regions are shown in \figref{VV_ScaleComp_APG} and \figref{VV_ScaleComp_FPG} respectively. We observe that $u_{\tau}$ scaling does not lead to a collapse of stress profiles in both APG and FPG regions. Using $u_{\gamma,i}$ for scaling the stress profile, we get a much better collapse and agreement with the theoretical estimates in the inner boundary layer region for both APG and FPG regions of the flow. These results indicate that including local pressure gradient and centrifugal force terms is important to account for the dynamics of these quantities in the scaling. Our scaling analysis leads to the derivation of scaling for $\langle u'_{s} u'_{n} \rangle$ and $\langle u'_{n} u'_{n} \rangle$. However, these derived scalings are specific to $\langle u'_{s} u'_{n} \rangle$ and $\langle u'_{n} u'_{n} \rangle$, and do not extend well to other Reynolds stress components: $\langle u'_{s} u'_{s} \rangle$ and $\langle u'_z u'_z \rangle$.

\subsubsection{Outer Region}
\label{sec:IntAnalysis_Outer}

\begin{figure}
    \centering
    \subfigure[\label{VV_ZS_APG}]{\includegraphics[width=0.49\textwidth]{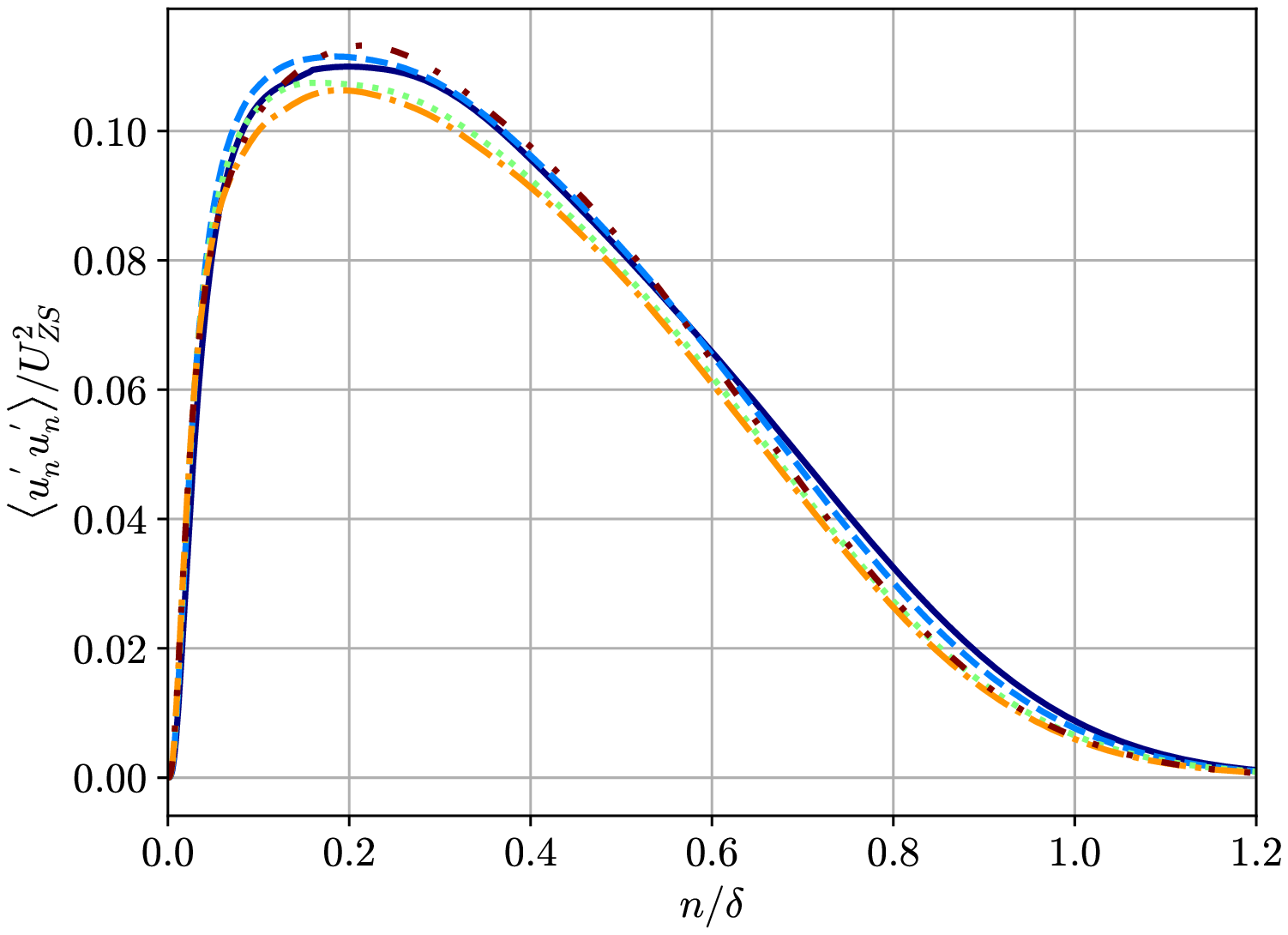}}
    \subfigure[\label{UV_ZS_APG}]{\includegraphics[width=0.49\textwidth]{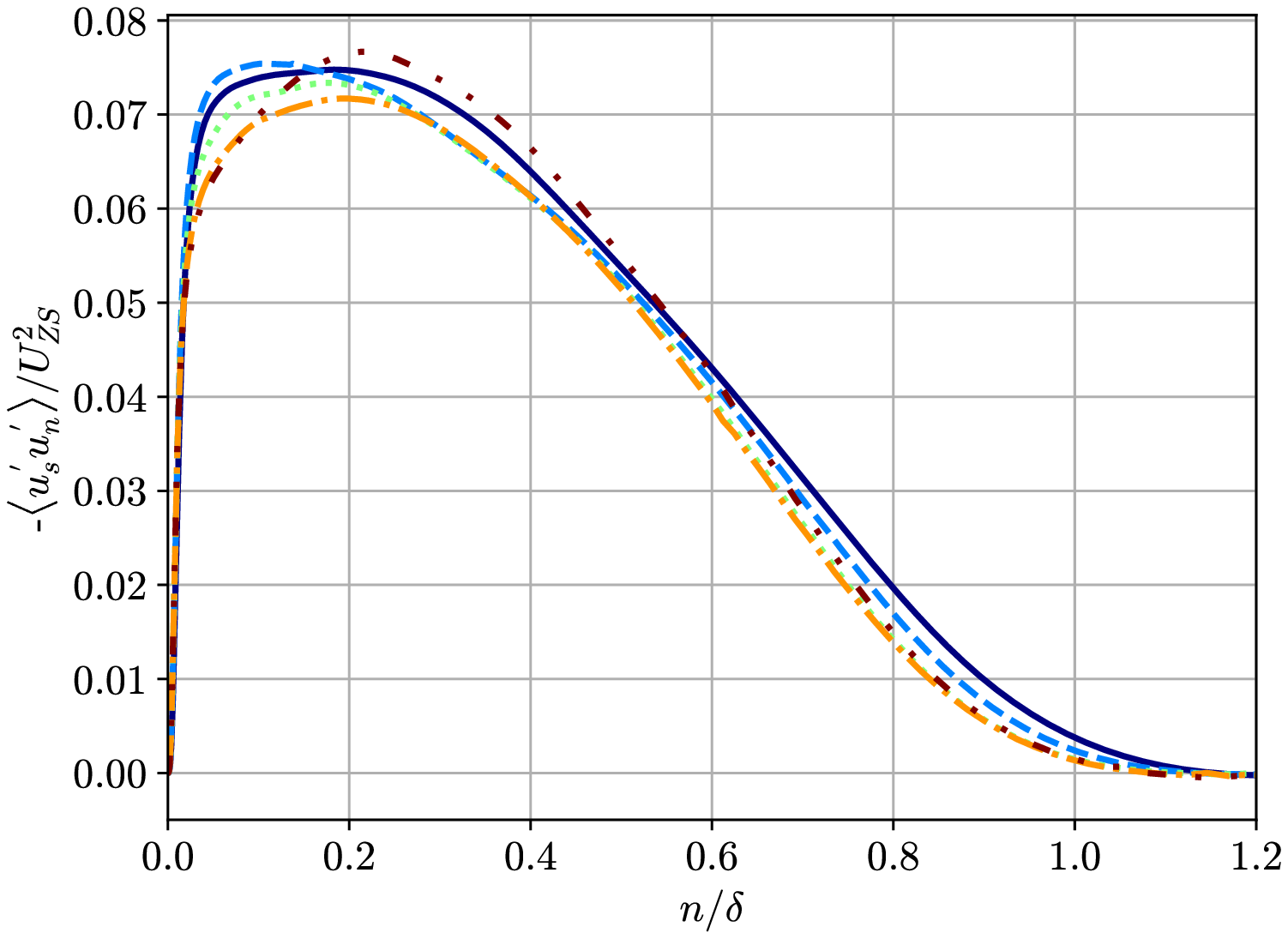}}

    \caption{Reynolds stress tensor components: (a) $\overline{u'_n u'_n}$ and (b) -$\overline{u'_s u'_n}$  scaled in outer layer units in the mild APG region of the boundary layer. Line types for APG region at $x/L$: ${\color{Violet}\bm{-}}$, $-0.5$; ${\color{Cerulean}{\bm{--}}}$, $-0.45$; ${\color{green}{\bm{\cdot}}}$, $-0.39$; ${\color{orange}{\bm{-\cdot-}}}$, $-0.35$; ${\color{Mahogany}{\bm{-\cdot\cdot-}}}$, $-0.3$.}
    \label{fig:RSS_OuterScaling_APG}
\end{figure}

\begin{figure}
    \centering
    \subfigure[\label{VV_ZS_FPG}]{\includegraphics[width=0.49\textwidth]{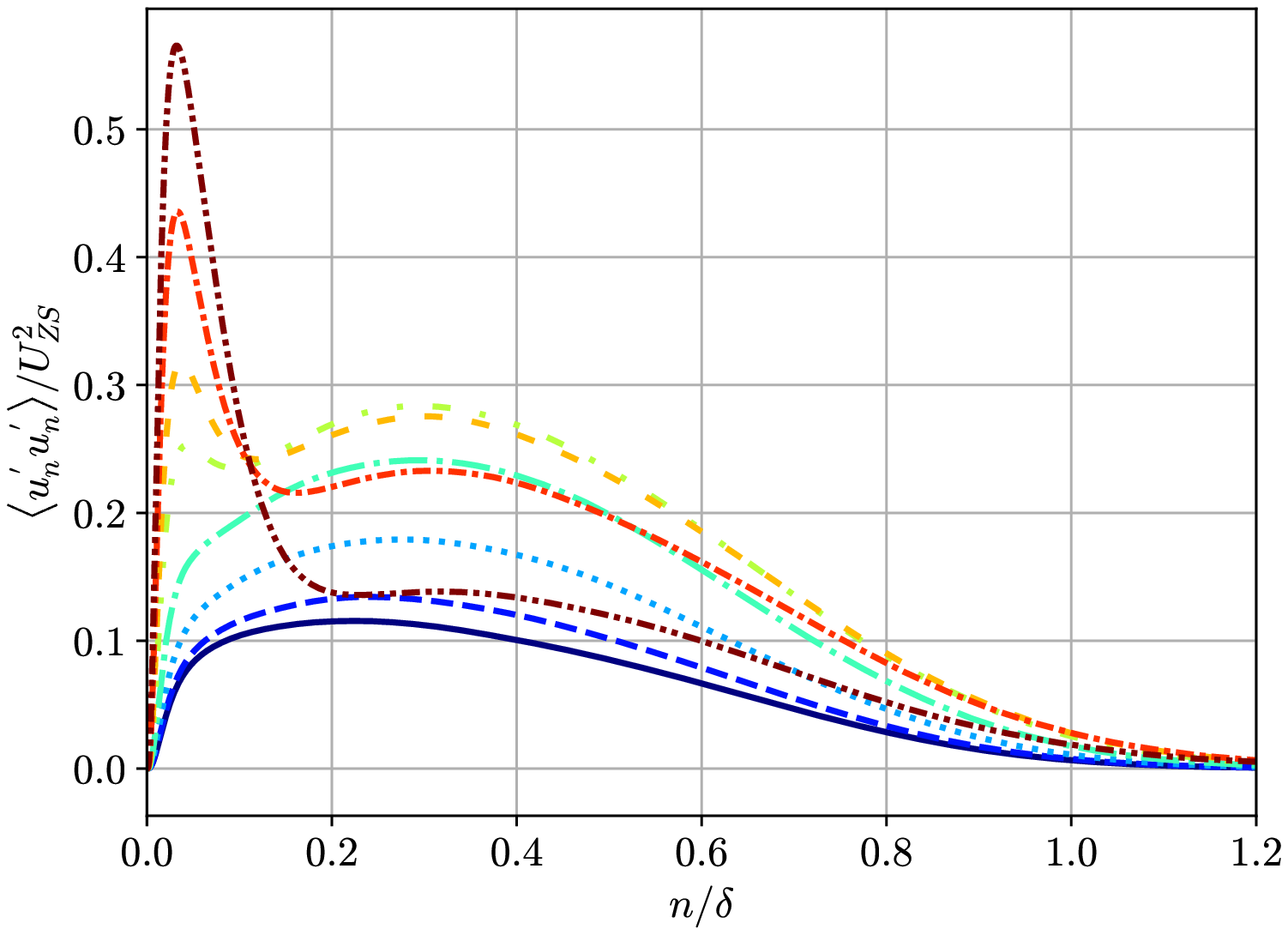}}
    \subfigure[\label{UV_ZS_FPG}]{\includegraphics[width=0.49\textwidth]{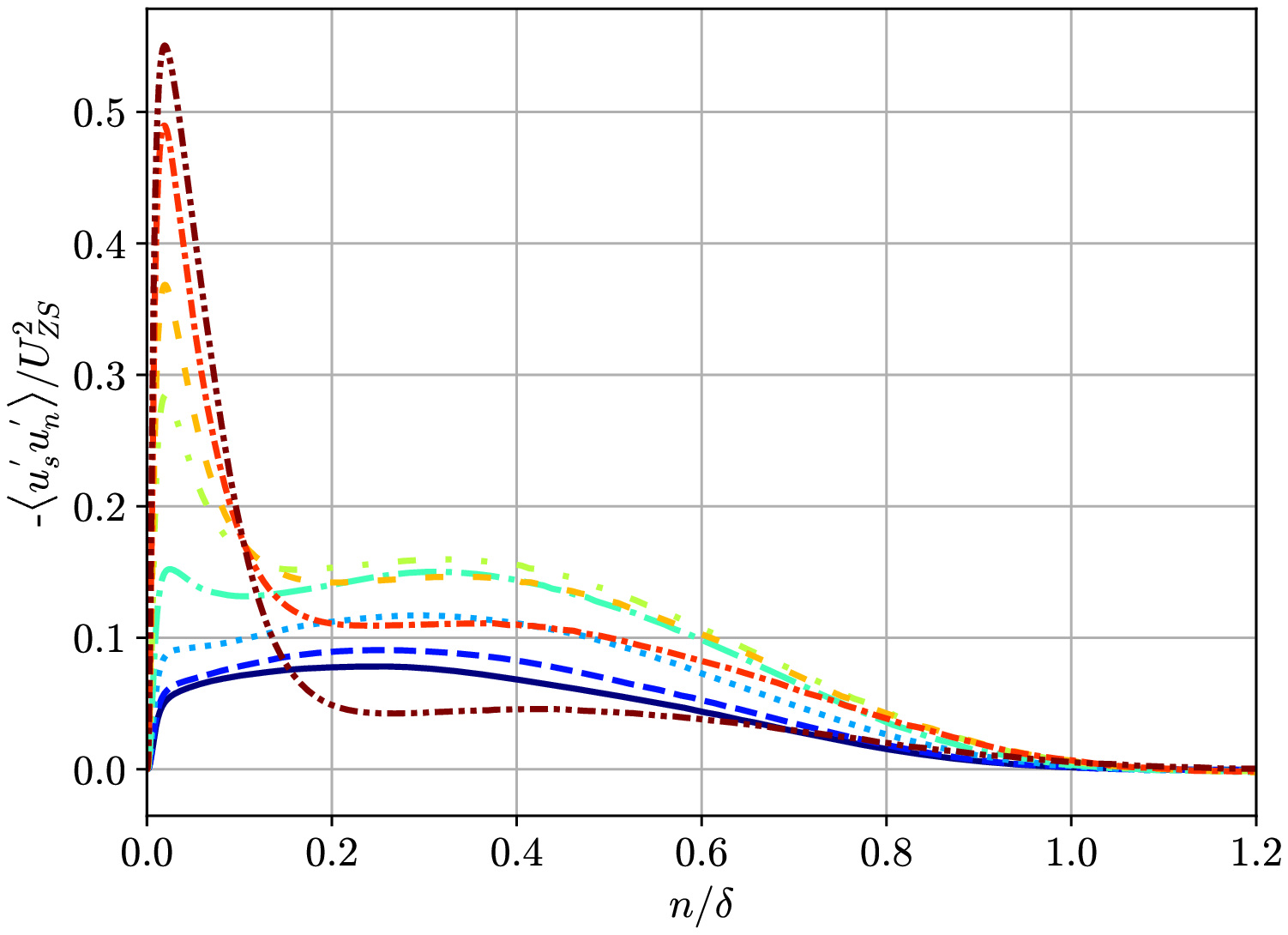}}

    \caption{Reynolds stress tensor components: (a) $\overline{u'_n u'_n}$ and (b) -$\overline{u'_s u'_n}$ scaled in outer layer units in the strong FPG region of the boundary layer. Line types for FPG region at $x/L$: ${\color{Violet}{{\bm{-}}}}$, $-0.29$; ${\color{RoyalBlue}{{\bm{--}}}}$, $-0.25$; ${\color{Cyan}{\bm{\cdot}}}$, $-0.2$; ${\color{SeaGreen}{\bm{-\; \cdot\; -}}}$, $-0.15$; ${\textcolor{green}{\bm{-\;\;\cdot\;\;\cdot\;\;-}}}$, $-0.1$ ;
    ${\color{YellowOrange}{\bm{-\;\;- }}}$, $-0.079$ ; ${\color{RedOrange}{\bm{-\cdot-}}}$, $-0.05$ ; ${\color{Mahogany}{\bm{-\cdot\cdot-}}}$, $-0.006$. }
    \label{fig:RSS_OuterScaling_FPG}
\end{figure}

Before deriving the scalings for Reynolds stresses in the outer region using integral analysis, we first define commonly used scaling. The scaling of Reynolds stresses in the outer region is often represented using the defect law for the Reynolds stress tensor \citep{Clauser1956, Castillo2001}:
\begin{equation}
    \frac{\overline{u_i u_j}}{R_{ij,o}} = f_{ij} \Big(\frac{n}{\delta}\Big),
\end{equation}
\noindent where $R_{ij,o}$ is the scaling for $ij^{th}$ component of the Reynolds stress tensor in the outer region of the boundary layer. If scaling for Reynolds stresses is based on the same velocity scale in both the inner and outer region of the boundary layer \citep{Clauser1954}, then friction velocity is used for scaling the stresses in the outer region and $R_{ij,o} = u_{\tau}^2$. Similarly, two velocity scale approaches can be taken for Reynolds stress, where $R_{ij,i} = u_{\tau}^2 \ne R_{ij,o}$. Using similarity analysis, \cite{Castillo2001} derived a scaling of Reynolds stresses in the outer region of the turbulent boundary layer in the s-n-z coordinate system as,
\begin{equation}
    R_{sn,o} = \bar{u}_s^2 \frac{\p \delta}{\p s}, \quad R_{ss,o} = \bar{u}_s^2, \quad R_{nn,o} = \bar{u}_s^2.
\end{equation}
\cite{Brzek2005} showed that this scaling does not collapse the stress profiles in the presence of pressure gradients. Another common outer region velocity scale is the Zagarola-Smits velocity scale, 
\begin{equation}
U_{ZS} = \bar{u}_{s,e}\frac{\delta^*}{\delta},   
\end{equation} 
\noindent where $\bar{u}_{s,e}$ is the velocity at the edge of boundary layer and $\delta^*$ is the displacement thickness. \cite{Brzek2005} showed that Zagarola-Smits velocity scale ($R_{ij,o} = U_{ZS}^2$) works better for scaling the Reynolds stresses in the presence of pressure gradients. For the bump flow, the scaled Reynolds stress profiles in the APG region of the flow are shown in \figref{RSS_OuterScaling_APG}. The Reynolds stress tensor profiles collapse well when scaled by $U_{ZS}$ in the mild APG of the boundary layer. As shown in \figref{RSS_OuterScaling_FPG}, this behavior is not observed for the strong FPG region of the flow where the stress profiles do not collapse for all stress components. In the strong FPG region, the outer region peak of stress components increases until $x/L = -0.1$, close to the maximum $\beta$ location. Downstream of this location, there is a decrease in the outer region peak. A case could be made for the inclusion of upstream effects. However, several studies \citep{Wosnik2000, George2006} have shown that using the $ZS$ velocity scale already incorporates these effects. These results indicate that a local scaling may not work well for strong FPG regions of the flow, therefore integral-analysis based scalings for Reynolds stresses need to be considered.

 The momentum budget in the SCS indicates that the viscous terms are negligible in the outer region of the boundary layer, and the $\psi$-momentum equation reduces to,
\begin{equation}
    \bar{u}_{\psi} \frac{\p \bar{u}_{\psi}}{\p \psi} = -\frac{1}{\rho} \frac{\p \bar{p}}{\p \psi} - \frac{\p \overline{u_{\psi}^2}}{\p \psi} - \frac{\p \overline{u_{\psi} u_{\phi}}}{\p \phi}.
    \label{eq:Budget_Psi_reduced_outer}
\end{equation}  
\noindent Rearranging the equation and integrating it from $\phi$ to freestream, we get,
\begin{equation}
    - \overline{u_{\psi} u_{\phi}} = \int_{\phi}^{\infty} \Big(  \bar{u}_{\psi} \frac{\p \bar{u}_{\psi}}{\p \psi} + \frac{1}{\rho} \frac{\p \bar{p}}{\p \psi} + \frac{\p \overline{u_{\psi}^2}}{\p \psi} \Big) d \phi'.
    \label{eq:UVscaleOuterInt}
\end{equation}
\noindent From the discussion in Section \ref{sec:IntAnalysis_Inner}, we observe that the $\hat{\phi}$ and $\hat{n}$ have some misalignment in the outer region of the boundary layer for the bump flow. Nevertheless, within the boundary layer, that misalignment may not be significant enough to spoil scaling, and thus the integral in \eref{UVscaleOuterInt} may be approximated using wall-normal coordinate as the integrand. This assumption reduces the expression to,
\begin{equation}
    - \overline{u_{\psi} u_{\phi}} \approx \int_{n}^{\infty} \Big(  \bar{u}_{\psi} \frac{\p \bar{u}_{\psi}}{\p \psi} + \frac{1}{\rho} \frac{\p \bar{p}}{\p \psi} + \frac{\p \overline{u_{\psi}^2}}{\p \psi} \Big) d n'.
    \label{eq:UVscaleOuterInt_2}
\end{equation}
\noindent Classically, $\frac{\p \overline{u^2_{\psi}}}{\p \psi}$ is often assumed to be insignificant relative to the other terms of the equation. The $\psi$-momentum equation budgets shown in \figref{PsiMom_FPG_Outer} indicated that this term is small compared to the momentum flux due to total pressure and Reynolds shear stress. However, as this term is non-zero, it may still impact the scaling behavior of the Reynolds shear stress terms. Therefore, we compare two scalings,
\begin{equation}
    \frac{-\overline{u_{\psi} u_{\phi}}}{u_{*,o}^2} \approx \frac{-\overline{u_{s} u_{n}}}{u_{*,o}^2} \approx 1, \; u_{*,o} = \int_{n}^{\infty} \Big(  \bar{u}_{\psi} \frac{\p \bar{u}_{\psi}}{\p \psi} + \frac{1}{\rho} \frac{\p \bar{p}}{\p \psi} + \frac{\p \overline{u_{\psi}^2}}{\p \psi} \Big) d n',
\end{equation}
\noindent and,
\begin{equation}
    \frac{-\overline{u_{\psi} u_{\phi}}}{u_{**,o}^2} \approx \frac{-\overline{u_{s} u_{n}}}{u_{**,o}^2} \approx 1,
 \; u_{**,o} = \int_{n}^{\infty} \Big(  \bar{u}_{\psi} \frac{\p \bar{u}_{\psi}}{\p \psi} + \frac{1}{\rho} \frac{\p \bar{p}}{\p \psi} \Big) d n',
\end{equation}
\noindent where $u_{*,o}$ includes the influence of $\frac{\p \overline{u_{\psi}^2}}{\p \psi}$ term, while $u_{**,o}$ neglects this influence.  This scaling varies locally with wall-normal locations due to the dependence of the lower integral limit on wall-normal distance. The upper limit of the integral is set as the inviscid core where the magnitude of the integrand is negligible.

\begin{figure}
    \centering
    \subfigure[\label{UV_IntStarScale_APG}]{\includegraphics[width=0.49\textwidth]{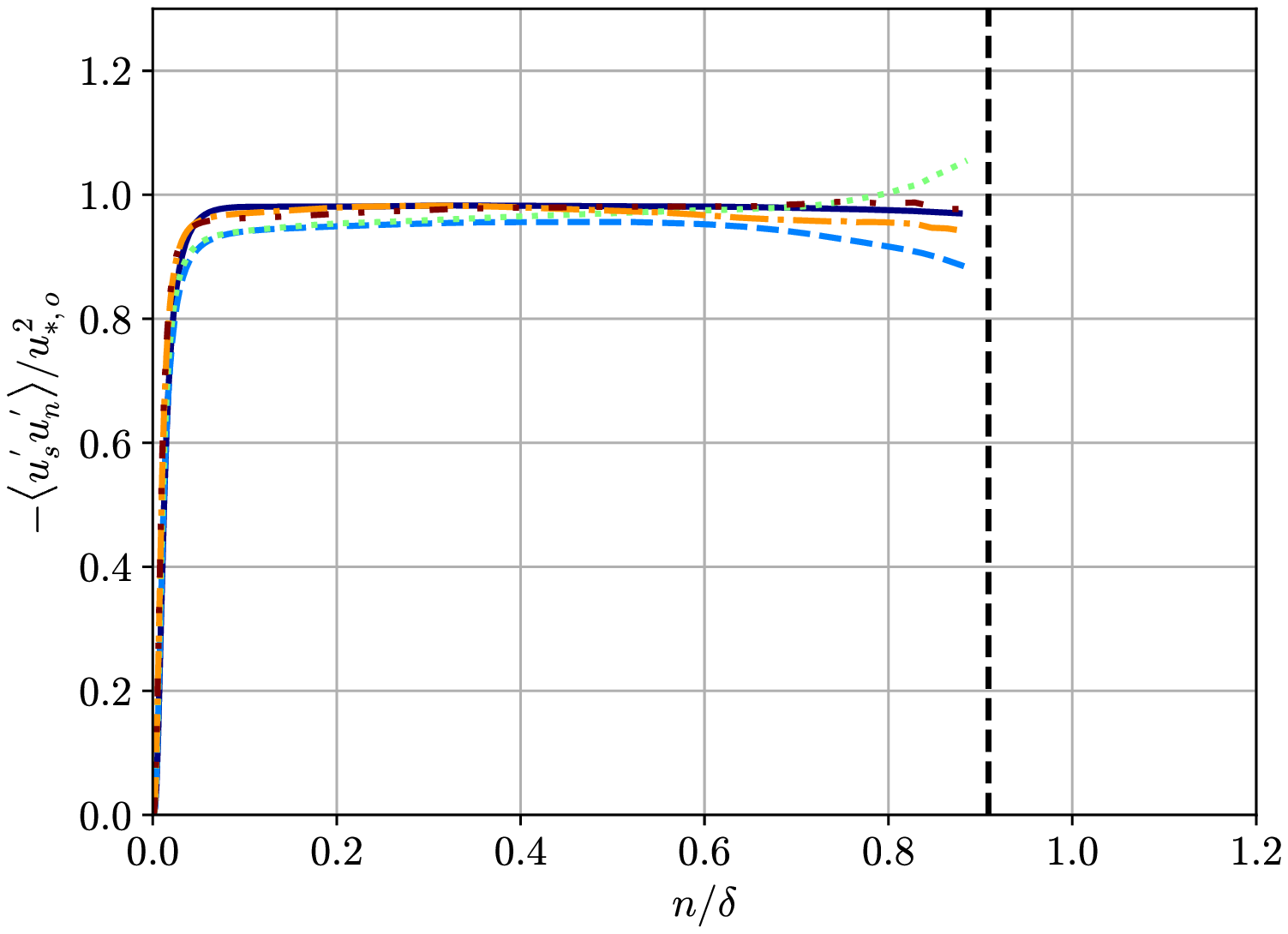}}
    \subfigure[\label{UV_IntStarStarScale_APG}]{\includegraphics[width=0.49\textwidth]{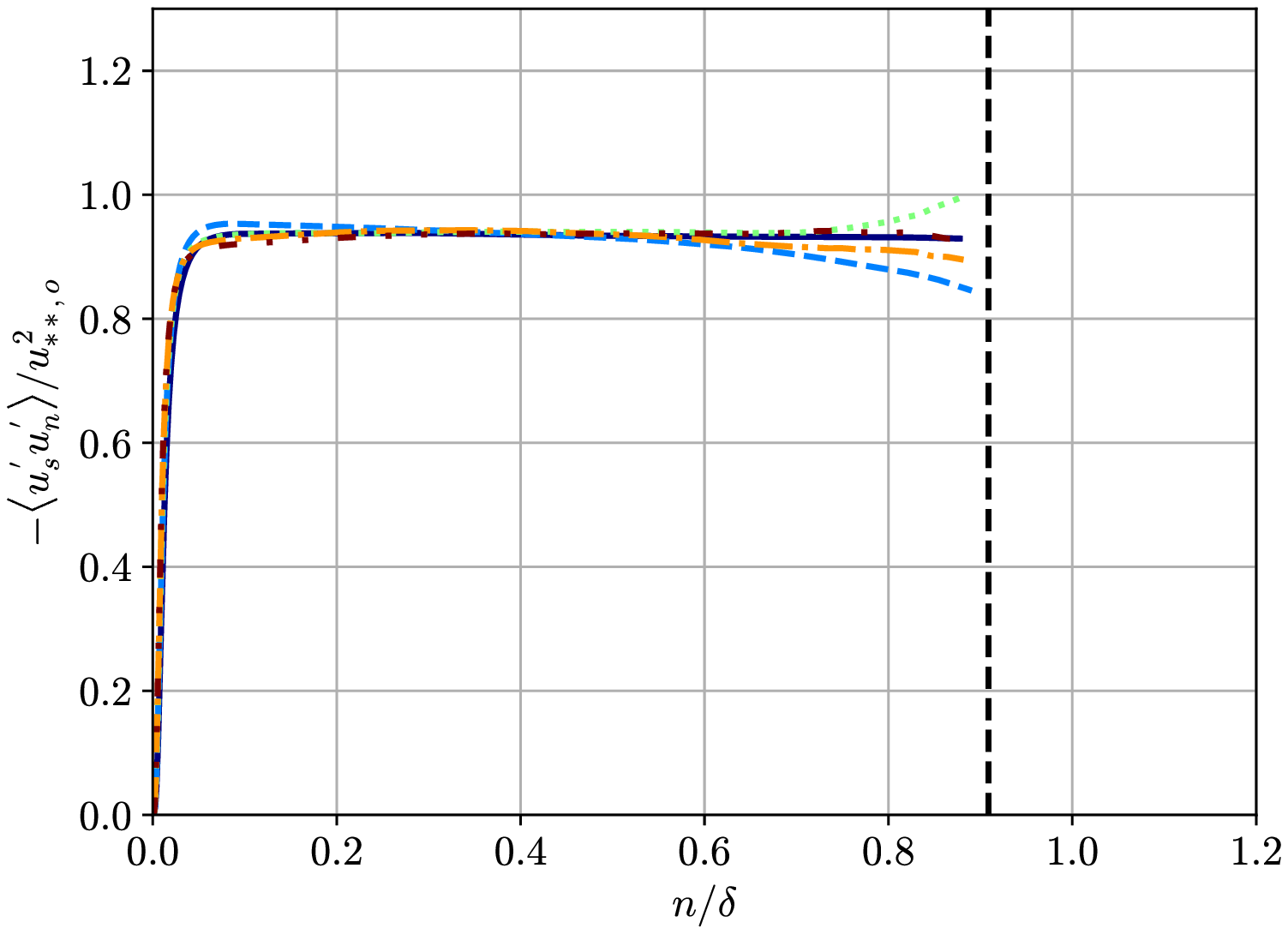}}    
    \caption{Reynolds shear stress in the APG region of the flow scaled by (a) $u_{*,o}$ and (b) $u_{**,o}$. Vertical lines indicate that the quantity asymptotically goes to $\pm \infty$. Line types for APG region at $x/L$: ${\color{Violet}\bm{-}}$, $-0.5$; ${\color{Cerulean}{\bm{--}}}$, $-0.45$; ${\color{green}{\bm{\cdot}}}$, $-0.39$; ${\color{orange}{\bm{-\cdot-}}}$, $-0.35$; ${\color{Mahogany}{\bm{-\cdot\cdot-}}}$, $-0.3$.}
    \label{fig:UV_Outer_ScaleComp_APG}
\end{figure}

\begin{figure}
    \centering
    \subfigure[\label{UV_IntStarScale_FPG}]{\includegraphics[width=0.49\textwidth]{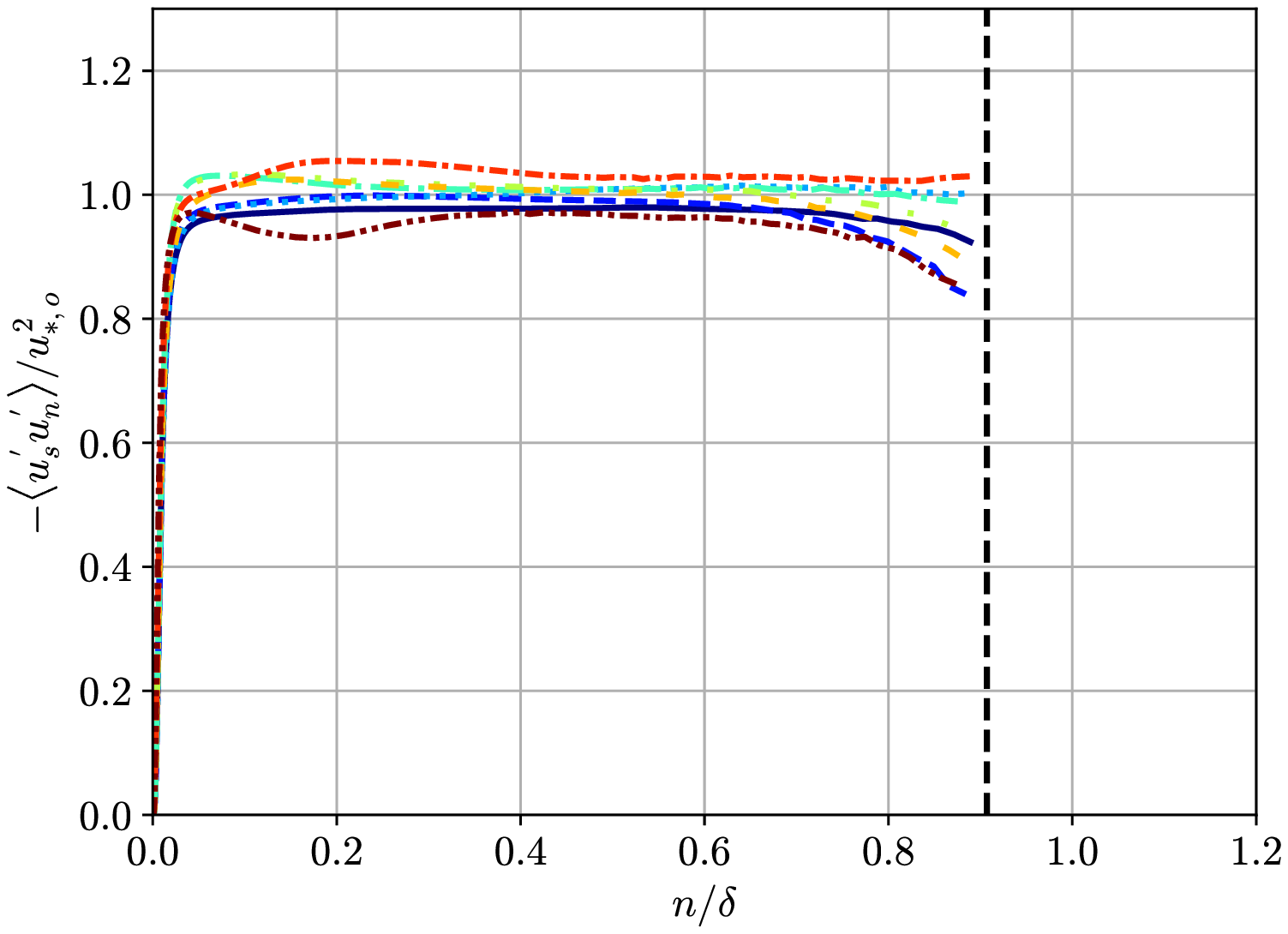}}
    \subfigure[\label{UV_IntStarStarScale_FPG}]{\includegraphics[width=0.49\textwidth]{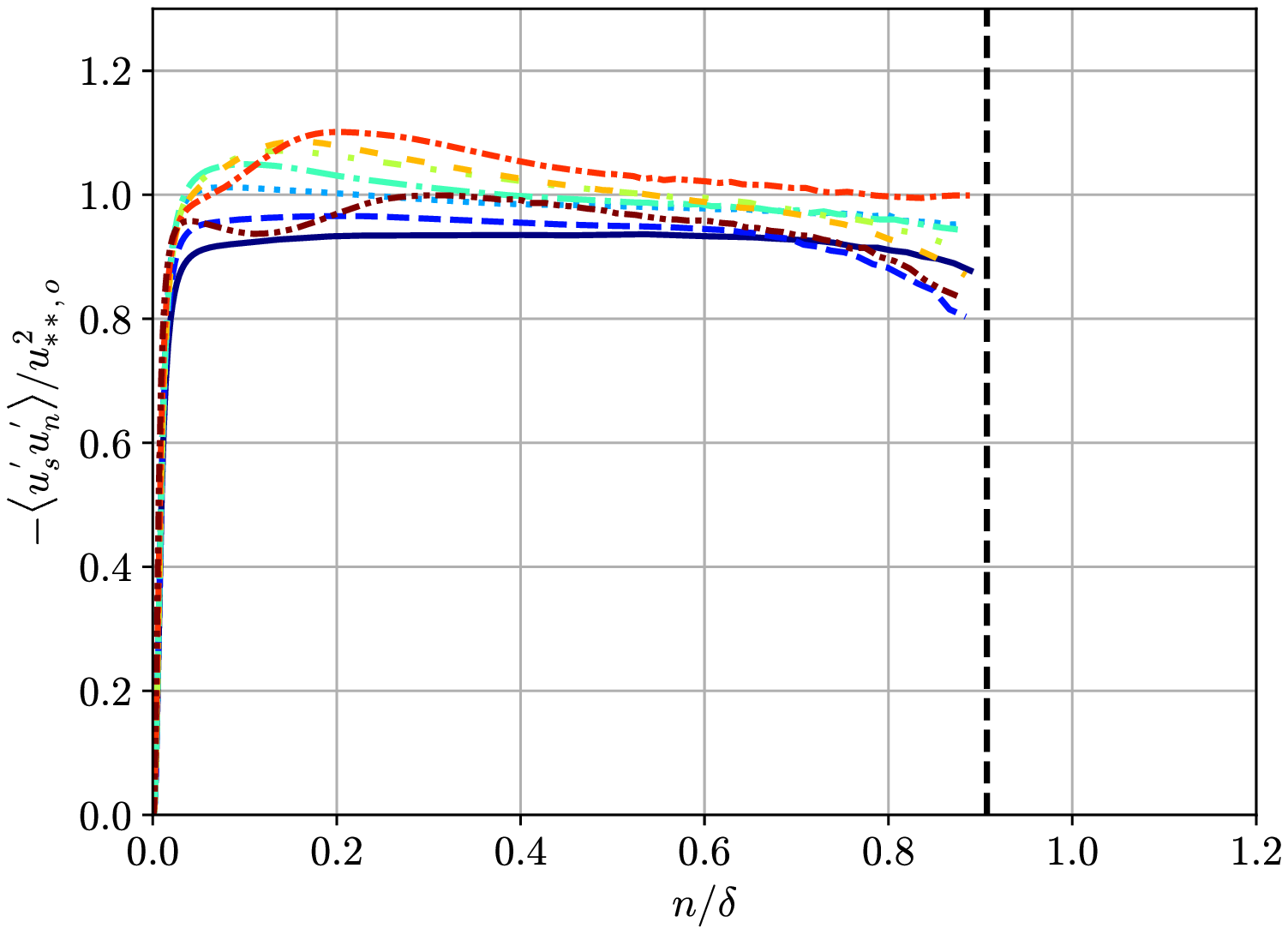}}    
    \caption{Reynolds shear stress in the FPG region of the flow scaled by (a) $u_{*,o}$ and (b) $u_{**,o}$. Vertical lines indicate that the quantity asymptotically goes to $\pm \infty$. Line types for FPG region at $x/L$: ${\color{Violet}{{\bm{-}}}}$, $-0.29$; ${\color{RoyalBlue}{{\bm{--}}}}$, $-0.25$; ${\color{Cyan}{\bm{\cdot}}}$, $-0.2$; ${\color{SeaGreen}{\bm{-\; \cdot\; -}}}$, $-0.15$; ${\textcolor{green}{\bm{-\;\;\cdot\;\;\cdot\;\;-}}}$, $-0.1$ ;
    ${\color{YellowOrange}{\bm{-\;\;- }}}$, $-0.079$ ; ${\color{RedOrange}{\bm{-\cdot-}}}$, $-0.05$ ; ${\color{Mahogany}{\bm{-\cdot\cdot-}}}$, $-0.006$. }
    \label{fig:UV_Outer_ScaleComp_FPG}
\end{figure}

The Reynolds shear stresses scaled by $u_{*,0}$ and $u_{**,o}$ in the APG region are shown in \figref{UV_Outer_ScaleComp_APG}. The profiles are plotted until $n/\delta = 0.9$ to avoid singularities observed at $n/\delta \approx 1.0$ as the scaling decrease to zero faster than the decrease in the Reynolds shear stresses. Both scalings result in a good collapse of stresses at different $x/L$ locations. Scaled Reynolds shear stresses shown in \figref{UV_Outer_ScaleComp_FPG} show a better collapse than the Zagarola-Smits scaling results shown in \figref{RSS_OuterScaling_FPG}. We observe that $u_{*,o}$ scaling collapses the Reynolds shear stress much better than the $u_{**,o}$  scaling indicating the influence of $\frac{\p \overline{u}^2_{\psi}}{\p \psi}$ is non-negligible, unlike many other simpler boundary layer flows. From a turbulence modeling perspective, the term $\frac{\p \overline{u^2_{\psi}}}{\p \psi}$ is generally unknown and often neglected while simplifying boundary layer equations. This term is included for RANS modeling but is frequently approximated using the linear eddy-viscosity hypothesis. The non-negligible influence of $\frac{\p \overline{u}^2_{\psi}}{\p \psi}$ indicates that the accurate modeling of this term will be required to obtain a better approximation of the flow solution.

\begin{figure}
    \centering
    \subfigure[\label{VV_IntScale_APG}]{\includegraphics[width=0.49\textwidth]{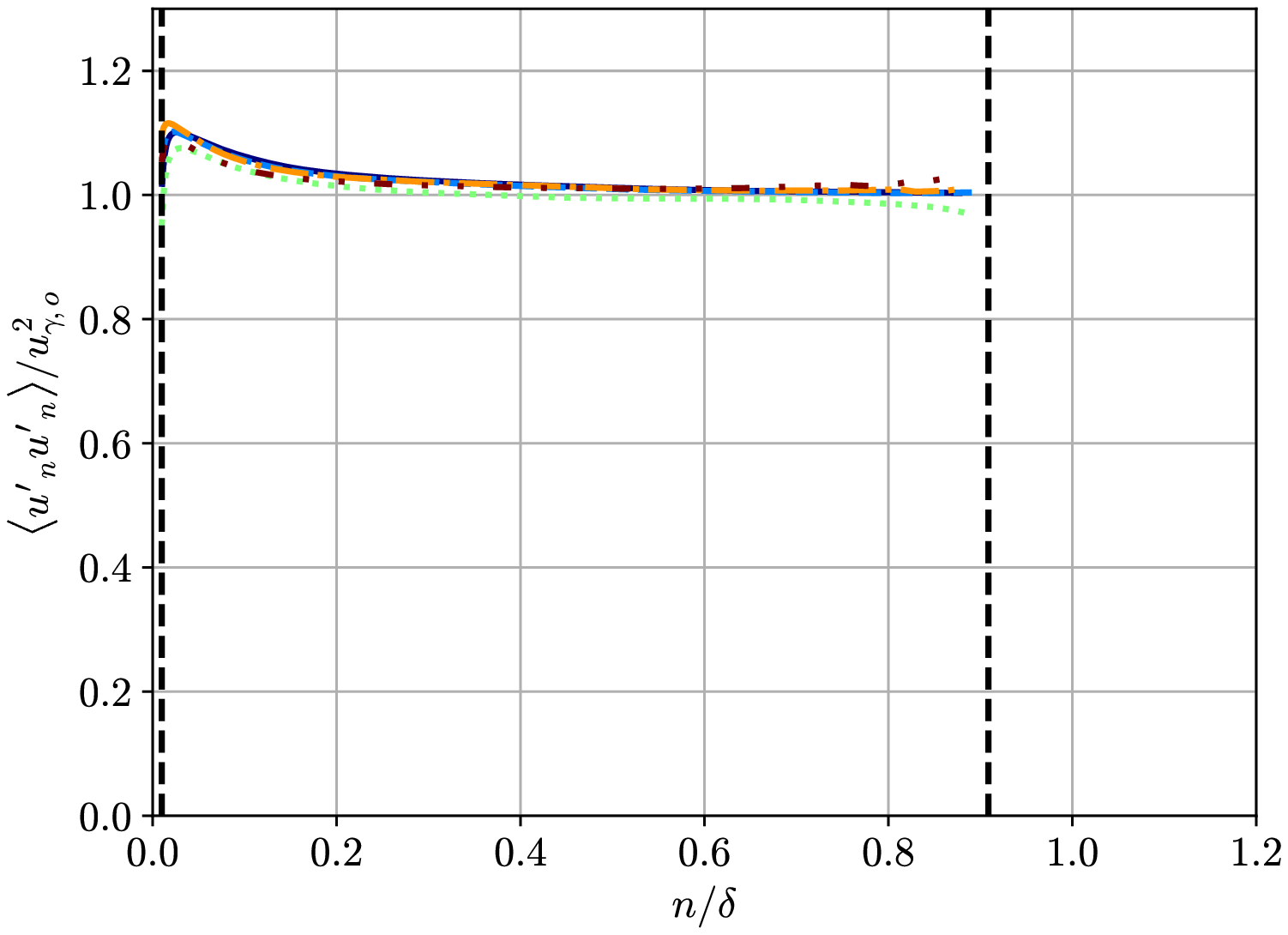}}
    \subfigure[\label{VV_IntScale_FPG}]{\includegraphics[width=0.49\textwidth]{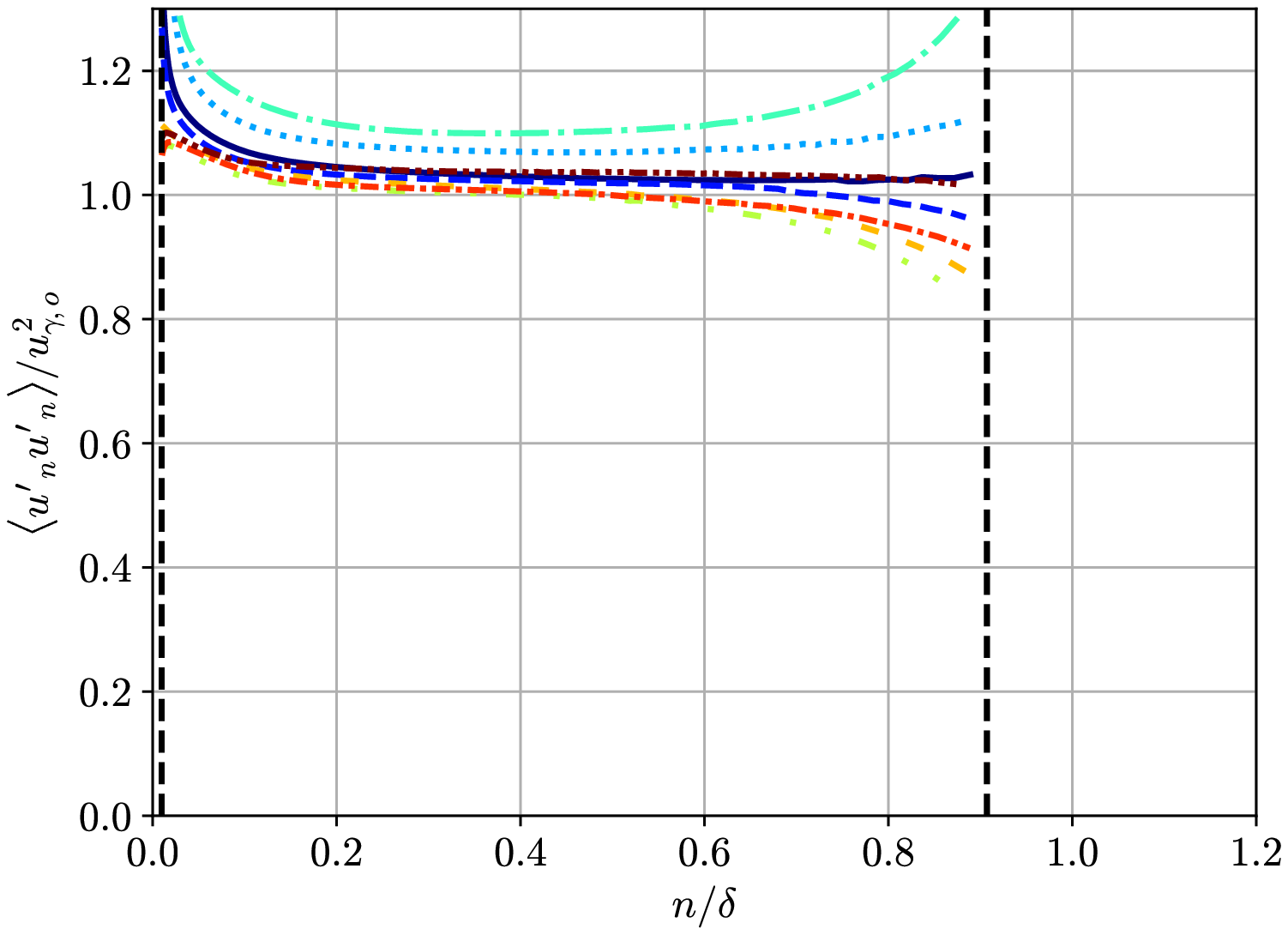}}    
    \caption{Wall-normal normal stress in the (a) APG region and (b) FPG region of the flow scaled by $u_{\gamma,o}$. Vertical lines indicate that the quantity asymptotically goes to $\pm \infty$. Line types for APG region at $x/L$: ${\color{Violet}\bm{-}}$, $-0.5$; ${\color{Cerulean}{\bm{--}}}$, $-0.45$; ${\color{green}{\bm{\cdot}}}$, $-0.39$; ${\color{orange}{\bm{-\cdot-}}}$, $-0.35$; ${\color{Mahogany}{\bm{-\cdot\cdot-}}}$, $-0.3$. Line types for FPG region at $x/L$: ${\color{Violet}{{\bm{-}}}}$, $-0.29$; ${\color{RoyalBlue}{{\bm{--}}}}$, $-0.25$; ${\color{Cyan}{\bm{\cdot}}}$, $-0.2$; ${\color{SeaGreen}{\bm{-\; \cdot\; -}}}$, $-0.15$; ${\textcolor{green}{\bm{-\;\;\cdot\;\;\cdot\;\;-}}}$, $-0.1$ ;
    ${\color{YellowOrange}{\bm{-\;\;- }}}$, $-0.079$ ; ${\color{RedOrange}{\bm{-\cdot-}}}$, $-0.05$ ; ${\color{Mahogany}{\bm{-\cdot\cdot-}}}$, $-0.006$. }
    \label{fig:VV_Outer_ScaleComp}
\end{figure}

A similar integral analysis conducted with the $\phi$-momentum equation results in the scaling,
\begin{equation}
    \frac{\overline{u^2_{\phi}}}{{u_{\gamma,o}^2}} \approx \frac{\overline{u^2_{n}}}{{u_{\gamma,o}^2}} \approx 1, \; u_{\gamma,o} = \int_{n}^{\infty} \Big( \frac{\bar{u}_{\psi}^2}{R} + \frac{1}{\rho} \frac{\p \bar{p}}{\p \phi} \Big) d n',
\end{equation}
\noindent where the upper limit of the integral is in the inviscid core. The results obtained using this scaling are shown in \figref{VV_Outer_ScaleComp}. We observe a singularity at $n/\delta \approx 0$ and $n/\delta \approx 1.0$ as the scaling reduces to zero faster than the decrease in the wall-normal normal stress. In addition, we observe that the scaling collapses the profiles well in the APG region. In the FPG region, the scaling collapses the profiles 
better than the Zagarola-Smits scaling (shown in \figref{RSS_OuterScaling_FPG}). However, this collapse is not as close as in the APG region, indicating that neglected budget terms in this calculation can combine to have first-order effects that are not included in the above scaling. 

The scalings obtained through the integral analysis for the inner and outer region of the turbulent boundary layer cannot be directly used to model near-wall turbulence behavior as it does not solve the closure problem and requires an accurate solution of the quantities until the point where the scaling is applied. However, the integral scaling analysis indicates which terms are relevant and essential to be included in the modeling. For specific scenarios, such as when advection and pressure gradient terms are small, these terms can be neglected, and the scaling reduces to the previously proposed scalings for Reynolds shear stresses. On the other hand, for complex flows such as the bump flow, where these terms cannot be neglected, appropriate empirical relations such as the log-linear behavior of advection could be invoked and used to develop a scaling model.

\section{Conclusions}
\label{sec:Conclusion}

Characterizing turbulent boundary layers is immensely important for understanding and modeling aerospace and geophysical flows. Turbulent boundary layers with pressure gradient and curvature effects are particularly interesting to the aerospace industry due to their relevance for smooth body flow separation, which could result in massive aerodynamic drag. The Gaussian (Boeing) bump is a recent benchmark problem specifically defined to replicate this flow behavior to get fundamental insights on the behavior of velocity and Reynolds stresses leading to flow separation. 

This article discusses insights on turbulent boundary layers in the presence of pressure gradient and curvature effects based on direct numerical simulation of the turbulent boundary layer over the Gaussian bump at $Re_L = 2$ million. The simulation results are compared to another DNS \citep{Uzun2021b} with a different simulation setup. Both agreements and differences in the results for these two DNSs are analyzed. The characteristics of the turbulent boundary layer are also compared to another DNS of the flow over the same geometry but at a lower Reynolds number ($Re_L = 1$ million), and differences between the flow physics are assessed. Momentum budgets for the flow are analyzed in a streamline-aligned coordinate system, and pressure gradient and curvature effects are isolated. These results simplify the momentum equations that are used for integral analysis. Integral analysis is used to formulate non-local Reynolds stress scalings in the inner and outer regions of the turbulent boundary layer. These scalings showed a great collapse of the stress profiles in the mild APG region. In contrast, in the strong FPG region, the collapse of profiles is not perfect but it was observed to be much better than the traditional scaling based on friction velocity. The integral analysis-based scalings suggest that relevant fluid flow dynamics must be accounted for in developing new scalings and modeling turbulent flow behavior in the presence of strong pressure gradients and curvature. 

\section{Acknowledgements}
\label{sec:acknow}
The authors would like to acknowledge the Transformational Tools and Technologies Project of the National Aeronautics and Space Administration (NASA) 80NSSC18M0147 for funding this work. Moreover, this research used computational resources of the NASA High-End Computing (HEC) Program through the NASA Advanced Supercomputing (NAS) Division at Ames Research Center and the Argonne Leadership Computing Facility, a DOE Office of Science User Facility supported under Contract DE-AC02-06CH11357. In addition, the authors thank Dr. Phillipe Spalart and Dr. Christopher Rumsey for their insightful comments and discussions. Finally, the authors thank Dr. Ali Uzun for providing reference DNS data for comparison.

\appendix 
\section {Comparison with results in \cite{Uzun2021b}}
\label{sec:AppA}

\subsection{Differences in simulation setup}
\label{sec:Setup_partc}

This section highlights the differences in the simulation setup compared to the DNS in \cite{Uzun2021b}. Even though the bump flow geometry and Reynolds number are the same, key differences between these two DNS arise from the following reasons: 1) Solving either incompressible or compressible Navier-Stokes equations, 2) Differences in the length of the spanwise domain ($0.156L$ vs. $0.08L$), 3) STG vs recycling inflow, and 4) Selection of different boundary condition for the top surface. To understand the impact of 1), we conducted compressible RANS simulations at the same Mach number used in \cite{Uzun2021b}. We did not observe significant differences in the mean flow profiles, coefficients of skin friction, and pressure between the incompressible and compressible RANS simulations. Although these results are not shown in the article for brevity, these results indicate that the effect of compressibility is small enough, and the solution of either incompressible or compressible Navier-Stokes equations should not affect the solution. 
In this article, we only investigate the region upstream of flow separation. In that region, using a larger domain is only expected to improve the convergence of statistics and not affect fully converged statistics. Likewise, though different, the inflow turbulence generation techniques had similar success. The most crucial difference between these two DNS is the selection of different boundary conditions at the top surface. As mentioned in Section \ref{sec:Setup_parta}, we account for the presence of a wall at the top surface of the domain, whereas the DNS in \cite{Uzun2021b} uses freestream air as the boundary condition on the top surface placed at $y/L = 1.0$. These differences in simulation setup appear to influence the DNS results. Therefore, it is of value to quantify the differences between these two DNS conducted for the same geometry and Reynolds number. As it is computationally infeasible to perform another DNS to understand these differences, we run a suite of RANS simulations to investigate the influence of these differences. In particular, we conduct RANS simulation with three different domains/boundary conditions: 1) the domain and boundary conditions are the same as the preliminary RANS simulation described in \ref{sec:Setup_parta}, 2) the domain and boundary conditions are the same as the DNS and 3) the domain and boundary conditions are similar to the preliminary RANS simulation, with the location of the top surface at $y/L = 1.0$, and freestream air boundary condition is prescribed at that surface.

\begin{figure}
    \centering
    \subfigure[\label{fig:Cp_RANS_comp}]{\includegraphics[width=0.49\textwidth]{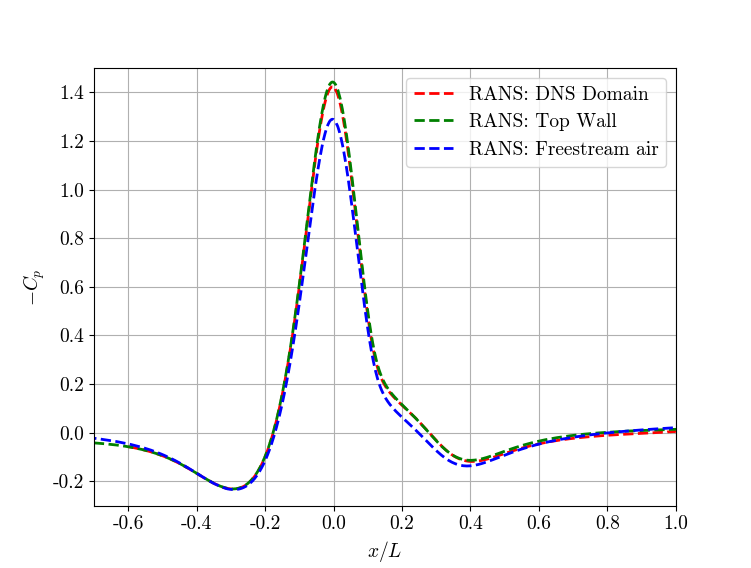}}
    \subfigure[\label{fig:Cf_RANS_comp}]{\includegraphics[width=0.49\textwidth]{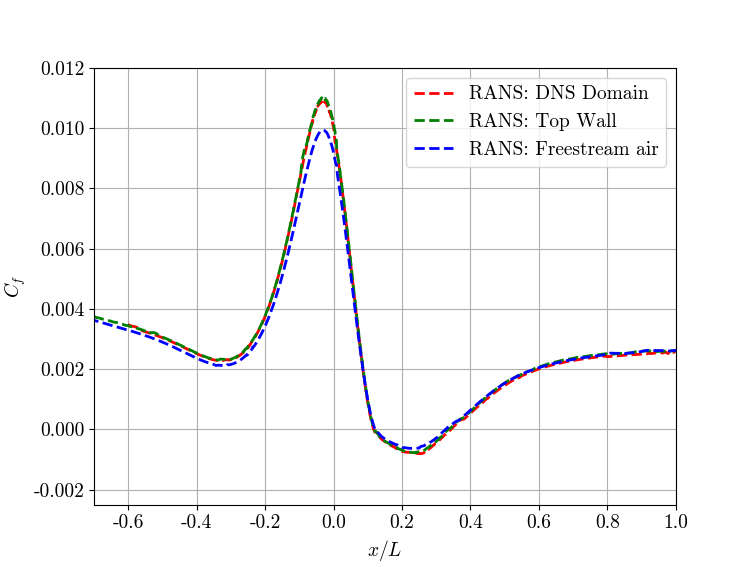}}
    \caption{(a) Coefficient of pressure ($C_p$) and (b) Skin-friction coefficient ($C_f$) for different RANS simulations.}
    \label{fig:RANS_comp}
\end{figure}

We compare the coefficient of pressure and skin-friction coefficient for different RANS simulations in \figref{RANS_comp}. We observe a clear difference in the $C_p$ at the bump peak between these RANS simulations. This indicates the effect of the top wall in decreasing $C_p$ at the bump peak compared to using freestream air boundary conditions at the top surface. A wall at the top surface suppresses the outward flow, which leads to a further reduction in pressure and an increase in the suction peak. Furthermore, differences in $C_f$ are observed for these two boundary conditions. Using a freestream air boundary condition appears to reduce skin friction compared to using a wall at the top surface. This difference in $C_f$ is about $5 \%$ in the mild APG region, and then it quickly rises close to the bump peak at $x/L = 0$. Similar differences were observed for $k-\omega$ SST RANS simulations as shown in \cite{Prakash2022}.  While it may seem odd to use RANS to confirm differences in the DNS, the pressure distribution from an attached flow is fully set by the height of the geometry plus the displacement thickness.  Even if RANS has errors, its relative difference due to boundary conditions closely matches the difference in the two DNS.  Further, while we would not trust the RANS prediction of the $C_f$ peak, its offset on the very weak pressure gradient $C_f$ also matches the DNS offset.  We regard this difference as a good result-- that both DNS are accurate representations of two flows with slightly different pressure gradients.  We have quantified  the pressure gradients in our confined case as 9.3\% more favorable and about 3.4\% more adverse than the free-air DNS. We expect that there will be a significant benefit to having two close cases for data-driven models that might be trained on one and tested on the other to assess the ability to handle this modest difference from a relatively small change in boundary conditions.

\begin{figure}
    \centering
    \subfigure{\includegraphics[width=0.32\textwidth]{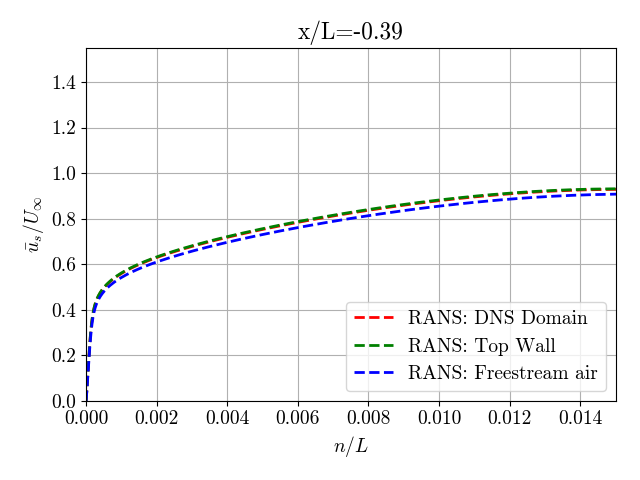}}
    \subfigure{\includegraphics[width=0.32\textwidth]{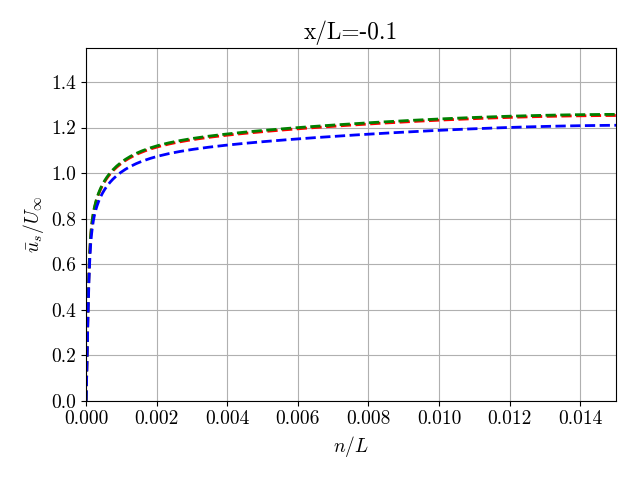}}    
    \subfigure{\includegraphics[width=0.32\textwidth]{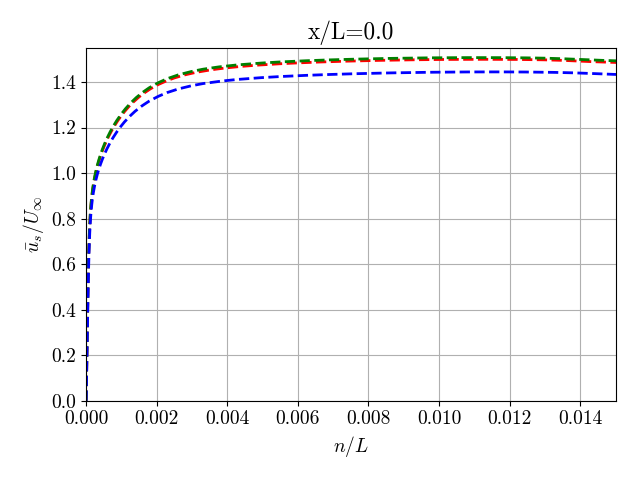}}
    
    \caption{Comparison of velocity profiles at several $x/L$ stations for different RANS simulations.}
    \label{fig:Up_RANS_comp}
\end{figure}

We compare the velocity profiles from different RANS simulations at various $x/L$ stations in \figref{Up_RANS_comp}. We observe that using a freestream air boundary condition at the top surface reduces the streamwise velocity at the core. Similar differences were also observed for Reynolds stress profiles, but the results are not discussed here for brevity. These results highlight that the selection of the top surface boundary condition could significantly influence the DNS results. Furthermore, the RANS simulation using the DNS domain and boundary conditions gives the same coefficient of pressure, skin-friction coefficient, and velocity profiles compared to the preliminary RANS simulation. This gives fidelity to selecting slip boundary conditions at the top surface instead of attempting to resolve the wall at the top surface.

\subsection{Velocity profiles}

\begin{figure}
    \centering
    \includegraphics[width=0.5\textwidth]{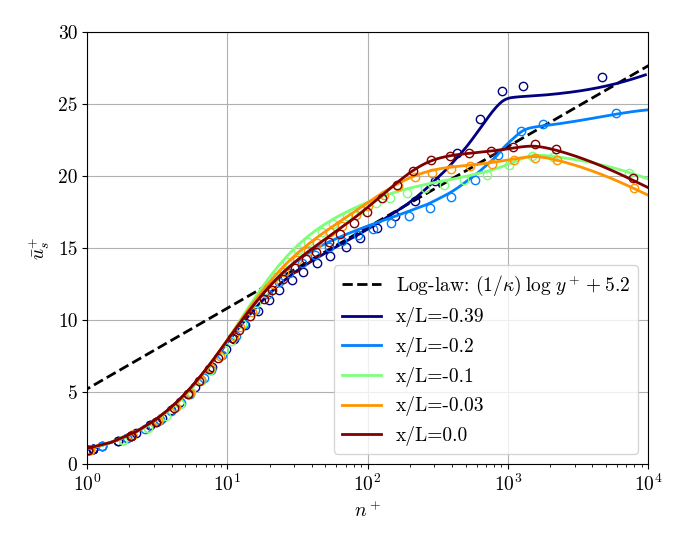}
    \caption{Velocity profiles at different $x/L$ stations. The DNS results in \cite{Uzun2021b} are shown in circles and our DNS results are shown in solid lines. Different colors denote different x/L locations: $x/L = -0.39$ is shown in dark blue, $x/L = -0.2$ is shown in light blue, $x/L = -0.1$ is shown in green, $x/L = -0.03$ is shown in orange and $x/L = 0.0$ is shown in red.}
    \label{fig:Up_DNS_scale}
\end{figure}

In the previous section, we observed that the two DNS exhibit differences in the coefficient of forces. The differences in simulation setup also result in variations in velocity and stress profiles between the two DNS runs. As the primary differences lie in the $C_f$ distribution, we can account for the offset by scaling the velocity and stress profiles accordingly. The velocity profiles scaled by local inner-layer units are shown in \figref{Up_DNS_scale}. This scaling results in an excellent agreement of the velocity profiles for these two DNS. As these two DNS were conducted independently with different codes and turbulent flow inflow generation techniques, these results add to the confidence of the simulations and results presented in this article and the one in \cite{Uzun2021b}. As discussed by the authors, the effect of pressure gradients is observed for the velocity profile indicated by the departure from the log-law. In the APG region, at $x/L = -0.39$, the velocity profile obeys the log-law until $n^+ = 200$ and departs from the log-law at a larger distance from the wall. The resilience of the log-law, as discussed in \cite{Johnstone2010}, is observed for the velocity profile for $x/L \leq -0.39$. The other downstream $x/L$ locations are in the FPG region and show a significant departure from the log-law in the inner region of the boundary layer. 


\

\subsection{Reynolds stress profiles}

\begin{figure}
    \centering
    \subfigure{\includegraphics[width=0.49\textwidth]{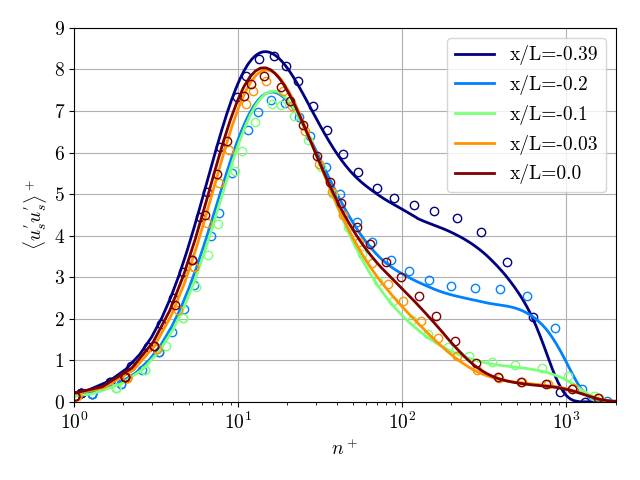}}
    \subfigure{\includegraphics[width=0.49\textwidth]{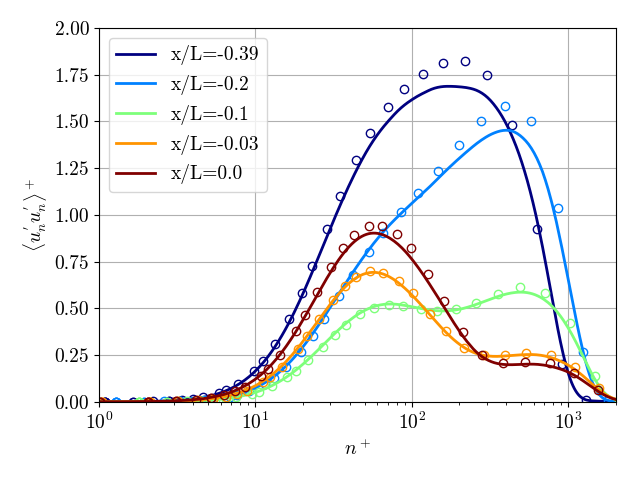}}
    
    \subfigure{\includegraphics[width=0.49\textwidth]{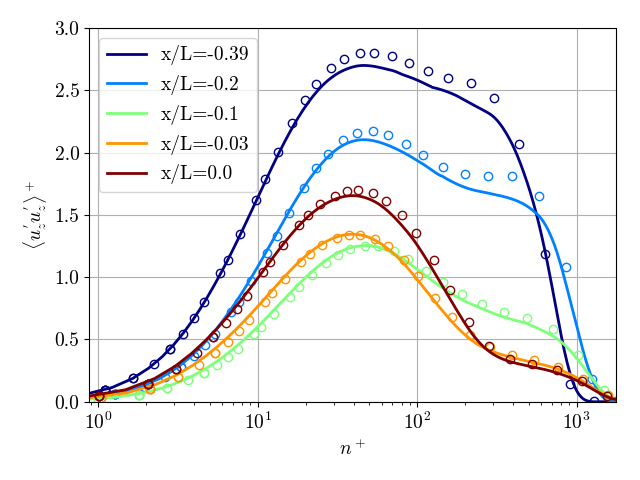}}    
    \subfigure{\includegraphics[width=0.49\textwidth]{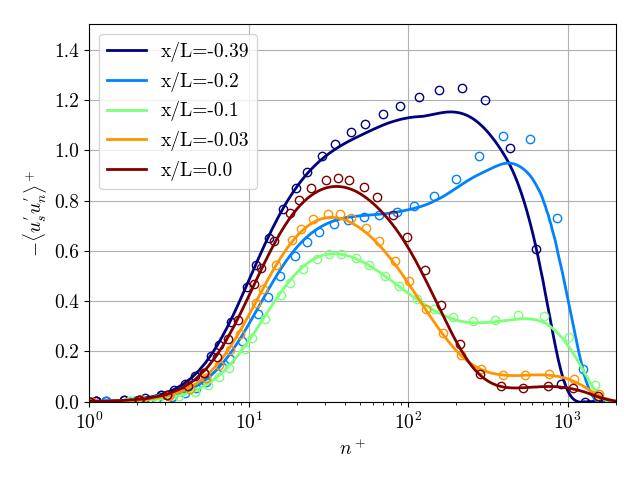}}    
    \caption{Comparison of DNS Reynolds stresses profiles at several $x/L$ stations. The DNS results in \cite{Uzun2021b} are shown in circles and our DNS results are shown in solid lines. The colors denote different $x/L$ stations: $x/L = -0.39$ is shown in dark blue, $x/L = -0.2$ is shown in light blue, $x/L = -0.1$ is shown in green, $x/L = -0.03$ is shown in orange and $x/L = 0.0$ is shown in red.}
    \label{fig:UiUj_DNS_scale}
\end{figure}

In \figref{UiUj_DNS_scale}, we compare the components of Reynolds stress tensor scaled by friction velocity to those shown in \cite{Uzun2021b}. We observe that the scaled Reynolds stresses from both DNS match well for all the stress components. Some differences in the outer region peak of Reynolds stresses are observed which could be due to the failure of friction velocity scaling to collapse the results in that specific region.

\bibliographystyle{jfm}
\bibliography{main.bbl}

\end{document}